\documentclass[usenatbib,useAMS]{mnras} 

\usepackage{graphicx}	
\usepackage{amsmath}	
\usepackage{amssymb}	
\usepackage{multicol}   
\usepackage{pdflscape}	
\usepackage{rotating}   
\usepackage{caption}    
\usepackage{longtable}  
\usepackage[normalem]{ulem} 

\usepackage{color}
\usepackage{wasysym}

\definecolor{darkgreen}{rgb}{0.0,0.75,0.0}

\newcommand{\tn}[1]{\textnormal{#1}}

\newcommand{\nh}{N$_{\rm HI}$}
\newcommand{\hi}{\ion{H}{i}}
\newcommand{\ha}{H$\alpha$}
\newcommand{\hb}{H$\beta$}
\newcommand{\oii}{[\ion{O}{ii}]$\lambda\lambda$3726,3729}
\newcommand{\oiiiaur}{[\ion{O}{iii}]$\lambda$4363}
\newcommand{\oiii}{[\ion{O}{iii}]$\lambda\lambda$4959,5007}
\newcommand{\oiiia}{[\ion{O}{iii}]$\lambda$4959}
\newcommand{\oiiib}{[\ion{O}{iii}]$\lambda$5007}
\newcommand{\nii}{[\ion{N}{ii}]$\lambda\lambda$6549,6584}
\newcommand{\niia}{[\ion{N}{ii}]$\lambda$6549}
\newcommand{\niib}{[\ion{N}{ii}]$\lambda$6584}
\newcommand{\sii}{[\ion{S}{ii}]$\lambda\lambda$6717,6731}
\newcommand{\hii}{\mbox{H~{\sc ii}}} 

\newcommand{\Te}{$T_e$}

\newcommand{\logOH}{$12+\log({\rm O/H})$}
\newcommand{\Rtwo}{$R_2$}
\newcommand{\Rthree}{$R_3$}
\newcommand{\Rtwothree}{$R_{23}$}
\newcommand{\Othreetwo}{O$_{32}$}
\newcommand{\Othree}{O$_3$}
\newcommand{\Ntwo}{N$_2$}
\newcommand{\OthreeNtwo}{O$_3$N$_2$}
\newcommand{\NtwoStwo}{N$_2$S$_2$H$\alpha$}

\usepackage[T1]{fontenc}
\usepackage{ae,aecompl}

\usepackage{newtxtext,newtxmath}

\title[Absorption vs emission line metallicities]{Comparing emission- and absorption-based gas-phase metallicities in GRB host galaxies at $z=2-4$ using JWST}

\author[P. Schady et al.]{P.~Schady$^1$\thanks{E-mail: \href{mailto:p.schady@bath.ac.uk}{p.schady@bath.ac.uk}},
R.~M.~Yates,$^2$ 
L.~Christensen,$^{3,4}$
A.~De~Cia,$^{5,6}$
A.~Rossi,$^7$
V.~D'Elia,$^{8,9}$
K.~E.~Heintz,$^{3,4}$\newauthor
P.~Jakobsson,$^{10}$
T.~Laskar,$^{11}$
A.~Levan,$^{12}$
R.~Salvaterra,$^{13}$
R.~L.~C.~Starling,$^{14}$
N.~R.~Tanvir,$^{14}$
C.~C.~Th\"one,$^{15}$\newauthor
S.~Vergani,$^{16,17}$
K.~Wiersema,$^{2}$
M.~Arabsalmani,$^{18,19}$
H.-W.~Chen,$^{20}$
M.~De Pasquale,$^{21}$
A.~Fruchter,$^{22}$\newauthor
J.~P.~U.~Fynbo,$^{3,4}$
R.~Garc\'ia-Benito,$^{23}$
B.~Gompertz,$^{24,25}$
D.~Hartmann,$^{26}$
C.~Kouveliotou,$^{27}$\newauthor
B.~Milvang-Jensen,$^{3,4}$
E.~Palazzi,$^{7}$
D.A.~Perley,$^{28}$
S.~Piranomonte,$^{9}$
G.~Pugliese,$^{29}$
S.~Savaglio,$^{30,7,31}$\newauthor
B.~Sbarufatti,$^{32}$
S.~Schulze,$^{33}$
G.~Tagliaferri,$^{32}$
A.~de~Ugarte~Postigo,$^{34}$
D.~Watson,$^{3,4}$
P.~Wiseman$^{35}$\\
(Affiliations can be found after the references)
}

\date{}

\pubyear{}

\begin{document}
\label{firstpage}
\pagerange{\pageref{firstpage}--\pageref{lastpage}}
\maketitle

\begin{abstract}
Much of what is known of the chemical composition of the universe is based on emission line spectra from star forming galaxies. Emission-based inferences are, nevertheless, model-dependent and they are dominated by light from luminous star forming regions. An alternative and sensitive probe of the metallicity of galaxies is through absorption lines imprinted on the luminous afterglow spectra of long gamma ray bursts (GRBs) from neutral material within their host galaxy. We present results from a JWST/NIRSpec programme to investigate for the first time the relation between the metallicity of neutral gas probed in absorption by GRB afterglows and the metallicity of the star forming regions for the same host galaxy sample. Using an initial sample of eight GRB host galaxies at $z=2.1-4.7$, we find a tight relation between absorption and emission line metallicities when using the recently proposed $\hat{R}$ metallicity diagnostic ($\pm0.2$~dex). This agreement implies a relatively chemically-homogeneous multi-phase interstellar medium, and indicates that absorption and emission line probes can be directly compared. However, the relation is less clear when using other diagnostics, such as \Rtwothree\ and \Rthree. We also find possible evidence of an elevated N/O ratio in the host galaxy of GRB~090323 at $z=3.58$, consistent with what has been seen in other $z>4$ galaxies. Ultimate confirmation of an enhanced N/O ratio and of the relation between absorption and emission line metallicities will require a more direct determination of the emission line metallicity via the detection of temperature-sensitive auroral lines in our GRB host galaxy sample.
\end{abstract}


\section{Introduction}
The chemical enrichment of galaxies across cosmic time encodes vital information on galaxy evolution, tracing the successive episodes of star formation that synthesise and recycle metals back into the galactic interstellar medium (ISM). Galaxy-scale outflows and accretion of pristine gas further redistributes and dilutes enriched material. Tracing the metallicity of the multi-phase ISM of galaxies thus enables the relative importance of these competing processes in enriching a galaxy to be studied.

The majority of gas-phase metallicity measurements of galaxies are based on emission lines, which trace the ionised gas within star forming regions. In such a case, the most direct method available to trace the gas phase metallicity is using metal recombination lines \citep[e.g.][]{ost89,Peimbert+93}, which are relatively unaffected by temperature fluctuations. However, they are extremely faint ($\sim{}10^{3}$ times fainter than the hydrogen recombination line, H$\beta$), limiting this method to only the highest-resolution spectra of nearby systems \citep[e.g.][]{Esteban+09,Esteban+14}. Alternatively, measurements of the electron temperature ($T_{\rm e}$) can also provide (semi-)direct metallicity estimates \citep[e.g.][]{Peimbert67,ost89}. This method requires metal auroral lines such as \oiiiaur{}, which although still faint ($\sim{}10^{2}$ times fainter than H$\beta$), are ten times brighter than metal recombination lines and are thus detectable in a wider range of systems in the nearby Universe, or in gravitationally lensed galaxies out to $z\approx 3.6$ \citep[e.g.][]{vcg+04,clr+12,ssk+16}. Oxygen auroral lines have now also been detected out to $z\sim{}8$ for a few galaxies with the \textit{James Webb Space Telescope} \citep[JWST, e.g.][]{Schaerer+22,Arellano-Cordova+22,Curti+23,Trump+23,hbg+23,noi+23,rwh+23}, but these remain the exception.

In the absence of metal recombination or auroral lines, indirect metallicity estimates of star forming regions must be used, which are obtained via strong emission line ratios \citep[e.g.][]{kd02,pp04,pg16}. Calibrations for such diagnostics are now becoming possible at high redshift \citep[e.g.][]{lmc+23,hcs23,sst+23}. However, strong-line metallicity estimates are known to vary by up to $\sim{}0.6$ dex depending on the line ratios chosen \citep{ke08, tjs+21}. Moreover, all such emission line methods are luminosity-weighted tracers of the star forming regions of galaxies, which at $z>2$ contain just 20\% of the baryon fraction \citep{fp04,bcw10,msd+10,pwt+14}.

A very different but complementary method of studying the cosmic build up of heavy elements is with absorption lines from the cold ISM, using the luminous light offered by background quasi-stellar objects (QSOs) and long gamma ray bursts (GRBs). Absorption from neutral hydrogen reveals copious quantities of neutral gas in these systems \citep[e.g.][]{tfd+19}, in the large majority of cases classifying them as damped Lyman-$\alpha$ (Ly~$\alpha$) absorbers (DLAs, defined as having $\log[N_{\rm HI}/{\rm cm}^{-2}]>20.3$; see \citealt{wgp05}), where ionisation corrections are negligible. Combining the neutral hydrogen abundance with the measured abundances of metals provides an accurate and largely model-independent measure of the neutral gas metallicity \citep{pgw+03,pcd+03,sff03,sav06,wgp05,fln+11,rwp+12,nwp+13,kfl+13}. Such data have enabled abundances to be measured out to $z>6$ \citep{kka+06,tfg+13,hmf+15,svd+23} and a few dex below what can be probed with emission lines \citep{ph20,wsb+17,blw+19,dls+13}. The combination of absorption and emission line probes can therefore provide a more complete understanding of the chemical enrichment of galaxies.

Extensive efforts have been made to identify the emission counterparts to QSO-DLAs, but the often large, projected offsets of tens of kpc between the QSO line of sight and the centre of the galaxy associated with the absorber (or `impact parameter') \citep[e.g][]{ckr+05,rtn+06,kfm+12,rpt+16} make it challenging to identify the intervening system in emission \citep[e.g.][]{fop+15}. Detecting emission counterparts at smaller impact parameters is also complex due to the bright light from the background QSO. Thus despite there now being on the order of a few hundred QSO-DLAs with measured absorption metallicities \citep{bep+15,dlm+16,dlp+18}, of these, the emission line metallicity (or limits) has only been reported for 20--30 QSO-DLA emission counterparts \citep{cmf+14,rpt+16,wpk+23}, most of which are at $z<2$. The measured emission line metallicities are generally larger than the absorption-based metallicities, although it remains unclear whether this offset is a result of a difference in the phase or in the location of the gas probed, or the presence of systematics. The QSO-DLA towards SBS~1544+5912 has an impact parameter of just 1~kpc \citep{rpt+16}, and in this case the emission and absorption line metallicities were consistent within the uncertainties \citep{srd+04,skm+05}.

\begin{table*}
\begin{minipage}[H]{1\textwidth}
\caption{GRB host galaxy sample and details on JWST/NIRSpec observations}\label{tab:obs}
\end{minipage}
\begin{tabular}{|lcccc}
\hline
\hline
GRB & E(B-V)$_{\rm Gal}$ & Grating/filter & On-source & Obs. date \\ 
 & & combination & exposure (s) & (DD-MM-YY) \\
\hline\hline
030323 & 0.042 & G140M/F100LP & 4204 & 13-06-23 \\
 & & G235M/F170LP & 1225 & \\
\hline
050505 & 0.019 & G235M/F170LP & 2101 & 18-03-24 -- 04-05-24$^\dag$ \\
 & & G395M/F290LP & 642 & \\
\hline
050820A$^\star$ & 0.039 & G140M/F100LP & 5310 & 28-11-22 \\
 & & G235M/F170LP & 905 & \\
\hline
080804 & 0.014 & G140M/F100LP & 934 & 28-10-22 \\
 & & G235M/F170LP & 525 & \\
\hline
090323 & 0.021 & G235M/F170LP & 934 & 19-06-23 \\
\hline
100219A & 0.066 & G235M/F170LP & 5952 & 20-01-24 \\
 & & G395M/F290LP & 934 & \\
\hline
120327A & 0.293 & G140M/F100LP & 934 & 07-03-23\\
 & & G235M/F170LP & 525 & \\
\hline
120815A & 0.099 & G140M/F100LP & 5952 & 24-08-23 \\
 & & G235M/F170LP & 1517 & \\
\hline
141109A & 0.032 & G140M/F100LP & 934 & 02-12-23 \\
 & & G235M/F170LP & 525 & \\
\hline
150403A$^\star$ & 0.047 & G140M/F100LP & 934 & 19-06-23 \\
 & & G235M/F170LP & 642 & \\
\hline

\end{tabular}
\begin{minipage}[H]{0.60\textwidth}
\vspace{-0.3cm}
$^\star$ Observed with the IFS\\
$^\dag$ Window given for when target due to be observed
\end{minipage}
\end{table*}

Unlike QSO-DLAs, long GRBs fade, enabling even faint host galaxies to be observed in emission. Furthermore, their association to the death of a massive star\footnote{The detection of a kilonova associated with the long GRB~211211A and GRB~230307A has shown that not all long GRBs are formed from the core collapse of a massive star \citep[e.g.][]{tfo+22,rgl+22,grn+23,lgs+23}. Nevertheless, the majority of long GRB at $z<1$ that are followed up show supernova (SN) features in their light curves and/or spectra \citep[e.g.][]{cwd+17}.} \citep{gvv+98,hsm+03,wb06} offer a sightline that pierces through the same star forming regions that dominate emission-line spectra \citep[e.g.][]{fls+06,wsv+07,kks+17}. The closest absorbing clouds have been found to typically lie at a distance of just a few hundred parsec from the GRB \citep{vls+07,vlr+13,dfg+14} \citep[although see][]{svd+23}, which is far smaller than typical QSO-DLA impact parameters and places the absorbing material within the galaxy ISM. Any difference between emission and absorption line metallicities in the case of GRB host galaxies would therefore reflect differences in the chemical enrichment of the multi-phase ISM.

Prior to the launch of JWST, only the host galaxy of GRB~121024A at $z=2.298$ \citep{fdk+15} had a well-measured absorption line metallicity together with an emission line metallicity. This is due to the need for restframe optical galaxy spectroscopy to capture the emission lines required for the metallicity diagnostic \citep[e.g.,][]{kmf+15,pvs+19,gsf23} as well as rest-frame UV GRB afterglow spectra for deriving the absorption metallicity. Ly~$\alpha$ absorption can only be detected from the ground for GRBs at $z\gtrsim 2$, but at such redshifts it becomes challenging to detect the weaker emission lines from the same host galaxies. In the case of GRB 121024A, the emission line metallicity was larger than the absorption metallicity by 0.2--0.7~dex \citep{fdk+15,kmf+15}. Deriving emission metallicities at higher redshifts ($z\gtrsim2$) from direct observations of the host galaxies requires near-infrared spectroscopy. 

It is only now, with the sensitivity and near-infrared coverage of JWST, that it is possible to obtain sensitive emission-line data at wavelengths out to \ha\ for a sample of GRB hosts with well-constrained absorption line metallicities. In this paper, we report results from a cycle-1 JWST NIRSpec program (PI: P. Schady, ID 2344) to measure emission line metallicities for a subset of 10 GRB host galaxies at $2.1 < z < 4.7$ that have accurately measured  ($<0.1$~dex) absorption line metallicities. In section~\ref{sec:sample} we describe our sample and provide details on our NIRSpec observations, followed by our data analysis in section~\ref{sec:analysis}. We present our results in section~\ref{sec:results}, and in section~\ref{sec:disc} we discuss the implications of our analysis on the relation between emission and absorption metallicity probes. All uncertainties are given as $1\sigma$ unless otherwise stated and we assume a standard Lambda cold dark matter ($\Lambda{\rm CDM}$) cosmological model with $\Omega_{\rm M}=0.31$, $\Omega_\Lambda=0.69$, and H$_0=67.8$~km~s$^{-1}$~Mpc$^{-1}$ \citep{Planck+16}.

\section{JWST GRB host galaxy sample}
\label{sec:sample}
There are around 30 GRBs with an afterglow absorption line metallicity measured with a statistical uncertainty of better than 0.25~dex \citep{wsb+17,blw+19,hdt+23}. From this parent sample we selected those host galaxies with measured UV/optical star-formation rates (SFRs) from UV continuum or emission line fluxes, but without the necessary spectra to measure a metallicity. This left us with a sample of 15 GRB host galaxies. We then further down-selected the sample to only include those GRB host galaxies with an estimated \ha\ flux brighter than $3.5\times 10^{-18}$~erg~cm$^{-2}$~s$^{-1}$ based on the SFR, leaving us with a final sample of 10 long GRB host galaxies. This flux limit was set by our requirement to measure strong emission lines at wavelengths spanning from \oii\ to \ha\ with S/N$>5$ in less than 4~hours (including overheads) according to the JWST exposure time calculator, above which the data volume exceeds the middle threshold set by JWST when using the {\sc nrsirs2rapid} readout pattern\footnote{\href{https://jwst-docs.stsci.edu/jwst-general-support/jwst-data-volume-and-data-excess\#JWSTDataVolumeandDataExcess-DataVolumeDatavolume}{jwst-docs.stsci.edu/jwst-general-support}}.

Our GRB host galaxy sample and the details of the JWST/NIRSpec observations are given in Table~\ref{tab:obs}. The detection of Ly~$\alpha$ absorption in the optical afterglow spectrum is necessary in order to be able to measure the metallicity in absorption (see \ref{ssec:Zabs} for details), and this requirement imposes a hard lower bound on the redshift of $z>2$ for the sample. It should be noted that our need for an accurate absorption line metallicity in our selection criteria biases our sample against more metal-rich and thus dusty host galaxies, which significantly attenuate the afterglow spectrum. In addition, the requirement that our host galaxies were previously detected in emission, either in imaging or spectra, introduces a preference for the brightest, and thus most star-forming galaxies of those with accurate absorption metallicities. The redshift range spanned by our final sample is $z=2.1$--4.7 (Table~\ref{tab:linefluxes}), and the absorption line metallicities range from 0.04\,Z$_\odot$ to 2.5\,Z$_\odot$ (Table~\ref{tab:hostprops}). 

\section{Observations and data analysis}
\label{sec:analysis}
Our sample of GRB host galaxies were all observed with NIRSpec \citep{jfa+22}, two using the integral field spectrograph (IFS) and the rest with the S400A fixed slit.

The host galaxies of GRB~050820A and GRB~150403A showed evidence of spatially extended emission in the available imaging data, and NIRSpec observations were therefore performed using the IFS, which has a $3\arcsec\times 3\arcsec$ field of view. For the same given object, longer exposures are required in IFS mode than with the fixed slit to reach the same integrated line flux sensitivity, which is why we only used the IFS in cases where there was evidence of extended emission. For both these GRB host galaxies the G140M/F100LP and G235M/F170LP grating and filter combinations were used, corresponding to a spectral resolution of $R\sim 1000$. In the case of GRB~050820A, previous \textit{Hubble Space Telescope} ({\sc HST}) imaging data showed that the host galaxy consists of at least two components separated by $< 1\farcs5$ (12.3~kpc physical size) \citep{chen12}, and the galaxy complex was therefore sufficiently compact to be able to perform a two-point nod to cover both components and additionally sample the sky background. For GRB~150403A, pre-imaging data were available with the GRB Optical and Near-infrared Detector \citep[GROND;][]{gbc+08} mounted on the 2.2~m Max Planck Institute telescope in La Silla, Chile. From these data the host galaxy appeared extended over $\sim 2$\arcsec (17.1~kpc physical size), and thus a four-point dither was used instead of nodding to avoid any of the galaxy falling out of the $3\arcsec\times 3$\arcsec\ NIRSpec field of view during a nod. The reduced and flux calibrated IFS data were downloaded from the Mikulski Archive for Space Telescopes (MAST) Data Discovery Portal\footnote{\href{https://mast.stsci.edu/portal/Mashup/Clients/Mast/Portal.html}{mast.stsci.edu/portal/Mashup/Clients/Mast/Portal.html}}. The \texttt{QFitsView}\footnote{\href{https://www.mpe.mpg.de/$\sim$ott/QFitsView/}{www.mpe.mpg.de/$\sim$ott/QFitsView}} software package was used to visualise the cubes and to extract stacked spectra and corresponding uncertainties from the regions of interest. For the remaining eight GRB host galaxies in our sample observed with the NIRSpec S400A fixed slit, a two-point nod pattern was used. The reduced and combined 2D spectra were similarly downloaded from MAST. All downloaded data were reduced with version 11.17.2 of the CRDS file selection software, using context jwst\_1140.pmap. The galaxies appear compact in the 2D spectra but in some cases they are resolved. The 1D spectra were extracted manually using the JWST \texttt{Extract1DStep} python function (v1.8.3), applying an extraction region centred on the detectable line emission and with a width 0\farcs7--0\farcs9.

The resolving power of NIRSpec is in the range $R=350-1400$ in the G140M grating, corresponding to a velocity dispersion $\sigma=90-360$~km/s, and in the G235M grating it is $R=630-1500$ or $\sigma=80-200$~km~s$^{-1}$. The typical intrinsic line velocities measured in our sample are $\sigma\sim 50-150$~km~s$^{-1}$ after accounting for the line spread function. The emission lines in the 1D spectra are generally well fit by a single Gaussian component, with the exception of the \nii\ and \ha\ emission lines from the host galaxy of GRB~090323, where there is evidence of additional emission that likely originates from unresolved additional components in velocity space (section~\ref{ssec:fluxes}). The current version of \texttt{Extract1DStep} fails to provide a flux uncertainty, which is due to the relevant flat field reference files not having an associated variance array. New flat fields are required to resolve this problem, and for now the recommendation provided on the JWST webpages\footnote{\href{https://jwst-docs.stsci.edu/jwst-calibration-pipeline-caveats/jwst-nirspec-mos-pipeline-caveats\#JWSTNIRSpecMOSPipelineCaveats-1-Dextractionconsiderations}{jwst-docs.stsci.edu/jwst-calibration-pipeline-caveats}} is to calculate the flux error by summing in quadrature the contribution from the Poisson noise and read-out noise alone, which are available in the extracted 1D spectral file. In the remainder of this section we describe the JWST observations and data analysis process for each of the GRB host galaxies in our sample, beginning with the two targets with IFS observations. 

\subsection{IFS observations}
\label{ssec:IFSobs}
\subsubsection{GRB~050820A}
\label{sssec:grb050820A}
\begin{figure}
    \includegraphics[width=1.0\linewidth]{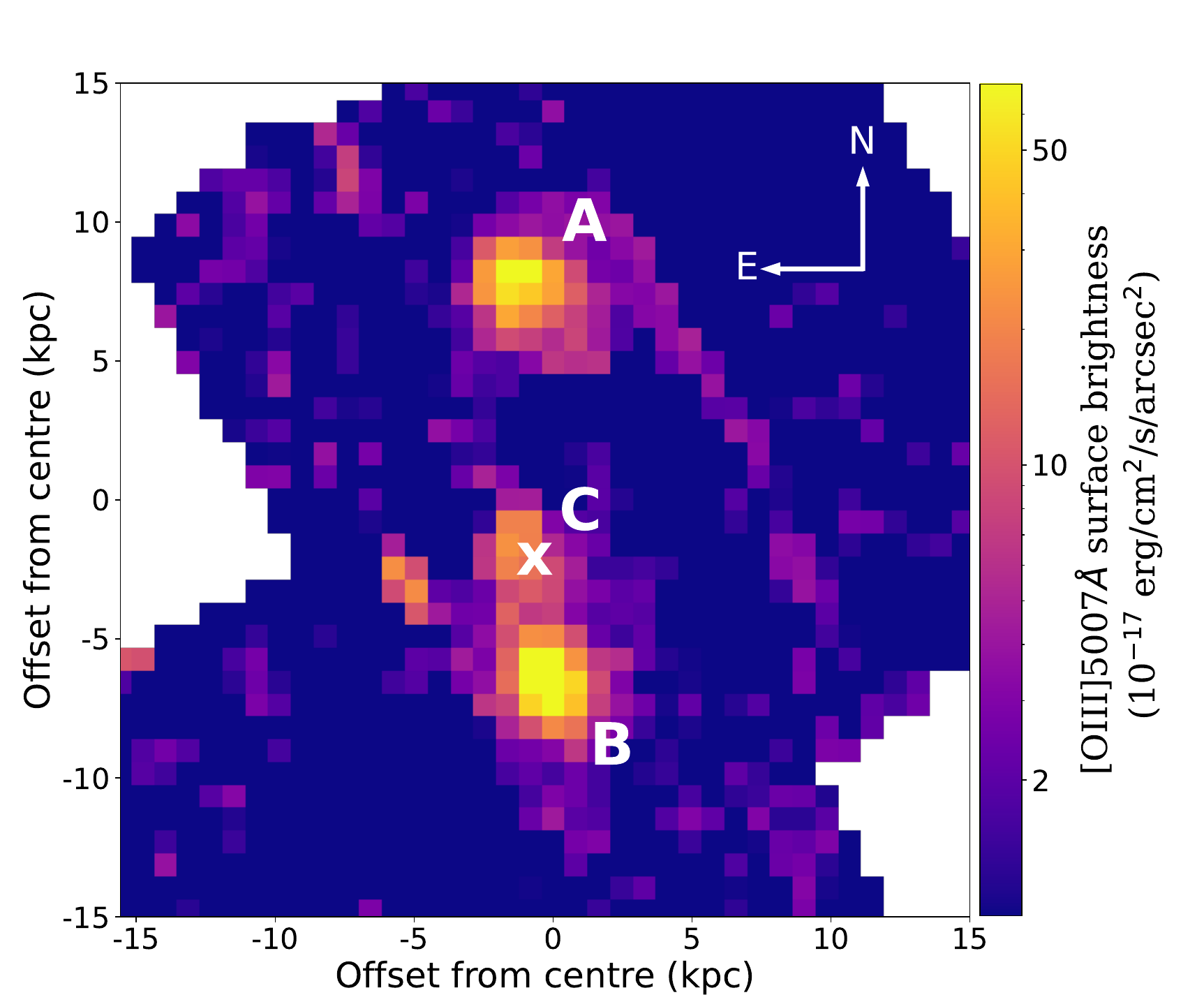}
    \caption{Surface brightness (SB) map of the G140M/L100LP NIRSpec IFS observations of the host galaxy of GRB~050820A at $z_{\rm abs}=2.615$ centred on \oiiib. A number of resolved emission regions are detected, including components A and B identified in \citet{chen12} and labelled in the image above. The position of the GRB afterglow is indicated with an `X', which lies close to a third emission component, labelled here as C. Additional emission can also be seen to the left of region C, which is only detected at 1.79$\mu$m, consistent with \oiiib\ at $z=2.615$. However no corresponding emission from \oiiia\ or \ha\ at this same redshift is detected at this location. The image is orientated with north up and east left. The pixel scale of the image is 0\farcs1, corresponding to 819~pc, and the offset from the image centre in kpc is indicated along the axes. Observations were taken with a two-point dither, which is why the shape of the field of view comprises two overlapping squares.}
    \label{fig:grb050820A_IFS}
\end{figure}

Level 3 data show clear emission from \ha, \hb, \oii\ and \oiii\ at an observer wavelength consistent with the GRB afterglow absorption redshift \citep[$z=2.6147$;][]{lve+05,pcd+07,flv+08} (see Fig.~\ref{fig:grb050820A}). An image of the G140M/F100LP NIRSpec data cube centered at the redshifted \oiiib\ line is shown in Fig.~\ref{fig:grb050820A_IFS}, where a galaxy complex made up of numerous emission components can be seen. Spectroscopic observations of this host galaxy were previously taken using the Folded port InfraRed Echellette (FIRE) spectrograph on the Magellan Baade Telescope, covering the wavelength range 0.8--2.5$\mu$m with a spectral resolution of $\sim 50$~km~s$^{-1}$ \citep{chen12}. In line with the naming convention used in \citet{chen12}, we refer to the upper, northern component as component A, and the lower, southern component as component B, which are separated by a projected distance of $\sim 13$~kpc, consistent with the HST observations \citep{chen12}. The GRB projected position was located between components A and B, and is indicated by an `X' in Fig.~\ref{fig:grb050820A_IFS}, which lies on a third emission component seen in Fig.~\ref{fig:grb050820A_IFS}, which we have identified as C. The absolute astrometry of the JWST image is limited by the JWST pointing accuracy, which is 0\farcs1, corresponding to a single NIRSpec pixel. Component C is not the brightest region of its host galaxy (see Table~\ref{tab:linefluxes}), but it nevertheless has a high SFR (see Section~\ref{ssec:fluxes} and Table~\ref{tab:hostprops}). This is in line with what is observed at $z<1$, with long GRBs tracing, on average, some of the brightest regions of their host galaxy \citep{fls+06}, even if not the brightest \citep[e.g. GRB~980425][]{cvs+08,kks+17}. The environmental properties at the position of the GRB will be presented in greater detail in Top\c{c}u et al. (in prep). An intervening absorption system was detected in the GRB afterglow at $z=2.3597$ \citep{lve+05,vpl+09}, but no emission lines of this intervening system are detected in the NIRSpec data cube. 

Spectra of the stacked pixels within each of the A, B and C components were extracted within \texttt{QFitsView}, as well as a spectrum of all emission host galaxy regions combined (see Table~\ref{tab:linefluxes}). Component A is the brightest of the host galaxy complex, and the velocity dispersion for all three components is consistent within $2\sigma$, although in the case of component C we only measure an upper limit ($<98$~km~s$^{-1}$, corresponding to the instrument resolution at $1.8\mu$m). 

After applying the barycentric correction to the measured radial velocities, the Gaussian fits to the \ha, \hb\ and \oiii\ doublet give a best-fit redshift of $z=2.6129\pm 0.0001$ for component A, $z=2.6133\pm 0.0001$ for component B, and $z=2.6136\pm 0.0002$ for component C, corresponding to a maximum velocity separation of $\Delta v\approx 58\pm 25$ km~s$^{-1}$ between the components. This is consistent with the redshifts for components A and B reported in \citet{chen12}.

\begin{figure}
    \includegraphics[width=1.0\linewidth]{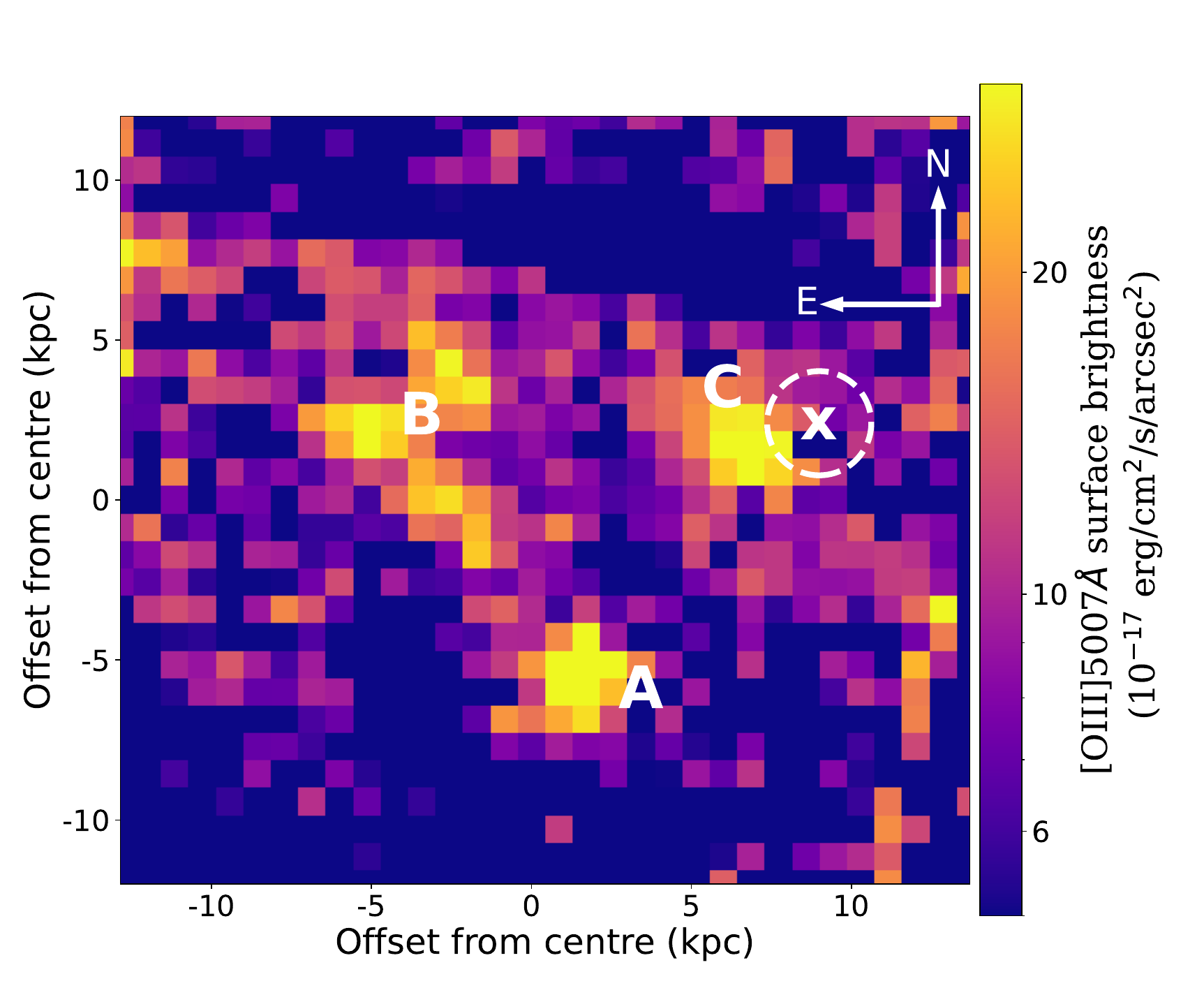}
    \caption{Surface brightness (SB) map of the G140M/L100LP IFS observations of the host galaxy of GRB~150403A centred on \oiiib\ at $z=2.057$. A number of resolved emission regions are detected, and the labels A, B and C indicate the regions where stacked spectra have been extracted. Region B is itself resolved into multiple components. The position of the GRB afterglow is just west of component C, marked with a `X', and the corresponding $1\sigma$ positional uncertainty is indicated with the white dashed circle. Note that no background subtraction has been applied and the colour bar thus does not go down to zero. The image is oriented with north up and east left. The pixel scale of the image is 0\farcs1, corresponding to 857~pc, and the offset from the image centre in kpc is indicated along the axes.}
    \label{fig:grb150403A_IFS}
\end{figure}

\begin{table*}
\begin{center}
\begin{minipage}[H]{1\textwidth}
\caption{GRB host nebular line fluxes corrected for Milky Way dust extinction}\label{tab:linefluxes}
\end{minipage}
\begin{tabular}{|l|c|c|c|c|c|c|c|c|c|}
\hline
\hline
GRB host & z$_{\rm abs}$ & z$_{\rm em}$ & \multicolumn{7}{|c|}{Line Flux ($10^{-17}$~erg~cm$^{-2}$~s$^{-1}$)} \\
\cline{4-10}
 & & & \hb & \ha & \oii & \oiiia & \oiiib & \niib & \sii \\
\hline\hline
030323 & 3.372$^a$ & 3.3710 & $0.17\pm 0.06$ & $0.43\pm 0.04$ & $0.11\pm 0.05$ & $0.27\pm 0.06$ & $0.73\pm 0.06$ & \ldots & \ldots \\
050820A & 2.615$^{b,c}$ & & & & & & & & \\
~~~galaxy-integrated & & 2.6133 & $2.49\pm 0.13$ & $8.81\pm 0.22$ & $4.56\pm 0.36$ & $5.71\pm 0.13$ & $16.06\pm 0.15$ & \ldots & \ldots \\
~~~component A & & 2.6129& $0.82\pm 0.05$ & $3.17\pm 0.11$ & $1.67\pm 0.08$ & $2.04\pm 0.06$ & $4.95\pm 0.06$ & \ldots & \ldots \\
~~~component B & & 2.6133& $1.20\pm 0.10$ & $4.12\pm 0.09$ & $2.05\pm 0.10$ & $2.84\pm 0.11$ & $8.49\pm 0.12$ & \ldots & \ldots \\
~~~component C & & 2.6136 & $0.31\pm 0.04$ & $1.02\pm 0.12$ & $0.74\pm 0.07$ & $0.44\pm 0.03$ & $1.55\pm 0.04$ & \ldots & \ldots  \\
080804 & 2.205$^d$ & 2.2065 & $0.40\pm 0.18$ & $2.35\pm 0.23$ & $1.01\pm 0.64$ & $1.02\pm 0.18$ & $2.59\pm 0.18$ & \ldots & \ldots \\
090323 & 3.57$^{e,f}$ 	& 3.5844 & $1.52\pm 0.10$ & $6.44\pm 0.13$ & $1.94\pm 0.48$ & $1.54\pm 0.10$ & $5.24\pm 0.11$ & $1.67\pm 0.12$ & $0.86\pm 0.12$ \\
100219A & 4.667$^g$ & 4.6698 & $0.06\pm 0.01$ & $0.09\pm 0.03$ & $<0.03$ & $0.06\pm 0.01$ & $0.20\pm 0.01$ & \ldots & \ldots \\
150403A & 2.057$^h$ & & & & & & & & \\
~~~galaxy-integrated & & 2.0570 & $2.25\pm 0.32$ & $6.07\pm 0.37$ & $3.99\pm 0.54$ & $2.99\pm 0.33$ & $9.14\pm 0.38$ & \ldots & \ldots \\
~~~component A & & 2.0570 & $0.94\pm 0.16$ & $1.77\pm 0.16$ & $1.11\pm 0.27$ & $1.10\pm 0.16$ & $2.88\pm 0.18$ & \ldots & \ldots \\
~~~component B & & 2.0567 & $1.13\pm 0.21$ & $3.93\pm 0.36$ & $2.64\pm 0.66$ & $1.07\pm 0.21$ & $4.01\pm 0.24$ & \ldots & \ldots \\
~~~component C & & 2.0576 & $0.25\pm 0.11$ & $1.30\pm 0.16$ & $0.75\pm 0.22$ & $0.76\pm 0.12$ & $2.27\pm 0.14$ & \ldots & \ldots \\
\hline

\end{tabular}
\begin{minipage}[H]{0.90\textwidth}
References: $^a$ \citet{vel+04}; $^b$ \cite{pcb+07}; $^c$ \citet{lvs+09}; $^d$ \citet{tdv+08}; $^e$ \citet{cpc+09}; $^f$ \citet{srg+12}; $^g$ \citet{tfg+13}; $^h$ \citet{smg+19} \\
\end{minipage}
\end{center}
\end{table*}

\subsubsection{GRB~150403A}
\label{sssec:grb150403A}
An image taken of the field of GRB~150403A almost six months after the GRB with GROND \citep{gbc+08} showed extended emission at the position of the GRB. This is confirmed with the NIRSpec IFS observations, where line emission from \ha, \hb, \oii\ and \oiii\ is clearly detected in three regions of the data cube at a redshift consistent with the GRB afterglow \citep[$z=2.06$;][]{pxt+15} (see Fig.~\ref{fig:grb150403A}). We have labelled the three brightest emission components A, B and C on an image taken from the NIRSpec G140M/L100LP data, centred on the observer frame \oiiib\ emission line (Fig.~\ref{fig:grb150403A_IFS}). The GRB afterglow position (labelled `X') appears to lie just to the west of component C, and not evidently on a star forming region. However, the astrometry of these data is limited by the combined pointing accuracy of JWST (within 1 pixel) and the accuracy of the acquisition target position, which was $\sim$ 0\farcs3. The estimated total $1\sigma$ uncertainty on the GRB afterglow position within the NIRSpec IFS field of view is indicated in Fig.~\ref{fig:grb150403A_IFS} with the dashed white circle. Accurate astrometry was possible in the case of the host galaxy of GRB~050820A because the A and B components were detected in previous HST imaging data, allowing the GRB position to be located accurately. However, for the host galaxy of GRB~150403A, there is no imaging data available that resolves the components shown in Fig.~\ref{fig:grb150403A_IFS}, so that further refined astrometry is not possible.

Component B has the brightest emission lines, followed by components A and then C. The largest velocity separation is between components B and C, which are offset by $\Delta v\approx 88$~km~s$^{-1}$, whereas component A lies somewhere between the other two components.

\subsection{Fixed slit observations}
\label{ssec:FSobs}
The remaining eight GRB host galaxies in the sample for which we had no evidence of extended emission were all observed with the NIRSpec S400A fixed slit (0\farcs4 slit width). In most cases this required observations in two grating/filter combinations with the exception of GRB~090323A, which at $z=3.57$ \citep{cpc+09,srg+12}, had all relevant nebular emission lines redshifted into the wavelength range of the F170LP filter. 

In four of the seven GRB host galaxy candidates that have been observed with the fixed slit\footnote{One remaining target is due to be observed between March and May 2024} (host galaxies of GRB030323, GRB080804, GRB090323 and GRB100219A), hydrogen Balmer and \oiii\ emission was detected at observer wavelengths consistent with the corresponding GRB afterglow absorption redshift. Emission from \oii\ was also detected from the host galaxy of GRB080804 and GRB~090323, and tentatively from the host of GRB~030323. However, no \oii\ emission was detected in the host galaxy spectrum of GRB~100219A. In the case of the host galaxy of GRB~090323, emission from \nii\ and \sii\ was also detected as well as the galaxy continuum. This is indicative of a luminous, metal-rich galaxy, consistent with the high absorption-based metallicity ($>2\,Z_\odot$).

In the case of two targets (GRB~120327A and GRB~120815A), the candidate host galaxies we observed were unfortunately found to be relatively dim foreground stars. These stars were not present in Gaia or other publicly available catalogues and the field is not covered by SDSS. The targeted sources could thus not previously be ruled out as the host galaxy. In the case of GRB~120815A, the offset between the identified host galaxy candidate and the GRB position was 1\farcs5, whereas in the case of GRB120327A the offset was 0\farcs3. However, in the latter case the observed source is three magnitudes brighter than expected, implying that the wrong source was targeted in the JWST NIRspec observations. No emission was detected in the NIRSpec spectrum at the position of the candidate of the host galaxy of GRB~141109A.

\begin{table*}
\begin{center}
\begin{minipage}[H]{1\textwidth}
\caption{Nebular line fluxes of intervening galaxy along line of sight to GRB~120815A at z=1.539. All lines have been corrected for a Milky Way dust reddening of E(B-V)=0.10 \citep{sf11}.}\label{tab:grb120815A-intervener}
\end{minipage}
\begin{tabular}{|c|c|c|c|c|c|c|}
\hline
\hline
 \hb & \ha & \oii & \oiiia & \oiiib & \niib & \sii \\
 \multicolumn{7}{c}{($10^{-17}$~erg~cm$^{-2}$~s$^{-1}$)}\\
\hline\hline
$0.23\pm 0.05$ & $2.10\pm 0.06$ & \ldots & $0.17\pm 0.05$ & $0.43\pm 0.05$ & $1.02\pm 0.06$ & $0.80\pm 0.06$ \\
\hline
\end{tabular}
\end{center}
\end{table*}

The host galaxy candidate of GRB120327A was identified by \citet{dfg+14} at the position of the GRB afterglow with AB magnitude $r'=24.50\pm 0.23$. However, the NIRSpec spectrum of this source does not reveal any emission lines at the expected observer frame wavelength, and instead resembles a black body with a temperature $\sim 3500$~K, suggesting that the source is in fact a foreground star (see Fig.~\ref{fig:grb120327A-grb120815A_spec}, left panel). 

In a late time observation of the field of GRB~120815A taken with the High Acuity Wide field K-band Imager (HAWK-I) a source is detected
1\farcs1 north-west of the GRB afterglow position, and this was considered to be the host galaxy (9.2~kpc). However, the NIRSpec data show the detection of two sources spatially offset by $\sim$1\farcs2 (corresponding to 10~kpc), neither of which correspond to a galaxy at the redshift of GRB~120815A, at $z=2.358$. The spectrum of the trace at the centre of the 2D spectrum resembles the Rayleigh-Jeans tail of a black body spectrum, implying that the targeted source is in fact a foreground star (see Fig.~\ref{fig:grb120327A-grb120815A_spec}, right panel). Precise astrometry was performed on the available imaging data of the GRB fields, and the lack of host galaxy detections in the case of GRB~120327A and GRB~120815A was therefore the result of the incorrect target being identified as the host galaxy in our pre-JWST imaging data, rather than due to an error in the positioning of the NIRSpec fixed slit.

In the case of GRB~141109A an extended source was detected at the location of the GRB in $r'$-band GROND observations taken three months after the GRB, and also in $3.6\micron$ observations taken with the {\em Spitzer Space Telescope} in March 2018 (program ID 13104; PI Perley), which we considered a to be a host galaxy candidate. The astrometry in our GROND images was typically good to within 0\farcs2, and we therefore accredit the lack of detected continuum or emission lines in our NIRSpec fixed slit observations to the host galaxy being dimmer than expected, rather than the slit having missed the target.

The second trace detected in the 2D spectra taken for GRB~120815A shows a bright continuum with strong Balmer and metal emission lines at observer wavelengths consistent with a galaxy at $z=1.539$. This galaxy is therefore likely the emission counterpart to a strong intervening absorption line system that was detected in the GRB afterglow spectrum at the same redshift \citep{klf+13}. At $z=1.539$ the \oii\ line doublet lies blueward of the NIRSpec F100LP spectral range. However, \ha, \hb\ and \oiii, as well as \nii\ and \sii\ are detected  and are given in Table~\ref{tab:grb120815A-intervener}.

Emission lines from the host galaxy of GRB~120815A were previously detected with {\em X-shooter} \citep{kmf+15}, and the lack of detection in our NIRSpec data is therefore due to a mistake in the target that we selected to observe rather than a lack of sensitivity. The \ha\ and \hb\ emission lines in the {\em X-shooter} data \citep{kmf+15} were detected at only $3\sigma$, and \oii\ was undetected. Nevertheless, the reported line fluxes are sufficient to be able to measure an emission line metallicity, albeit with large uncertainties (see Table~\ref{tab:hostprops}).

\subsection{Emission line flux measurements}
\label{ssec:fluxes}
Host galaxy emission lines in the 1D spectra were fit with Gaussian functions, with the velocity width of all lines in a given galaxy tied (corrected for the instrument resolution), and the position of the lines kept at a constant redshift (see Figs.~\ref{fig:grb050820A}--\ref{fig:grb090323A-2Dspec}). In the case of \oiii\ the line doublet was fixed to have a one-to-three line flux ratio \citep{ost89}. The model was generally a good fit to the data with the exception of the fits to the \nii\ and \ha\ lines from the host galaxy of GRB~090323 (see Fig.~\ref{fig:grb090323A}). The best fit Gaussian slightly under predicts the observed \ha\ line flux amplitude, and the fit to \niib\ is too narrow. Furthermore, the best-fit \nii\ doublet line ratio is \niib/\niia=1.6, which is far smaller than the expected line ratio of 3 \citep{sz00}. Forcing the line ratio to be 3 results in the fit to \niia\ significantly underestimating the observed emission of this line. The afterglow of GRB~090323 had strong absorption from two systems separated by just $\Delta v\approx 660$~km~s$^{-1}$, which are proposed to be the signature of two interacting galaxies \citep{srg+12}. It is therefore possible that our NIRSpec spectrum contains the combined emission from two galaxies, which may explain why our single Gaussian fits cannot fully describe the data. Although no spatial offset is evident around the \ha\ and \nii\ lines in the 2D spectrum (Fig.~\ref{fig:grb090323A-2Dspec}, left), we tried fitting the lines with a two-component model. The fit to the \ha\ line was marginally improved when using a two components model (Fig.~\ref{fig:grb090323A-2Dspec}, right), but the fit to \niib\ still appears too narrow. The Akaike and Bayesian information criterion increases by 10--15 when applying a two component fit, suggesting that the additional component does not significantly improve the fit, and thus for the rest of our analysis we use the flux measurements from the single Gaussian. The line fluxes from our single component fits are reported in Table~\ref{tab:linefluxes}.

Measured line fluxes for all host galaxies were corrected for Milky Way dust extinction along the GRB line of sight using the \citet{sf11} E(B-V) reddening maps (values given in Table~\ref{tab:obs}) and assuming a \citet{ccm89} dust extinction curve with an average total-to-selective dust extinction value $R_V=3.08$. The dust reddening from the GRB host galaxy was calculated from the \ha/\hb\ Balmer decrement assuming an intrinsic ratio \ha/\hb = 2.86 \citep{ost89}, which is appropriate for star-forming regions with temperature $\sim 10^4$~K and electron densities $n_e=10$--100~cm$^{-3}$. We corrected the line fluxes for host galaxy dust extinction using the average extinction law from the Small Magellanic Clouds (SMC) \citet{pei92}, which has a total-to-selective extinction $R_V=2.96$. We note that the majority of the host galaxies in the sample have uncertain host galaxy dust reddening such that only the host galaxy of GRB~090323 has a reddening measured at $>2\sigma$ confidence ($E(B-V)=0.34\pm 0.06$). In section~\ref{ssec:Zabs-vs-Zem} we investigate the effect of these uncertain dust reddening corrections on our results.

The Milky Way dust-corrected fluxes for all lines detected at a SNR$>2$ are reported in Table~\ref{tab:linefluxes}, and the fits to the lines are shown in the appendix (Figs.~\ref{fig:grb050820A}-\ref{fig:grb090323A-2Dspec}). In Table~\ref{tab:hostprops} we give the measured host galaxy dust reddening and the SFR based on the Galactic and host galaxy dust-corrected \ha\ luminosity using the \citet{Kennicutt98b} relation adopting a \citet{Chabrier03} IMF, which reduces the predicted SFR by a factor of $\sim 1.6$ compared to a Salpeter IMF \citep{sal55}. The errors on the SFR include the uncertainty on the dust reddening correction. In Table~\ref{tab:hostprops} we also provide the line velocity widths (corrected for the intrinsic instrument resolution) and the absorption and emission line metallicities based on several diagnostics (described in section~\ref{ssec:Zem}).

For the host galaxies of GRB~080804 and GRB~090323, lower signal to noise emission line flux measurements were already reported by \citet{kmf+15} from {\em X-shooter} data, most of which are consistent at $1\sigma$ with our NIRSpec measurements. The lack of any \hb\ detection from the host galaxy of GRB~090323 in \citet{kmf+15} ($<1.5\times 10^{-17}~{\rm erg}~{\rm cm}^{-2}~{\rm s}^{-1}$ at $3\sigma$) is only marginally consistent ($3\sigma$) with the strong emission line that we measure in the NIRSpec spectra, in contrast to the very good agreement in the \oii\ line flux. It is therefore possible that the uncertainties in the {\em X-shooter} \hb\ measurement for the host of GRB~090323 are underestimated. 

\subsection{Absorption line metallicities}
\label{ssec:Zabs}
The GRB absorption line metallicities used in this paper were taken from the literature, and they are given in Table~\ref{tab:hostprops} together with references. Three of the GRBs in the sample were observed with low resolution spectrographs ($R\sim 1000-2000$); either the Focal Reducer and low dispersion Spectrograph 2 (FORS2) mounted on the 8~m Very Large Telescope (VLT) \citep[GRB~030323 and GRB~090323][]{vel+04,srg+12}, or the Low Resolution Imaging Spectrometer (LRIS) mounted on the Keck~I 10~m telescope (GRB~050505). Two GRBs were observed with high resolution echelle spectrographs ($R\sim 30,000-50,000$); GRB~080804 with the VLT/Ultraviolet and Visual Echelle Spectrograph (UVES) \citep{tdv+08,fjp+09}, and GRB~050820A with UVES as well as the Keck/Higher Resolution Echelle Spectrometer (HiRES) \citep{pcb+07}. The remaining GRBs in the sample were observed with the medium resolution VLT/X-Shooter ($R\sim 6,000-10,000$) \citep{smg+19}.

\renewcommand{\arraystretch}{1.2}
\begin{table*}
\begin{center}
\begin{minipage}[H]{1\textwidth}
\caption{GRB host galaxy emission line properties. The host galaxies of GRB~120815A and GRB~121024A, listed separately at the bottom of the table, were not observed with NIRSpec, and instead line flux measurements were available from {\em X-shooter} observations in \citet{kmf+15}.}\label{tab:hostprops}
\end{minipage}
\begin{tabular}{|l|c|c|c|c|c|c|c|c|c|}
\hline
\hline
GRB host & \multicolumn{5}{c}{$12+\log({\rm O/H})$} & $z_{\rm em}$ & E(B-V)$_{\rm host}$ & SFR$\ddag$ & $\sigma\S$ \\
\cline{2-6}
 & abs$\dag$ & NOX22 \Rtwothree & NOX22 \Rthree & LMC23 $\hat{R}$ & DKS16 & & (mag) & $M_\odot/yr$ & (km/s) \\
\hline\hline
030323 & $7.48\pm 0.20^a$ & $7.65\pm 0.30$ & $7.65\pm 0.28$ & $7.54\pm 0.21$ & \ldots & 3.3710 & $0.00^{+0.29}_{-0.00}$ & $2.2 \pm 1.8$ & $80\pm 15$ \\
050820A & $8.20\pm 0.10^b$ & & & & & & & & \\
~~~galaxy-integrated & & $8.03^\star$ & $7.97\pm 0.03$ & $8.12^\star$ & \ldots & 2.6133 & $0.18\pm 0.05$ & $36.0\pm 5.2$ & $47\pm 1$ \\
~~~component A & & $8.03^\star$ & $8.02\pm 0.08$ & $8.12^\star$ & \ldots & 2.6129 & $0.26\pm 0.06$ & $15.3\pm 2.8$ & $50\pm 2$ \\
~~~component B & & $8.03^\star$ & $7.94^\star$ & $8.12^\star$ & \ldots & 2.6133 & $0.15\pm 0.08$ & $11.5\pm 2.6$ & $53\pm 2$ \\
~~~component C & & $8.17\pm 0.14$ & $8.16\pm 0.07$ & $8.12^\star$ & \ldots & 2.6136 & $0.12\pm 0.14$ & $2.8\pm 1.2$ & $<98$ \\
080804 & $8.33\pm 0.17^c$ & $8.21\pm 0.18$ & $8.12\pm 0.27$ & $8.12\pm 0.14$ & \ldots & 2.2065 & $0.62\pm 0.39$ & $4.3\pm 5.2$ & $148\pm 23$ \\
090323 & $9.10\pm 0.11^d$ & $8.45\pm 0.05$ & $8.34\pm 0.03$ & $8.43\pm 0.05$ & $8.91\pm 0.07$ & 3.5844 & $0.34\pm 0.06$ & $78.0\pm 13.5$ & $190\pm 5$ \\
100219A & $7.53\pm 0.11^e$ & $<7.42$ & $7.47\pm 0.11$ & $<7.40$ & \ldots & 4.6698 & $0.00^{+0.36}_{-0.00}$ & $1.0\pm 1.1$ & $66\pm 9$ \\
150403A & $7.77\pm 0.05^e$ & & & & & & & & \\
~~~galaxy-integrated & & $8.35\pm 0.11$ & $8.26\pm 0.06$ & $7.56\pm 0.10$ & \ldots & 2.0570 & $0.00^{+0.13}_{-0.00}$ & $9.4\pm 3.4$ & $55\pm 6$ \\
~~~component A & & $8.50\pm 0.08$ & $8.36\pm 0.06$ & $7.56\pm 0.10$ & \ldots & 2.0570 & $0.00^{+0.16}_{-0.00}$ & $2.8\pm 1.2$ & $65\pm 9$ \\
~~~component B & & $8.37\pm 0.14$ & $8.31\pm 0.06$ & $7.81\pm 0.17$ & \ldots & 2.0567 & $0.16\pm 0.17$ & $6.1\pm 3.0$ & $<116$ \\
~~~component C & & $8.10\pm 0.07$ & $8.04\pm 0.10$ & $8.12\pm 0.00$ & \ldots & 2.0576 & $0.51\pm 0.40$ & $2.0\pm 2.4$ & $<116$ \\
\hline
120815A & $7.46\pm 0.03^e$ & $8.26\pm 0.23$ & $8.13\pm 0.20$ & $7.86\pm 0.50$ & \ldots & 2.3587 & $0.00^{0.36}_{0.00}$ & $1.7 \pm 1.8$ & $28\pm 5$ \\
121024A & $8.01\pm 0.07^e$ & $8.38\pm 0.06$ & $8.23\pm 0.04$ & $7.68\pm 0.06$ & \ldots & 2.3012 & $0.00^{+0.09}_{-0.00}$ & $37.3\pm 9.6$ & $88\pm 4$ \\
\hline

\end{tabular}
\begin{minipage}[H]{0.90\textwidth}
$^\dag$ Absorption-based metallicity relative to solar and corrected for dust depletion. To convert to units of $\mbox{[M/H]}$, more commonly used in GRB absorption line studies, need to subtract the solar metallicity value 12+log(O/H)=8.69 \citep{ags+09}. \\
$^\ddag$ SFR derived from the Galactic and host galaxy dust extinction corrected \ha\ luminosity using the \citet{Kennicutt98b} relation adopting a \citet{Chabrier03} IMF. The large uncertainties in some cases are dominanted by the uncertainty on the dust correction.\\
$^\S$ Velocity width has been corrected for the intrinsic instrument resolution.\\
References: $^a$ \citet{vel+04}; $^b$ \citet{tdv+08}; $^c$ \citet{cfr+15}; $^d$ \citet{wsb+17}; $^e$ \citet{blw+19} \\
$\star$ In these cases the measured line ratio in question is larger than the maximum value covered by the diagnostics (\Rtwothree>0.96, \Rthree>0.78, or $\hat{R}=0.47R_2+0.88R_3>0.81$ at $1\sigma$), and the resulting best-fit metallicity thus corresponds to the value at the turn over point between the lower and upper branch (see Fig.~\ref{fig:ZabsvslogR}). The host galaxy of GRB~080804 and region C of the host galaxy of GRB~150403A also have a large value of $\hat{R}$, but with an uncertainty that lies within the range of values considered in the LMC23 metallicity diagnostic. 
\end{minipage}
\end{center}
\end{table*}
\renewcommand{\arraystretch}{1.0}

In all cases broad Ly-$\alpha$ absorption was detected in the afterglow spectra, corresponding to absorption from neutral hydrogen. Narrow metal absorption lines from species such as Zn, Si, S and Mg are also common \citep[e.g.][]{fjp+09}. Assuming solar relative abundances, it is then possible to obtain a largely model-independent measure of the neutral gas metallicity by combining the measured hydrogen absorption column density with metal column densities using unsaturated absorption lines, ideally from non-refractory metals. At the large column densities of neutral material in DLAs, hydrogen self-shielding make ionisation corrections negligible, and thus the greatest uncertainty in such a technique is the corrections for depleted metals that are in the dust phase, as well as the assumption of solar relative abundances, which we discuss further in section~\ref{ssec:Zabs-vs-Zem}.

For seven GRBs in the sample the dust-depletion corrected metallicities were determined following the method described in \citet{dls+13}, where the abundances of numerous singly-ionised metals are fitted simultaneously with a dust depletion model in order to measure a consistent, dust depletion-corrected metallicity \citep{wsb+17,blw+19}. For the remaining three GRBs, a single, non-refractory element was used to determine the metallicity; either sulphur (GRB~030323 and GRB~050505), or zinc (GRB~080804) \citep{cfr+15}. In these cases, dust depletion corrections were applied following the method of \citet{dlp+18}, where the \mbox{[S/Fe]} or \mbox{[Zn/Fe]} relative abundance can be used to determine the level of dust depletion, giving results that are consistent to when multi abundances are fit \citep[e.g.][]{hdt+23}. The \ion{Fe}{ii} column density along the line of sight to GRB~080804 is only constrained to lie within the range $\log N_{\rm Fe}/{\rm cm}^2=14.66-15.14$ (C. Ledoux private communication), which corresponds to a dust depletion correction $\delta_{Zn}$ between $-0.26$ and $-0.39$. We therefore use the mid-range dust depletion corrected metallicity and propagate through the uncertainty of 0.07~dex on the dust depletion correction to our absorption metallicity accordingly. In the case of GRB~030323 and GRB~050505, where only low resolution spectra were available, the published absorption metallicities should be considered lower limits. We nevertheless give the published metallicities in Table~\ref{tab:hostprops}, but will discuss the corresponding uncertainty on the absorption metallicity in section~\ref{ssec:Zabs-vs-Zem}. Although the optical afterglow spectrum of GRB~090323 was also low resolution, we consider the measured metallicity more robust due to the numerous metal abundances (Zn, S, Si, Cr, Fe) that were used to measure the dust depletion and metallicity, which reduce the effect of saturation in any single line \citep{wsb+17}. The uncertainties on the absorption metallicity given in Table~\ref{tab:hostprops} correspond to the statistical uncertainty on the metal and \hi\ column densities, and do not include systematic uncertainties, such as from the dust depletion correction.

\section{Results}
\label{sec:results}
\subsection{Emission line metallicities}
\label{ssec:Zem}
No emission was detected from the temperature-sensitive [\ion{O}{iii}]$\lambda$4363 line in any of the GRB host galaxies in our sample. For the majority of the sample, the $3\sigma$ upper limit on the [\ion{O}{iii}]$\lambda$4363/\oiii\ line flux ratio is $<0.05$, corresponding to a limit on the temperature of $T_{\rm e}(\ion{O}{iii})< 35,000$~K. More stringent constraints on the average electron temperature of our GRB host galaxy sample can be placed from a stacked spectrum, resulting in a $3\sigma$ upper limit on the [\ion{O}{iii}]$\lambda$4363 line flux of $1.3 \times 10^{-18}~{\rm erg}~{\rm cm}^{-2}~{\rm s}$, and an auroral-to-nebular line flux ratio of $<0.04$, corresponding to a limit on the temperature of $T_{\rm e}(\ion{O}{iii})\lesssim 27,000$~K. The electron temperatures measured in other $z=2-3$ galaxies is generally lower than this upper limit that we obtain \citep[e.g][]{clr+12,pcr+18,lmc+23,sst+23,srt+23}, and we therefore need to rely on strong emission line diagnostics to obtain gas-phase metallicities for our sample of galaxies.

The majority of strong emission line metallicity diagnostics are calibrated to galaxies and \hii\ regions within the local Universe \citep{kd02,pp04,kk04,mng+08,am13,dks+16,pg16,ccm+17}, whereas high-$z$ galaxies have higher radiation fields and/or ionisation parameters \citep[e.g.][]{kdl+13,srs+14}. Attempts have been made to calibrate strong emission line diagnostics to the conditions present in distant galaxies, either by using local analogues to high-$z$ galaxies \citep[e.g.][]{bkd+18,nox+22}, or by using recent, small samples of high-$z$ galaxy spectra with metallicities measured from the temperature-sensitive [\ion{O}{iii}]$\lambda$4363 line detected with JWST \citep[e.g.][]{hcs23,sst+23}.

The emission line diagnostics that are available for our GRB host galaxy sample are generally limited to those that use hydrogen and oxygen lines, since we do not detect emission from \nii\ or the \sii\ doublet in the majority of our sample. The most common of such line ratios used to trace metallicity are \Rtwo\ (\oii/\hb), \Rthree\ (\oiiib/\hb) (also referred to as \Othree), \Othreetwo\ (\oiiib/\oii) and \Rtwothree\ ((\oii\ + \oiii)/\hb). Given that our galaxy sample is at $z>2$, we only considered diagnostics that have been calibrated to the conditions present in the high-$z$ Universe, either using high redshift galaxy samples, or local analogues. We chose to use the \citet{lmc+23} (LMC23 from hereon) so-called $\hat{R}$ diagnostic, which is based on a combination of \Rtwo\ and \Rthree, and the metallicity diagnostics calibrated by \citet{nox+22} (NOX22 from hereon) and \citet{sst+23} (SST23 from hereon). NOX22 and SST23 provide diagnostics that include the hydrogen and oxygen line ratios listed above, but NOX22 additionally includes \Ntwo\ and \OthreeNtwo\ diagnostics. The LMC23 and SST23 diagnostics are calibrated against samples of high-$z$ galaxies ($2<z<9$) that have $T_{\rm e}$-based metallicities, whereas NOX22 used a combination of local SDSS galaxies and extremely metal-poor galaxies to extend the metallicity range down to far lower metallicities ($>1$~dex) than is covered in standard calibration samples \citep[e.g.][]{ccm+17}, making their diagnostics more appropriate for high-$z$ galaxies. We note that the \Rtwothree\ and \Rthree\ from \citet{mng+08} and NOX22 are very similar (SST23; see their fig.~6), and \citet{pcr+18} found that the \citet{mng+08} line diagnostics provided the best agreement (within 0.1~dex) to their measured \Te-based metallicities for a sample of 16 galaxies at $z=1.4-3.6$. This gives some support to the applicability of the NOX22 diagnostics to high-$z$ galaxies.

The NOX22 sample has a large scatter (up to an order of magnitude) in metallicity for a given \Rthree\ or \Othreetwo\ line ratio at the low metallicity end ($12+\log{\rm (O/H)}\lesssim 8$). They find this scatter to be dependent on the \hb\ equivalent width, EW(\hb), whereby galaxies with higher average EW(\hb) have a lower \Rtwo\ but a higher \Othreetwo\ at a fixed metallicity. We only detect the galaxy continuum in the host galaxy of GRB~090323, for which we measure a rest-frame EW(\hb)=$44\pm 4$\AA, placing it in the NOX22 low EW(\hb) bin. For the rest of the sample we can only place lower limits on the equivalent width, the most constraining being EW(\hb)$>50$\AA\ for the host galaxy of GRB~030323. We therefore cannot determine which NOX22 EW(\hb) bin the majority of our sample should lie in (low: $<100$\AA, intermediate: 100--200\AA, or high: $>200$\AA). For this reason we do not consider the NOX22 \Rtwo\ and \Othreetwo\ diagnostics in our analysis, considering only the NOX22 \Rtwothree\ and \Rthree\ diagnostics. Although these two latter diagnostics also show an EW(\hb) dependence, but it is much weaker than for \Rtwo\ and \Othreetwo\ ($\sim 0.5$~dex in metallicity at a fixed line ratio). To compute the NOX22 \Rtwothree\ and \Rthree, we initially use the high EW(\hb) calibrations, which NOX22 argue are most appropriate for high-$z$ galaxies, but in cases where we measure \logOH $>8$, we use the EW-averaged value, which are valid up to \logOH =8.9. For those diagnostics that are double branched, such as the NOX22 \Rtwo\ and \Rtwothree, and the LMC23 $\hat{R}$ diagnostic, we use the absorption line metallicity to select between the two solutions.

The SST23 diagnostics generally show a weaker dependence between metallicity and the relevant line ratios, than the NOX22 diagnostics for example (see Fig.~\ref{fig:ZabsvslogR}), and the SST23 metallicities thus cover a smaller dynamical range. We therefore move results based on the ST23 diagnostics to the appendix.

In the case of the host galaxy of GRB~090323, the \sii\ and \nii\ line doublets were detected and thus the \citet{dks+16} (DKS16 from hereon) \NtwoStwo\ diagnostic could be used, which has the advantage that it is independent of dust reddening, relatively independent of the ionisation parameter, and the authors claim it can be used over the full abundance range encountered at high-$z$. The \NtwoStwo\ best fit metallicity is $12+\log{\rm (O/H)} = 8.89\pm 0.08$, corresponding to $\log(Z/Z_\odot)=0.20\pm 0.08$, which is within $2\sigma$ of the measured absorption line metallicity \citep[$0.41\pm 0.11$;][]{wsb+17}.

In Table~\ref{tab:hostprops} we list the computed NOX22 \Rtwothree\ and \Rthree, the LMC23 $\hat{R}$, and the DKS16 \NtwoStwo\ metallicities. In this table we also provide the afterglow absorption metallicities and redshifts, and the emission line redshift and velocity dispersion corrected for the instrument resolution. The SST23 \Rtwothree, \Rthree, \Rtwo\ and \Othreetwo\ metallicities are given in Table~\ref{tab:sst23metals}.

\begin{figure*}
    \includegraphics[width=0.49\linewidth]{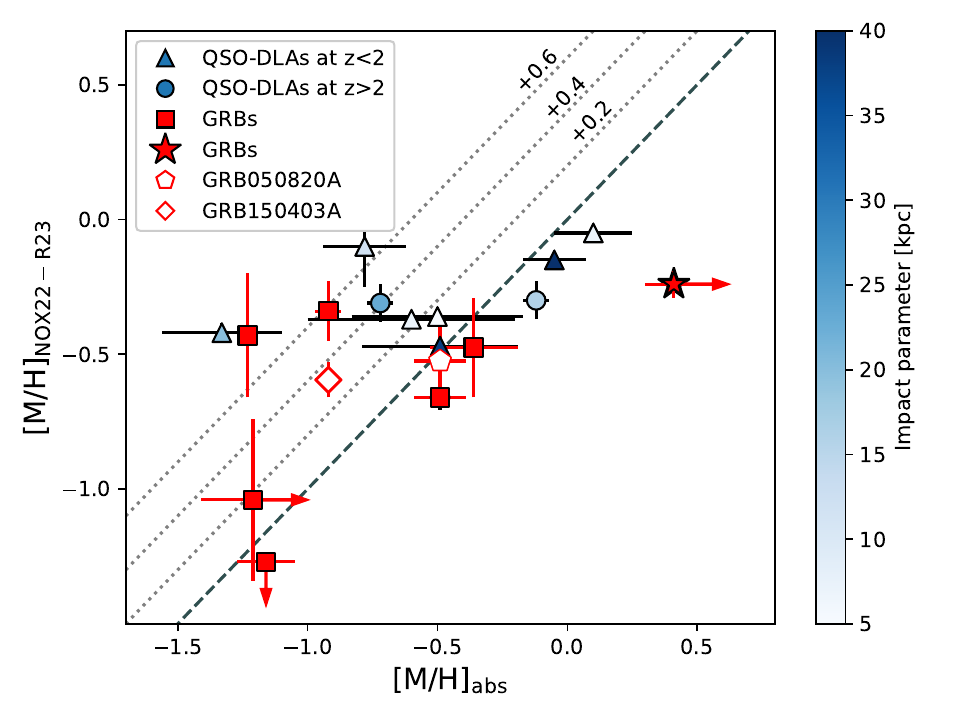}
    \includegraphics[width=0.49\linewidth]{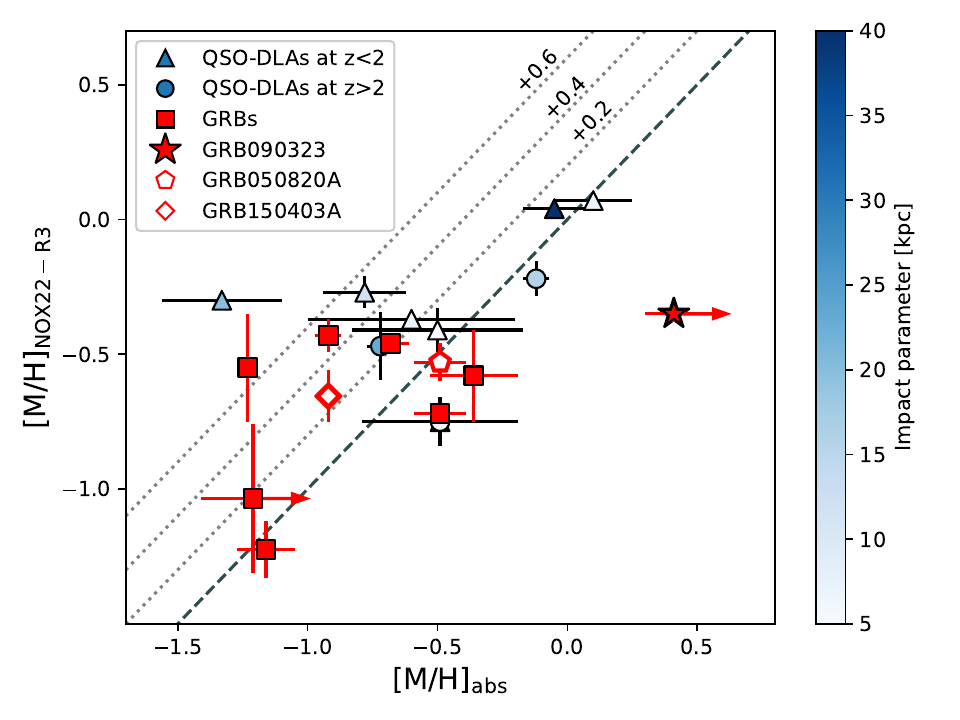}
    \caption{The NOX22 \Rtwothree\ (left) and NOX22 \Rthree\ (right) emission line metallicities against absorption line metallicities for our sample of GRB host galaxies and a compilation of QSO-DLAs and emission counterparts taken from \citet{rpt+16}. The emission line metallicities are in units of $\tn{[M/H]}=\tn{log(O/H)}-\tn{log(O/H)}_\odot$. For the host galaxies of GRB050820A and GRB150403A, metallicities for the resolved component closest to the projected GRB afterglow position (components C in both cases; see Figs.~\ref{fig:grb050820A_IFS} and \ref{fig:grb150403A_IFS}) are plotted as an open pentagon and diamond symbol, respectively. For GRB~030323 and GRB~090323 the absorption metallicity is considered a lower limit because they were measured with low resolution spectra, and this is represented in the figures with right-pointing arrows. GRB~090323 is plotted with a star symbol. The host galaxy of GRB~100219A had no \oii\ detection and the \Rtwothree\ emission line metallicity is therefore shown as an upper limit. The QSO-DLA data points are colour-coded by the impact parameter. The dashed line indicates where the absorption and emission line metallicities are equal, and the dotted lines represent the emission line metallicity offset from the absorption metallicity by the amount shown on the label. The number of QSO-DLA and GRB data points in each panel varies due to the need for different line ratios in each metallicity diagnostic.}
    \label{fig:ZabsvsZem-NOX22}
\end{figure*}

\begin{figure}
    \includegraphics[width=1\linewidth]{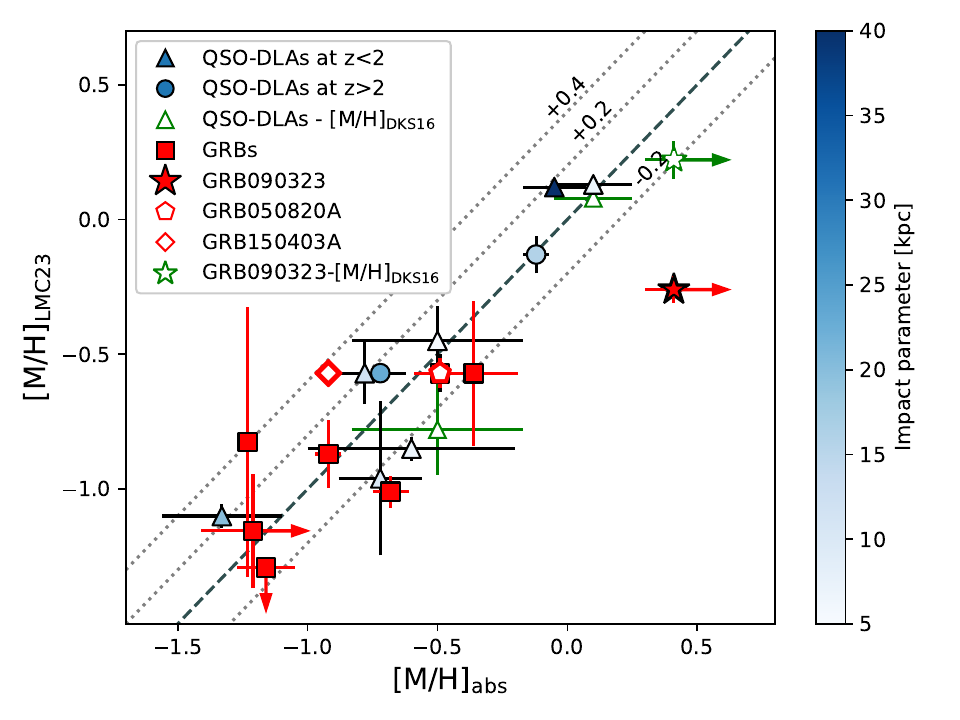}
    \caption{Same as Fig.~\ref{fig:ZabsvsZem-NOX22} but now with the emission line metallicity derived from the LMC23 $\hat{R}$ diagnostic. We additionally show the DKS16 \NtwoStwo\ emission line metallicity for GRB~090323 (green star) and for two QSO-DLAs (green triangles), which are the only GRB host galaxy and QSO-DLA emission counterparts in the sample with detected \niib\ and \sii\ emission lines necessary for this diagnostic.}
    \label{fig:ZabsvsZem-LMC23}
\end{figure}

\subsection{Nebular versus neutral gas-phase metallicity}
\label{ssec:Zabs-vs-Zem}
In Fig.~\ref{fig:ZabsvsZem-NOX22} we plot the absorption line metallicity against the NOX22 \Rtwothree\ (left) and NOX22 \Rthree\ (right) emission line metallicities for our GRB host galaxy sample (red) and for a compilation of QSO-DLAs (shades of blue) from \citet{rpt+16} with absorption and emission line metallicities of the intervening galaxy counterpart. The sample of QSO-DLAs are generally at $z<0.7$, but three lie at $z>2$ \citep[blue circle data points from][]{bmk+13,fgc+13,kfl+13} consistent with our GRB host galaxy sample (see Table~\ref{tab:QSOsample} for details on the QSO sample). In our GRB host galaxy sample we also include the hosts of GRB~120815A and GRB~121024A, which have absorption and emission lines detected from ground-based observations \citep{klf+13,fdk+15}. In Fig.~\ref{fig:ZabsvsZem-LMC23} we show the absorption against the LMC23 $\hat{R}$ emission metallicity, as well as against the DKS16 \NtwoStwo\ emission line metallicity for the host galaxy of GRB~090323 (green open star) and two QSO-DLAs (green open triangles; J0441-4313 at $z=0.1010$ and J1544+5912 at $z=0.0102$), all of which have the necessary \niib\ and \sii\ emission line detections to compute the \NtwoStwo\ metallicity. In both Figs.~\ref{fig:ZabsvsZem-NOX22} and \ref{fig:ZabsvsZem-LMC23} the emission and absorption metallicities are in units of [M/H], assuming a solar metallicity value $12+\log({\rm O/H})=8.69$ \citep{aag21}\footnote{$\tn{[M/H]}=\tn{log(O/H)}-\tn{log(O/H)}_\odot$, assuming that the relative abundance of oxygen is solar.}. The QSO-DLA data points are colour-coded by their impact parameter, ranging from $1-50$~kpc, although the association with the emission counterparts at large impact parameters is less secure. The offset from the galaxy centre of our sample of GRB sightlines is generally unknown, but when measured, they are typically small \citep[e.g. average offset of 1~kpc in sample of 68 GRB host galaxies observed with HST;][]{bbf16}. All GRB data points in Fig.~\ref{fig:ZabsvsZem-NOX22} and Fig.~\ref{fig:ZabsvsZem-LMC23} are therefore plotted with the same red colour. The sample of plotted QSO-DLA data points is smaller than the sample in \citet{rpt+16} because we did not consider limits, and we have the additional requirement that \hb\ and either \oii\ or \oiii\ have to be detected in order to apply the the emission line diagnostics that we consider in this work.

For the two galaxies observed with the IFS (for GRB~050820A and GRB~150403A) we plot the emission line metallicity of the resolved component closest in projection to the GRB afterglow, which in both cases is component C in Figs.~\ref{fig:grb050820A_IFS} and \ref{fig:grb150403A_IFS}. In the case of GRB~050820A the \Rtwothree, \Rthree\ and $\hat{R}$ line ratios of all components are larger than the maximum value covered by the NOX22 and LMC23 diagnostics, and the computed emission line metallicity in all cases is thus the maximum value possible with these diagnostics. The resolved components considered in the host galaxy of GRB~150403A, on the other hand, do vary in metallicity by up to 0.3~dex within the same diagnostic. The absorption line metallicity is most consistent with the emission line metallicity of component C in the case of NOX22 \Rtwothree\ and \Rthree, but it is most consistent with the $\hat{R}$ metallicity of component B. A more detailed analysis on the resolved spectroscopic properties of these two GRB host galaxies will be presented in a follow-up paper (Top\c{c}u et al., in prep). In Fig.~\ref{fig:ZabsvsZem-SST23} we show the results for the four SST23 diagnostics.

The well-known discrepancy between emission line metallicities \citep[e.g.][]{ke08,am13} is evident in Fig.~\ref{fig:ZabsvsZem-NOX22} and Fig.~\ref{fig:ZabsvsZem-LMC23}, with NOX22 and LMC23 giving notably different results. In the former case, the NOX22 metallicities are larger than the absorption metallicities by $\sim 0.2$~dex on average (Fig.~\ref{fig:ZabsvsZem-NOX22}), although this offset is predominantly at \mbox{[M/H]$_{\rm abs}<-0.5$}, where there appears to be a relatively weak relation between the absorption and emission metallicity. The standard deviation in the NOX22 emission line metallicities relative to the line of equality (dashed black line) is 0.4--0.5~dex. Applying a Spearman's rank test returns a rank coefficient $\rho=0.2-0.3$ with p-value of $\sim 0.2$, indicating a weak and non-significant correlation. The LMC23 metallicities, on the other hand, are more evenly distributed on both sides of the line of equality (Fig.~\ref{fig:ZabsvsZem-LMC23}), with LMC23 generally lying within $\pm 0.2$~dex of the absorption metallicities. In this case the Spearman's rank coefficient is $\rho=0.8$ with p-value=$2\times 10^{-5}$, demonstrating that there is a strong and significant positive correlation. The DKS16 \NtwoStwo\ metallicity is also within $\pm0.2$~dex of the absorption metallicity for the three cases where this diagnostic could be applied (Fig.~\ref{fig:ZabsvsZem-LMC23}, green data points). Of note is the good agreement between the absorption and the DKS16 metallicity for the host galaxy of GRB~090323 (consistent within $2\sigma$), which for NOX22 and LMC23 differed by a factor of four.

Given the large uncertainty on the host galaxy dust reddening corrections, we repeated the analysis in which we only applied a host galaxy dust correction if dust reddening was detected at more than $3\sigma$ significance. This only applies to the two most metal-rich absorbers in our sample; the host galaxy of GRB~090323 and the emission counterpart of QSO-DLA J0441-4313. We found that the results were only marginally affected, with the points having similar offsets and standard deviations as shown in Fig.~\ref{fig:ZabsvsZem-NOX22} and \ref{fig:ZabsvsZem-LMC23}.

Systematic uncertainties also exist in the absorption line metallicities, primarily from the possible saturation of metal absorption lines in low- and even mid-resolution spectra, and also in the dust depletion corrections. All QSO-DLA absorption line metallicities shown in Fig.~\ref{fig:ZabsvsZem-NOX22} and \ref{fig:ZabsvsZem-LMC23} are measured from mid- or high-resolution spectra, and systematic effects should therefore be less of a problem \citep{wsb+17}. However, the absorption metallicities for GRB~030323 and GRB~090323 are measured from low resolution data, and these metallicities should therefore be considered as lower limits \citep{pcb+06}. We have accordingly plotted these data points with right-pointing arrows, although, as described in section~\ref{ssec:Zabs}, we consider the absorption metallicity for GRB~090323 more robust due to the multiple metal lines used to constrain the dust depletion and metallicity.

Dust depletion corrections can be well constrained when multiple metals are used \citep[e.g][]{dls+13,dlm+16,wsb+17}, although \citet{dlp+18} also found that corrections based on just the \mbox{[Zn/Fe]} and \mbox{[S/Fe]} relative abundance can give very similar results. An additional advantage of using multiple lines to measure the absorption line metallicity is that the method is also sensitive to $\alpha$-element enhancements, whereby enhanced elements will have relative abundances that lie above the best-fit dust depletion curve provided there are enough Fe-group elements to constrain the depletion curve. Of those GRBs in our sample with absorption metallicities measured in this way \citep[i.e. from][]{wsb+17,blw+19}, only GRB~121024A shows a tentative Si overabundance of $\sim 0.5$~dex although no corresponding enhancement is observed in other $\alpha$-elements, such as O or S \citep{blw+19}. The Si abundance therefore appears as an outlier in the fit and does not contribute to the best-fit absorption metallicity. The absorption metallicity for the host galaxies of a further two GRBs was determined from the \mbox{[S/H]} abundance, which if they have an $\alpha$-element enhancement, may overestimate the metallicity by $~\sim 0.2-0.3$~dex \citep{bsr+12,dlm+16,drk+24}, further increasing the apparent discrepancy between absorption and emission line metallicities. The QSO-DLA absorption metallicities are similarly measured using a range of metals, three of which relied on alpha-element abundances (S or Si), although they all had measured metallicities \mbox{[M/H]}$>-0.5$, where the agreement between the NOX22 and absorption metallicities is relatively good. Any uncertainties in $\alpha$-element enhancements are therefore likely to bring absorption and emission line metallicities further apart, and although dust depletion corrections introduce some uncertainty to the absorption-based metallicities, these are unable to explain the extent of the disparity with the NOX22 metallicities at the low metallicity end.

We do not see any clear dependence in Fig.~\ref{fig:ZabsvsZem-NOX22} and \ref{fig:ZabsvsZem-LMC23} on the emission and absorption metallicity offset of the QSO-DLA data points with impact parameter (i.e. colour of data point), as would be expected if QSO-DLAs with large impact parameters probe correspondingly less enriched material. The QSO-DLA with greatest difference between the emission and absorption line metallicity is QSO-DLA J0958+0549, which has an impact parameter of 20~kpc \citep{rpt+16}, whereas the QSO-DLA with the largest impact parameter of 50~kpc (J1436-0051) has an emission line metallicity that is within 0.2~dex of the absorption line metallicity in Fig.~\ref{fig:ZabsvsZem-NOX22} and Fig.~\ref{fig:ZabsvsZem-LMC23}. This is consistent with the relatively shallow (but negative) metallicity gradient reported in the literature out to 20--40~kpc \citep[e.g.][]{ckr+05,pbk+13,cmf+14,rpt+16,rcm+18}.

The agreement between the absorption and the LMC23 emission line metallicities (and also the DKS16 metallicities) for the GRB and QSO-DLA sample is quite remarkable given that absorption and emission lines probe different phases of the gas, and the measured metallicities are averaged over different regions of the galaxy; either luminosity-weighted over the whole galaxy in emission, or density-weighted along a single sightline through the galaxy in absorption. The standard deviation of the full GRB and QSO-DLA sample is 0.24~dex, which is comparable (if not slightly better) to the scatter that has been observed between \Te- based and strong line metallicities in recent high-$z$ galaxy samples observed with JWST \citep{lmc+23,sst+23}. The NOX22 metallicities, on the other hand, are systematically larger than the absorption metallicities, even though there may be some weak relation with the absorption line metallicity (Fig.~\ref{fig:ZabsvsZem-NOX22}). However, the uncertainties on what is the most appropriate emission line diagnostic to use clearly dominates over any statistical uncertainty, and limits the conclusions that can be reached on the relation between the metallicity of the neutral gas ISM probed in absorption and that of ionised star-forming regions probed in emission, or on the effect of single sightline versus galaxy-integrated measurements. 

\subsection{GRB host stellar masses and implied metallicities}
\label{MZR}
Several GRB host galaxies in our sample have stellar masses reported in the literature, which we use to investigate where our GRB host galaxy sample lies on the mass-metallicity relation (MZR). The host galaxies with mass estimates or limits are GRB~030323 \citep[$\log M_\star/M_\odot<9.23$;][]{lbc11}, GRB~050820A \citep[$\log M_\star/M_\odot=9.29$;][]{cpp+09}, GRB~080804 \citep[$\log M_\star/M_\odot=9.28$;][]{pth+16}, GRB~090323 \citep[$\log M_\star/M_\odot=10.2$;][]{ks17}, and GRB~121024A \citep[$\log M_\star/M_\odot=9.9$;][]{fdk+15}. Although the stellar masses are taken from several different references, the majority are determined from fits to the optical through to NIR galaxy spectral energy distribution (SED), and are thus relatively insensitive to assumptions made on the SFR history and dust attenuation prescriptions \citep[e.g.][]{pvs+19}. The exception is in the case of the host galaxy of GRB~080804, where the stellar mass is based on a single mid-infrared data point, which can over-estimate $M_\star$ by $\sim 0.4$~dex compared to stellar masses from SED fitting. This stellar mass can therefore be considered an upper limit.

\begin{figure*}
\begin{minipage}[H]{1\textwidth}
    \includegraphics[width=0.95\linewidth]{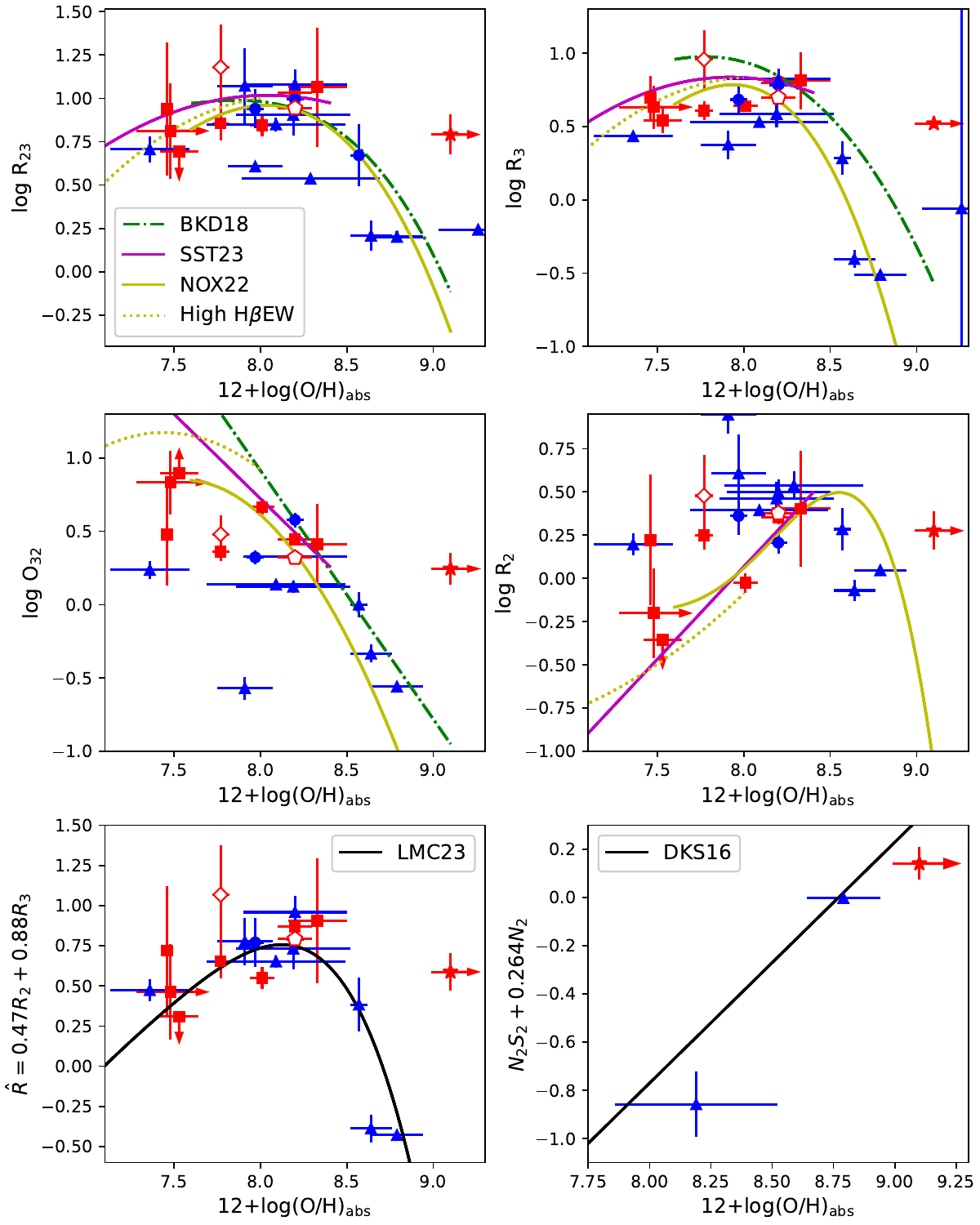}
    \caption{GRB host galaxy and QSO-DLA galaxy counterpart absorption line metallicity against the logarithmic line ratios \Rtwothree\ (top left), \Rthree\ (top right), \Othreetwo\ (middle left), \Rtwo\ (middle right), the LMC23 combined \Rtwo\ and \Rthree\ line ratio diagnostic ($\hat{R}$) (bottom left) and the \citet{dks+16} (DKS16) \NtwoStwo\ diagnostic (bottom right). All data points are the same as in Fig.~\ref{fig:ZabsvsZem-NOX22} and \ref{fig:ZabsvsZem-LMC23}. The best-fit curves from \citet{bkd+18} (BKD18; green dot-dashed), SST23 (solid magenta), and the NOX22 \hb\ EW-averaged (yellow solid) and high \hb\ EW (yellow dotted) relations are overplotted in the top four panels for reference. Note that \citet{bkd+18} did not provide an \Rtwo\ diagnostic.}
    \label{fig:ZabsvslogR}
\end{minipage}
\end{figure*}

Using the best-fit relation of \citet{mng+08} and \citet{mcm+09} for galaxies at $z\sim 2.2$ and $z=3-4$, respectively, we estimate the expected (emission line) metallicity given the host galaxy stellar mass. The emission line metallicity diagnostics used in these papers combined a number of line ratios, but the authors state that the results are dominated by the \Rtwothree\ or \Rthree\ diagnostic \citep{mng+08}, which is thus comparable to our analysis. More recently \citet{ssj+21} measured the MZR for a sample of galaxies at $z\sim 2.3$ and $z\sim 3.3$ from the MOSDEF survey \citep{ksr+15}. However, in their analysis they used the \citet{bkd+18} metallicity diagnostics, which can differ significantly from the NOX22 calibrations that we used (e.g. see \Rthree\ and \Othreetwo\ panels in Fig.~\ref{fig:ZabsvslogR}), which will cause there to be a systematic difference in the derived metallicities.

For the five host galaxies in our sample with stellar mass estimates, the \citet{mng+08} and \citet{mcm+09} MZRs predict metallicities that are on average within $\sim 0.2$~dex of the NOX22 and LMC23 metallicities computed in section~\ref{ssec:Zem}. The largest discrepancy is in the $z=3-4$ MZR, and if we only consider those host galaxies at $z=2-3$, the predicted and measured metallicities are consistent within 0.01~dex for the $\hat{R}$ LMC23 metallicities, and within 0.1~dex for the NOX22 diagnostics. GRB host galaxies have also been found to agree well with the fundamental metallicity relation (FMR) \citep{msc+11,pvs+19}, which adds a SFR-dependency to the MZR \citep[FMR;][]{mcm+10}. To check this with our sample of GRB hosts with $M_\star$ estimates, we use our \ha-based SFRs to determine the metallicity predicted by the FMR, but we find that this increases the disagreement between the expected and measured emission line metallicities, with the FMR predicting metallicities that are on average $\sim 0.4$~dex larger than we measure using strong emission line diagnostics. A larger sample of host galaxies and a more consistent approach in measuring $M_\star$ is required to investigate further where GRB host galaxies lie in high-$z$ MZR and FMR relations.

\subsection{Emission line ratios}
\label{ssec:logR}
Given that absorption line metallicities are less model-dependent compared to emission line metallicities, we investigate the relation between common emission line ratios and the GRB afterglow absorption metallicity. In Fig.~\ref{fig:ZabsvslogR} we plot the absorption line metallicity for our GRB host galaxy (red) and QSO-DLA (blue) sample (using the $12+\log{\rm (O/H)}$ scale) against the logarithm of the $R_3$, $R_{23}$, O$_{32}$ and $R_2$ line ratios in the top two panels, and against the LMC23 $\hat{R}$ (left) and the DKS16 \NtwoStwo\ line ratios in the bottom two panels. For reference, we also show the best-fit metallicity diagnostics from a number of papers, indicated in the top left figure legend. We show the NOX22 metallicity diagnostics calibrated across the full metallicity range of their sample (yellow solid), as well as the relations for their sample of high \hb\ EW galaxies (yellow dotted), which are valid for $12+\log{\rm (O/H)}<8.0$. The \citet{bkd+18} (BKD18; green dot-dashed) and SST23 (solid magenta) relations are generally shifted upwards relative to the NOX22 curves, and the greatest differences are in the \Othreetwo\ diagnostics, where the SST23 and \citet{bkd+18} relations do not turn over at low metallicities. Note that \citet{bkd+18} did not provide an \Rtwo\ diagnostic calibration.

The data are in general good agreement with the models for the \Rtwothree\ and \Rthree\ diagnostics, especially the GRB data points. However, greater offsets are present in the \Othreetwo\ and \Rtwo\ panels, where the curves predict metallicities that are typically larger than the absorption metallicity for a given \Othreetwo\ or \Rtwo\ line ratio. Much of the good agreement in the top two panels of Fig.~\ref{fig:ZabsvslogR} could in part be due to the GRB and QSO-DLA data points lying on the fairly flat portion of the \Rtwothree\ and \Rthree\ diagnostics, where the empirical relations between the line ratios and metallicity is fairly weak. As such, an increase in metallicity of 0.3--0.5~dex would shift the GRB data points closer to the \Othreetwo\ and \Rtwo\ diagnostics while still maintaining a relatively good agreement with the \Rtwothree\ and \Rthree\ diagnostics.

There is better agreement between the data points and the LMC23 best-fit relation (Fig.~\ref{fig:ZabsvslogR} bottom left panel; black solid curve) than seen for the \Rtwothree, \Rthree, \Rtwo\ and \Othreetwo\ diagnostics, as expected given the general agreement between the absorption metallicity and LMC23 emission metallicity shown in Fig.~\ref{fig:ZabsvsZem-LMC23}. Although, as was the case for the \Rtwothree\ and \Rthree\ line ratios, the majority of the data points lie close to the LMC23 diagnostic turn over point. A greater sample of data points with low ($12+\log{\rm (O/H)}<7.7$) and high ($12+\log{\rm (O/H)}>8.5$) absorption line metallicities are therefore required to determine how closely the absorption metallicities and galaxy emission line ratios trace each other.

A clear outlier in the top two rows of Fig.~\ref{fig:ZabsvslogR} and in the $\hat{R}$ line ratio is the super-solar metallicity data point corresponding to the afterglow of GRB~090323. However, as seen in the bottom right panel, the combined \NtwoStwo\ line ratio of this host galaxy is as expected if the absorption metallicity is representative of the nebular gas-phase metallicity. There are two QSO-DLAs with absorption line metallicities that also have detected \nii\ and \sii\ emission lines, which are shown in this figure, and their \NtwoStwo\ line ratios are also as expected given their absorption line metallicities. This diagnostic has the additional advantage that it is not double branched. Nevertheless, an important potential limitation of the diagnostic is that is assumes a fixed O/H--N/O relation based on observations in the local Universe that may not be applicable at high-$z$ \citep[e.g.][]{fvh+03,clr+12,bsc+23,csb+23,msk+24}. We discuss this diagnostic, and specifically the host galaxy of GRB~090323, in greater detail in section~\ref{ssec:grb090323}.

\section{Discussion}
\label{sec:disc}
\subsection{Emission line diagnostics}\label{ssec:Zemdiags}
We find remarkable agreement between our sample of GRB and QSO-DLA absorption line metallicities and the LMC23 metallicity diagnostic, with a scatter that is comparable to that seen in high-$z$ calibration samples \citep{lmc+23,sst+23}. This agreement has two important implications for the use of absorption and emission line probes to study the cosmic chemical enrichment. Firstly, it presents the possibility of combining absorption and emission line probes to study the cosmic chemical evolution out to higher redshifts and down to lower mass galaxies than is possible with emission line metallicities alone, even with JWST. Current JWST mass-metallicity samples at $z>4$ extend down to $M_\star\sim 10^7 M_\odot$ and $12+\log({\rm O/H})\sim 7.5$ \citep{cmc+23,noi+23}, whereas absorption line metallicities go down to equivalent oxygen abundances $12+\log({\rm O/H})< 7$ \citep[e.g.][]{blw+19,hdt+23}, corresponding to $M_\star\sim 10^5$~M$_\odot$ when extrapolating the best-fit $z=4-10$ MZR from \citet{noi+23}. Secondly, it suggests that star forming regions and the interstellar neutral gas have a very similar chemical composition, implying that the multi-phase ISM is well mixed within the galaxy. Furthermore, the similar scatter in the GRB and the QSO-DLA samples implies that the neutral gas is chemically homogeneous out to large distances from the galaxy centre. This would require enriched material within star forming regions to be efficiently distributed through outflows into the CGM in agreement with recent semi-analytic models of galaxy evolution \citep[{\sc L-Galaxies} 2020;][]{yhf+21}.

Although the scatter is larger in the NOX22 results, there is nevertheless an indication that the emission and absorption metallicities still trace each other. This is especially true when using the \Rtwothree\ diagnostic. These results thus still offer the possibility of combining emission and absorption line probes to study the cosmic chemical evolution in high-$z$ galaxies, as long as the relation between emission and absorption metallicities can be quantified with a larger sample.

The NOX22, SST23 and LMC23 diagnostics all yield metallicities for the host galaxy of GRB~090323 (star symbol in Figs.~\ref{fig:ZabsvsZem-NOX22} and \ref{fig:ZabsvsZem-LMC23}) that are a factor of $\sim 5$ smaller than the absorption line metallicity. In contrast, the DKS16 \NtwoStwo\ diagnostic gives a metallicity that are in very good agreement with the absorption metallicity for GRB~090323, and the \NtwoStwo\ metallicities for the two QSO-DLA counterparts with \nii\ and \sii\ line detections (QSO-DLAS J0441-4313 and J1544+5912) and also consistent with the absorption metalicities (see Fig.~\ref{fig:ZabsvslogR}, bottom right panel). We discuss the \NtwoStwo\ diagnostic and the case of GRB~090323 in more detail in the next section.

\subsection{The host galaxy and afterglow of GRB090323}
\label{ssec:grb090323}
Aside from appearing as an outlier in the emission and absorption metallicity parameter space, the host galaxy of GRB~090323 was also unusual in its high absorption line metallicity (\mbox{[M/H]=$0.41\pm 11$}) compared to the rest of the GRB host galaxy sample, and to the population of GRB-DLAs \citep[e.g.][]{cfr+15,dlp+18}, which at $z=3.57$, is all the more extraordinary. There is one QSO-DLA in our comparison sample that also has a super-solar absorption line metallicity (J0441-4313), although it is at much lower redshift than GRB~090323 ($z=0.10$).

The difference in the NOX22 and LMC23 emission line metallicities considered here and the absorption metallicity of GRB~090323 is driven by the significantly higher \Rtwo\ and \Rthree\ values than expected given its absorption line metallicity (see Fig.~\ref{fig:ZabsvslogR}). In contrast, the N/O ratio determined using the strong line diagnostic from \citet{teh96} is $\log({\rm N/O})=-0.61\pm 0.13$ or $\log({\rm N/O})=-0.57\pm 0.10$ when using the diagnostic from \cite{pc09}, which although high, is consistent with the N/O-O/H relation from \citet{am13} when extrapolated to the super-solar metallicity measured in absorption. This GRB host galaxy therefore has non-standard line ratios with either enhanced \Rthree\ and \Rtwo\ if the absorption line metallicity is assumed, or, if the \Rtwothree\ and \Rthree\ NOX22 emission line metallicities are true, then a significantly enhanced N/O ratio. Below we consider these two possibilities and corresponding implications separately.

The former scenario, where the \Rtwo\ and \Rthree\ lines ratios are enhanced, would have been possible in the presence of an AGN, which can boost the \Rthree\ ratio. However, the \mbox{\oiiib/\hb}, \mbox{\niib/\ha} and \mbox{\sii/\ha} line ratios place the host galaxy of GRB~090323 within the star-forming region of the Baldwin-Phillips-Terlevich, or so-called `BPT' diagram \citep{bpt81}, and of the [\ion{S}{ii}] variant of the BPT diagram \citep{Veilleux&Osterbrock87}, indicating that the observed line emission is excited by stars only, and not an AGN. Instead, the apparent enhanced \oiiib\ and \oii\ line fluxes may be a result of selection effects in the current samples used to calibrate the metallicity diagnostics, which do not include high-$z$ galaxies with $12+\log({\rm O/H})\gtrsim 8.4$. This would not affect the majority of our galaxy sample, with the host galaxy of GRB~090323 being the only case where the metallicity may be $12+\log({\rm O/H})\gg 8.4$. It is also worth noting that, for its given stellar mass and SFR, the metallicities obtained from the NOX22 diagnostics for the host galaxy of GRB090323 are in reasonable agreement with those inferred from the \Rtwothree{}-dominated MZR at $z=3.5$ ($12+\log({\rm O/H})=8.3$; \citealt{mcm+09}) and FMR ($12+\log({\rm O/H})=8.55$; \citealt{mcm+10}). This at least indicates that the oxygen emission lines in the host galaxy of GRB090323 are comparable to other star-forming galaxies at high-$z$ of the same stellar mass and SFR. One way to investigate possible selection effects at high metallicity in diagnostic calibration samples could be to study the dependence between \Rtwo\ and \Rthree\ as a function of stellar mass at high-$z$, which would reduce the selection effects present at the high mass end since no weak spectral lines are needed. 

The alternative scenario that the N/O (and thus N/S) ratio is enhanced, would cause the DKS16 metallicity to be over-estimated, since a fixed N/O-O/H relation is assumed in this diagnostic. It would then just be a coincidence that the GRB absorption metallicity and host galaxy $\log({\rm N/O})$ ratio are consistent with the locally observed N/O-O/H relation, which is the reason for the good agreement between the absorption and the DKS16 metallicity. A metallicity of $12+\log({\rm O/H})\sim 8.4$ as measured with the NOX22 diagnostics would put the N/O ratio a factor of 10 above what is observed in the local Universe. Although a small sample of nearby GRB host galaxies ($z<0.1$) with \Te-based metallicity measurements have previously been found to have enhanced N/O ratios \citep{wsv+07}, the ratios were not as large as is observed in the host galaxy of GRB~090323. The N/O ratio that we report here is based on strong emission lines, and as such is diagnostic-dependent. The \citet{teh96} diagnostic can over-predict N/O by up to 0.3~dex compared to measurements using auroral lines \citep[e.g.][]{kon+17}, although this is a worse case scenario. Even when using auroral lines to measure the N/O ratio, there is dispersion in the O/H-N/O relation, which has been found to correlate with EW(\hb) \citep{ism+06} and SFR \citep{am13}. The dispersion has further been found to decrease when considering the relation between N/O and galaxy stellar mass \citep{pc09,am13,mfc16}. However, even when taking this into account, the N/O ratio observed in the host galaxy of GRB~090323 continues to be $\sim 0.4$~dex higher than expected.

N/O ratios enhanced by $>1$~dex have been observed in other galaxies at $z\approx$3--10 with metallicities in the range $12+\log({\rm O/H})=7.5-8.0$ and N/O ratios $\log({\rm N/O})=-0.7- -0.1$ \citep{fvh+03,clr+12,bsc+23,csb+23,msk+24}. Such non-standard relative abundances have been observed in globular clusters (GCs), leading to the suggestion that high-$z$ galaxies with an enhanced N/O may contain proto-GCs \citep{csb+23,sps+23,msk+24}. Suggested origins for high N/O ratios at high-$z$ are very-massive or super-massive stars that efficiently pollute their environment with hydrogen-burning elements \citep{gch+18,csb+23,nu23,won+23,Vink+23}, and the enhancement of nitrogen from the stellar winds of young Wolf-Rayet (WR) stars that are in the hydrogen burning phase (WN phase), with fast rotation further extending this phase \citep{vdo+18,csb+23,kf23,sps+23,msk+24}. These effects may also occur alongside specific star-formation histories \citep{kf23}. Inflows have also been used to explain the N/O enhancement whereby accretion of pristine gas dilutes the metallicity without affecting the N/O ratio \citep[e.g.][]{kh05,am13,kon+17}.

The fact that the NOX22 metallicities place the host galaxy of GRB~090323 in a similar region of the N/O-O/H parameter space as other high-$z$ galaxies could be taken as indication that the NOX22 metallicities are correct, but an explanation is then needed for why the absorption metallicity is so much larger. One possibility is that the GRB line of sight crossed a particularly metal-rich but non-representative cloud within the host galaxy, which could arise if the galaxy had a very inhomogeneous metallicity distribution. As already mentioned in Section~\ref{ssec:fluxes}, two strong absorbers were detected in the afterglow of GRB~090323, which were attributed to absorption from two interacting systems separated by $\Delta v=660$~km~s$^{-1}$ \citep{srg+12}. Both absorbers were found to have super-solar metallicities \citep{srg+12}, but it may be possible that these metal-rich absorbers are mixed in with other regions of metal-poor gas. For example, there is evidence that bursts of star formation and induced pristine gas accretion in strongly interacting galaxies can introduce significant variation in the metallicity of the interacting system \citep{mla+08,pmt+11,tck+12,gfc15,tvm+19,swd+22}. The absorption metallicity given in Table~\ref{tab:hostprops} is the metallicity measured from the total absorption profile \citep{wsb+17}. The NOX22 and LMC23 diagnostics considered in this paper are calibrated against stacked SDSS spectra from \citet{ccm+17} at the high metallicity end ($12+\log({\rm O/H})\gtrsim 8.4$), and thus the absorption and emission line metallicities both correspond to averaged measurements. To check that the possible contribution to the observed spectrum from two emission components is not the cause of the unusual line ratios, we measured the emission metallicity and N/O ratio using the results from our two component fits to the \nii\ and \ha\ lines (Section~\ref{ssec:fluxes} and Fig.~\ref{fig:grb090323A-2Dspec}). The results remained very comparable, with NOX22 metallicities of $12+\log({\rm O/H})\sim 8.3$ and slightly lower N/O ratios in the range $\log({\rm N/O})\sim -0.7 - -0.8$, but still well above the typical values expected from the locally observed N/O-O/H relation.

It is therefore not clear what the origin is of the large disagreement between the absorption and NOX22 and LMC23 emission line metallicities for the host galaxy of GRB~090323, with both the possibility of either enhanced \Rthree\ and \Rtwo\ line ratios, or of an enhanced N/O ratio, implying an atypical ISM in a potentially merging galaxy. The \NtwoStwo\ diagnostic has also been found to give good agreement with the metallicities of \hii\ regions predicted by the BPASS \citep[Binary Population and Spectral Synthesis;][]{seb16,esx+17,xse18}, and metallicity gradients predicted by the modified {\sc L-Galaxies} 2000 galaxy evolution model \citep{yhf+21}. Although this may be a byproduct of assumptions made on the N/O-O/H relation in these simulations. It should be possible to further verify the robustness of the \NtwoStwo\ diagnostic with future, more sensitive JWST observations capable of detecting the weaker \nii\ and \sii\ line doublets in all but the most metal right galaxies in our sample.

\begin{figure*}
\begin{minipage}[H]{1\textwidth}
    \includegraphics[width=0.33\linewidth]{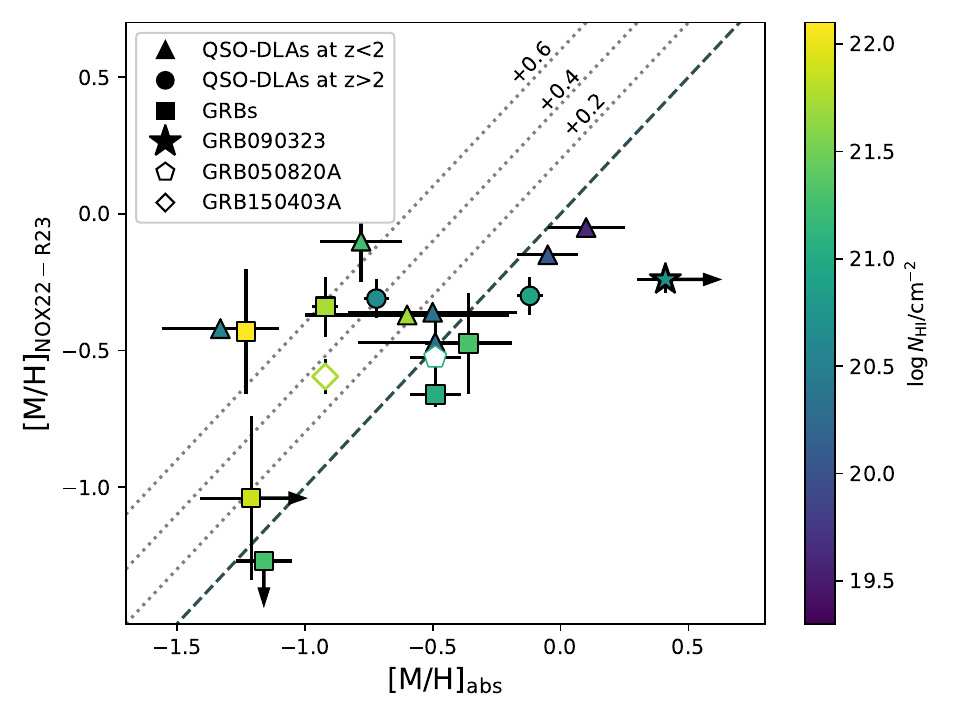}
    \includegraphics[width=0.33\linewidth]{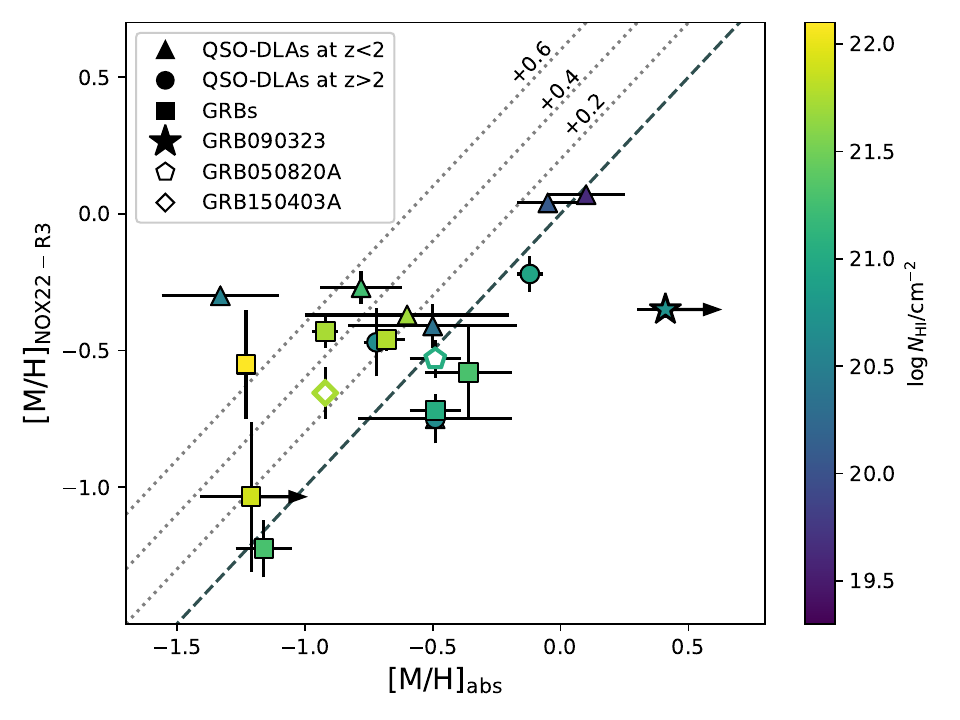}
    \includegraphics[width=0.33\linewidth]{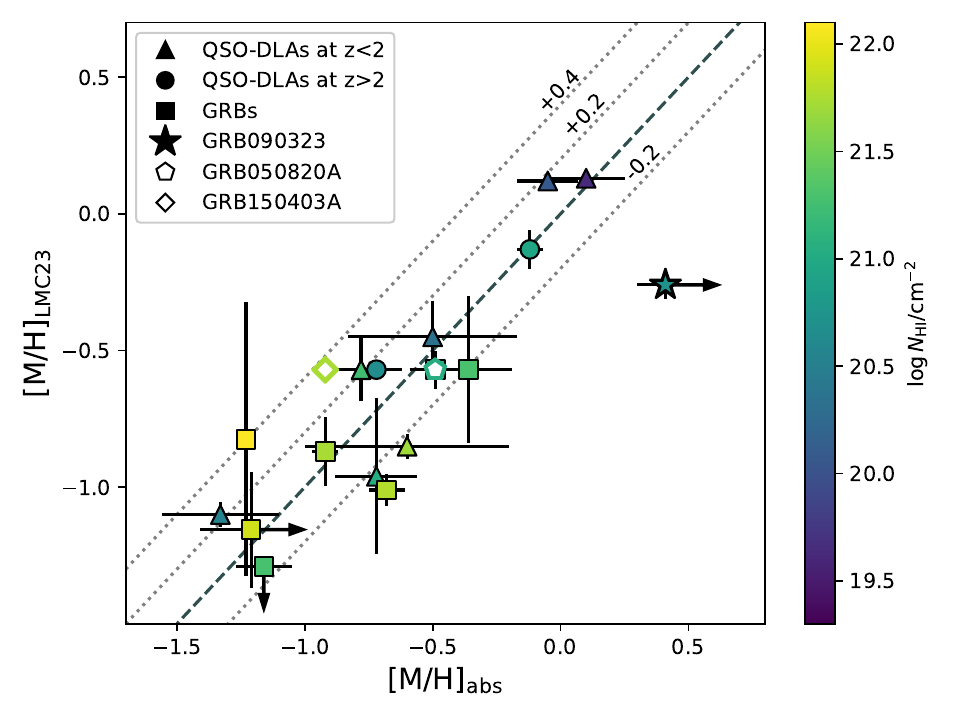}
    \caption{The NOX22 \Rtwothree\ and \Rthree\ and the LMC23 $\hat{R}$ emission line metallicities against absorption line metallicities, as in Fig.~\ref{fig:ZabsvsZem-NOX22}, but now colour-coded by the HI column density, \nh.}
    \label{fig:ZabsvsZem_NHcolor}
\end{minipage}
\end{figure*}

\subsection{Comparison with simulations}\label{ssec:smlts-comp}
Several efforts have been made to quantify the relation between single sightline absorption metallicities and galaxy-integrated emission line metallicities using cosmological hydrodynamical simulations \citep{mt20,mct21,mt23,agw+23}. These have generally found absorption line metallicities to be lower than emission line metallicities, although the difference is smaller for more metal-rich host galaxies and sightlines. The results from our GRB host galaxy sample are in qualitative agreement with this predication when using the NOX22 \Rtwothree\ and \Rthree\ diagnostics, with the emission metallicities being generally larger than the absorption metallicities, but with increased agreement for higher metallicity absorbers (see Fig.~\ref{fig:ZabsvsZem-NOX22}). \citet{agw+23} used the Evolution and Assembly of GaLaxies and their Environments (EAGLE) simulations \citep{scb+15} to investigate third parameter dependencies on the absorption-emission line metallicity relation. They found that absorber sightlines with very small impact parameters or offsets from the galaxy centre, which thus probe a higher column density of material within the galaxy disc, had absorption metallicities that were in better agreement with the average metallicity of the star forming regions (probed by emission lines). In our GRB and QSO-DLA sample we do not find any dependence on the difference in absorption and emission line metallicities with impact parameter (Fig.~\ref{fig:ZabsvsZem-NOX22} and \ref{fig:ZabsvsZem-LMC23}). However, the impact parameters of our QSO-DLA sample are generally large ($>6$~kpc), whereas the good agreement between absorption and emission line metallicities predicted by the EAGLE simulations is generally for impact parameters $b<0.05*R_{200}$ where $R_{200}$ is the radius from the galaxy centre where the average density is 200 times the critical density at the respective redshift. Typical values of $R_{200}$ in EAGLE are 40--90~kpc, and $0.05*R_{200}$ thus corresponds to 2.0--4.5~kpc. Nevertheless, although we do not know the offset of the GRB position from the host galaxy centre for the majority of our sample, on average we would expect GRBs to probe the very central regions of their host galaxies. The fact that our GRB sample of data points in Fig~\ref{fig:ZabsvsZem-NOX22} and \ref{fig:ZabsvsZem-LMC23} do not appear to lie closer to the line of equality (black dashed) than the distribution of QSO-DLA data points thus seems contrary to the predictions from the EAGLE simulations. Nevertheless, there is of course the complication that more massive galaxies will be physically larger, making comparisons between absolute offsets of impact parameters less meaningful. A fairer comparison may thus be to use the impact parameter normalised by the galaxy effective radius, although at $z>2$ the emission counterparts to QSO-DLAs appear to have comparable effective radii to GRB host galaxies \citep{rkc+21,bbf16}.

Smaller impact parameters have been found to have larger HI column densities, \nh, \citep{cl03,ckr+05,cwr+07,pqr+16,kmf+17,kbs+22,agw+23}, which can be understood by the fact that the sightline probes more central, and thus denser regions of the host galaxy \citep[e.g][]{kmc+20}. One may therefore expect greater agreement between absorption and emission line metallicities for those sightlines with larger HI column density. \nh\ is a parameter that is measured for our complete GRB and QSO-DLA sample by selection, and in Fig.~\ref{fig:ZabsvsZem_NHcolor} we thus plot the emission against absorption line metallicity for the NOX22 and LMC23 diagnostics, as in Figs.~\ref{fig:ZabsvsZem-NOX22} and \ref{fig:ZabsvsZem-LMC23}, but now with the data points colour-coded by the \nh. There is a possible indication that data points in the lower left of the plots in Fig.~\ref{fig:ZabsvsZem_NHcolor} have the largest column densities, and \nh\ decreases as we go to larger metallicities, at the top right corner of the plots. However, the relative offsets between the absorption and emission line metallicities do not show any clear dependence with \nh. The results from \citet{agw+23} imply some level of dependency on the relation between absorption and emission metallicities with \nh, but they found large scatter in the relation due to the intrinsic inhomogenities that exist in galaxies. It is therefore possible that we do not see a clear absorption metallicity-emission metallicity-\nh\ dependence in our observations due to our small sample size. Nevertheless, by investigating the dependence of the relation between the emission and absorption metallicities with \nh\ we can at least include the GRB host galaxy sample (unlike when considering impact parameter).

The lack of any clear, third-parameter dependency on the relation between absorption and NOX22 emission line metallicities along GRB and QSO-DLA sightlines may also imply that metals within the neutral phase ISM (and possibly also CGM) are poorly mixed, and the scatter observed in Figs.~\ref{fig:ZabsvsZem-NOX22} and \ref{fig:ZabsvsZem_NHcolor} could thus be indicative of the large intrinsic variation in the regions of the galaxy probed in absorption depending on the line of sight. Sightlines with the same impact parameter and \nh\ may probe very different regions of the galaxy depending on the galaxy orientation, the stellar mass, and the level of mixing, which in turn is dependent on feedback processes. Moreover, whereas QSO-DLA sightlines will cross through the full radial extent of the intervening galaxy, GRBs may lie at the front side of their host galaxy, and thus on average their sightlines will cross through 50\% of the galaxy along the radial direction. However, the good agreement between absorption and emission line metallicities when using the $\hat{R}$ diagnostic implies that the metallicity of the ionised and neutral material is relatively homogeneous, and that the effect of any metallicity gradients are averaged out along the radial and longitudinal direction in absorption and emission line probes. To understand the lack of an \nh-dependency in this case may therefore require analysis using zoom-in hydrodynamic simulations that include the relevant physics on the formation of molecular clouds and stellar feedback that is necessary to capture the smaller scale inhomogeneities present in star forming regions and the ISM.

It will also be important to increase the samples of QSO-DLA and GRBs with both emission and absorption line metallicites in order to be able to average out the intrinsic scatter that is likely introduced by the pencil-beam sightline offered by GRBs and QSO-DLAs. Knowing the characteristic properties of the GRB host galaxies and QSO-DLA emission counterparts is also important to be able to study the emission-absorption metallicity relation in bins of stellar mass, normalised impact parameter and \nh, for example.

\section{Conclusions}
\label{sec:concl}
In this paper we present the first investiation on a sample of GRB host galaxies on the relation between the gas-phase metallicity in star forming regions and in the neutral cold interstellar gas. This is probed through emission and absorption using the incredible IR sensitivity of NIRSpec as part of our cycle-1 JWST programme. We find good agreement between the absorption metallicities and the emission metallicities determined with the LMC23 $\hat{R}$ diagnostic ($\sigma =0.2$~dex). The DKS16 \NtwoStwo\ diagnostic also shows promise, but a larger sample of galaxies with \sii\ detections is required to verify the consistency with the \NtwoStwo\ diagnostic. Although our results are dependent on the emission line metallicity diagnostic, we find that, when considering only the most reliable emission diagnostics, there is a relation between the two metallicity probes (even if not one-to-one). This opens the possibility of combining both emission and absorption line probes in the future to study the cosmic chemical evolution down to lower mass galaxies than is currently possible through emission line studies alone. At high-$z$ GRBs are likely to have small and faint host galaxies that could be significant sources of ionising photons \citep{smc+13}, but for which spectra cannot be taken even with NIRSpec. The combination of emission and absorption line probes could therefore provide a less biased view of the chemical enrichment of galaxies at high-$z$.

The first results on new strong-line metallicity diagnostics for high-$z$ galaxies based on sensitive JWST data have started to be published, but significant progress is still required in the size of the calibration samples currently available, and especially in the range of galaxy properties covered by the calibration samples. A more conclusive analysis on the relation between emission and absorption line metallicities may thus require GRB host galaxies with direct, \Te-based emission line metallicities. In order to detect the temperature-sensitive, weak auroral lines in our sample of GRB host galaxies, of which [\ion{O}{iii}]$\lambda 4363$ is the strongest, each galaxy would need to be observed for several hours, even with JWST. Nevertheless, the resulting observations would provide possibly the only definitive result on how absorption and emission line metallicities compare, paving a way for both probes to be used in unison, as well as enabling the relation between the ionised and neutral phase gas in the ISM of high-$z$ galaxies to be investigated.

\section*{Acknowledgements} \label{sec:Acknowledgements}
We thank C. Ledoux for sharing the results from their absorption line analysis on the afterglow UVES spectrum of GRB~080804. PS acknowledges support from the UK Science and Technology Facilities Council, grant reference ST/X001067/1. LC is supported by DFF/Independent Research Fund Denmark, grant-ID 2032–00071. The Cosmic Dawn Center (DAWN) is funded by the Danish National Research Foundation under grant DNRF140. ADC acknowledges support from the Swiss National Science Foundation under grant 185692. AR acknowledges support from the INAF project Premiale Supporto Arizona \& Italia. R.G.B. acknowledges financial support from the Severo Ochoa grant CEX2021-001131-S funded by MCIN/AEI/10.13039/501100011033 and to PID2022-141755NB-I00.

\section*{Data availability}
The data underlying this article are available in the MAST Data Discovery Portal at https://stdatu.stsci.edu/datadownloads.html, and can be accessed with proposal ID 2344.

\bibliographystyle{mnras}
\bibliography{GRBabs.bib,robyates.bib}

\section*{Affiliations}
$^1$ Department of Physics, University of Bath, Claverton Down, Bath BA2 7AY, UK \\
$^2$ Centre for Astrophysics Research, University of Hertfordshire, Hatfield, AL10 9AB, UK \\
$^3$ Cosmic Dawn Center (DAWN) \\
$^4$ Niels Bohr Institute, University of Copenhagen, Jagtvej 128, 2200 Copenhagen N, Denmark \\
$^5$ European Southern Observatory, Karl-Schwarzschild Str. 2, 85748 Garching bei München, Germany \\
$^6$ Department of Astronomy, University of Geneva, Chemin Pegasi 51, 1290 Versoix, Switzerland \\
$^7$ INAF – Osservatorio di Astrofisica e Scienza dello Spazio, Via Piero Gobetti 93/3, 40129 Bologna, Italy \\
$^8$ Space Science Data Center (SSDC) - Agenzia Spaziale Italiana (ASI), 00133 Roma, Italy \\
$^9$ INAF - Osservatorio Astronomico di Roma, Via Frascati 33, 00078 Monte Porzio Catone, Italy \\ 
$^{10}$ Centre for Astrophysics and Cosmology, Science Institute, University of Iceland, Dunhagi 5, 107 Reykjavik, Iceland \\
$^{11}$ Department of Physics \& Astronomy, University of Utah, Salt Lake City, UT 84112, USA \\
$^{12}$ Department of Astrophysics/IMAPP, Radboud University, 6525 AJ Nijmegen, The Netherlands \\
$^{13}$ Istituto Nazionale di Astrofisica (INAF) Istituto di Astrofisica Spaziale e Fisica Cosmica, Via Alfonso Corti 12, I-20133, Milano, Italy \\
$^{14}$ School of Physics and Astronomy, University of Leicester, University Road, Leicester, LE1 7RH, UK \\ 
$^{15}$ Astronomical Institute, Czech Academy of Sciences, Fri\v cova 298, Ond\v rejov, Czech Republic \\
$^{16}$ GEPI, Observatoire de Paris, Université PSL, CNRS, 5 Place Jules Janssen, 92190 Meudon, France \\
$^{17}$ Institut d'Astrophysique de Paris and Sorbonne Universit\'e, 98bis Boulevard Arago, 75014, Paris, France \\
$^{18}$ Excellence Cluster ORIGINS, Boltzmannstra{\ss}e 2, 85748 Garching, Germany \\
$^{19}$ Ludwig-Maximilians-Universit\"at, Schellingstra{\ss}e 4, 80799 M\"unchen, Germany \\
$^{20}$ Department of Astronomy \& Astrophysics, The University of Chicago, 5640 S Ellis Ave., Chicago, IL 60637, USA \\
$^{21}$ Mathematics, Informatics, Physics and Earth Science Department of Messina University, Papardo campus, Via F. S. D'Alcontres 31, 98166, Messina, Italy \\
$^{22}$ Space Telescope Science Institute, 3700 San Martin Dr, Baltimore, MD 21218, USA \\
$^{23}$ Instituto de Astrof\'isica de Andaluc\'ia. CSIC. Apartado de correos 3004. 18080, Granada, Spain \\
$^{24}$ School of Physics and Astronomy, University of Birmingham, Birmingham B15 2TT, UK \\
$^{25}$ Institute for Gravitational Wave Astronomy, University of Birmingham, Birmingham B15 2TT, UK \\
$^{26}$ Clemson University, Department of Physics and Astronomy, Clemson, SC 29634, USA \\
$^{27}$ The George Washington University, Department of Physics, 725 21st street NW, Washington DC 20052, USA \\
$^{28}$ Astrophysics Research Institute, Liverpool John Moores University, 146 Brownlow Hill, Liverpool L3 5RF, UK \\
$^{29}$ Anton Pannekoek Institute for Astronomy, University of Amsterdam, P.O. Box 94249, 1090GE Amsterdam, The Netherlands \\
$^{30}$ Physics Department, University of Calabria, 87036 Arcavacata di Rende, CS, Italy \\
${31}$ INFN – Laboratori Nazionali di Frascati, Via Enrico Fermi 54, 00044 Frascati, RM, Italy  Italy \\
$^{32}$ INAF, Osservatorio Astronomico di Brera, via E. Bianchi 46, 23807, Merate, Italy \\
$^{33}$ The Oskar Klein Centre, Department of Physics, Stockholm University, AlbaNova, SE-106 91 Stockholm, Sweden \\
$^{34}$ Artemis, Observatoire de la Côte d’Azur, Université Côte d’Azur, CNRS, 06304 Nice, France \\
$^{35}$ School of Physics and Astronomy, University of Southampton, Southampton, SO17 1BJ, UK

\appendix
\section{Gaussian fits to spectral lines}
\begin{figure*}
\begin{minipage}[H]{1\textwidth}
    \includegraphics[width=0.33\linewidth]{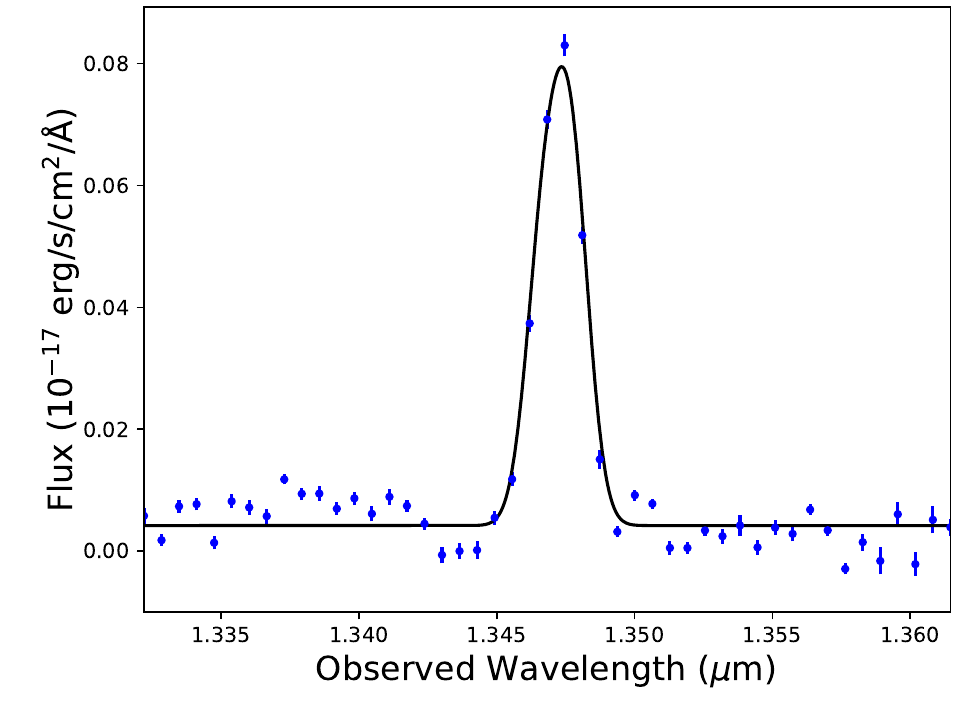}
    \includegraphics[width=0.33\linewidth]{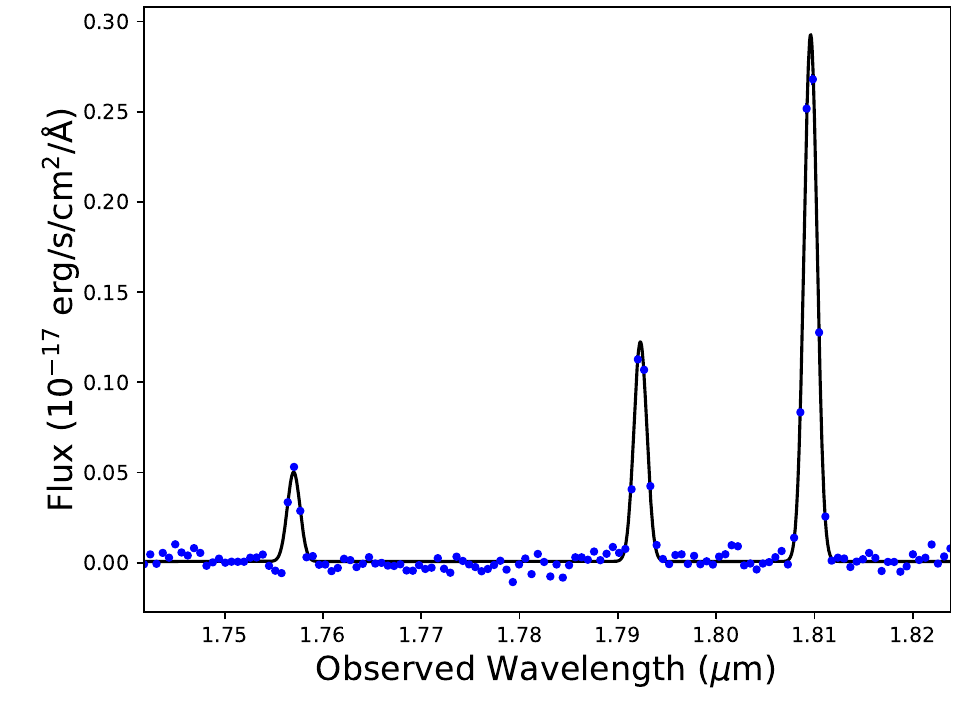}
    \includegraphics[width=0.33\linewidth]{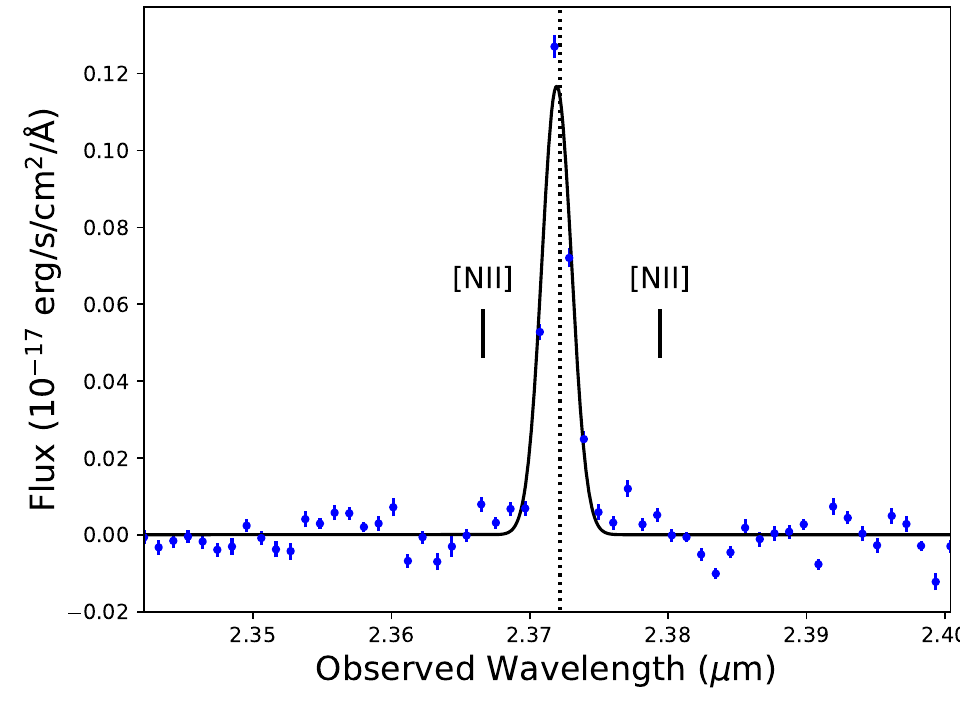}
    \includegraphics[width=0.33\linewidth]{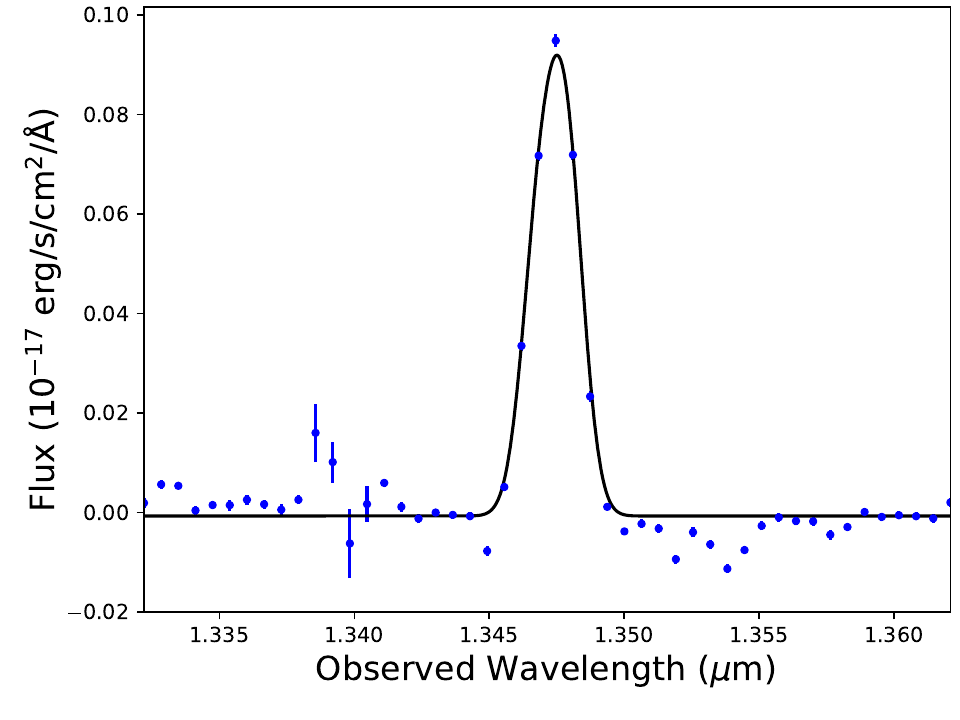}
    \includegraphics[width=0.33\linewidth]{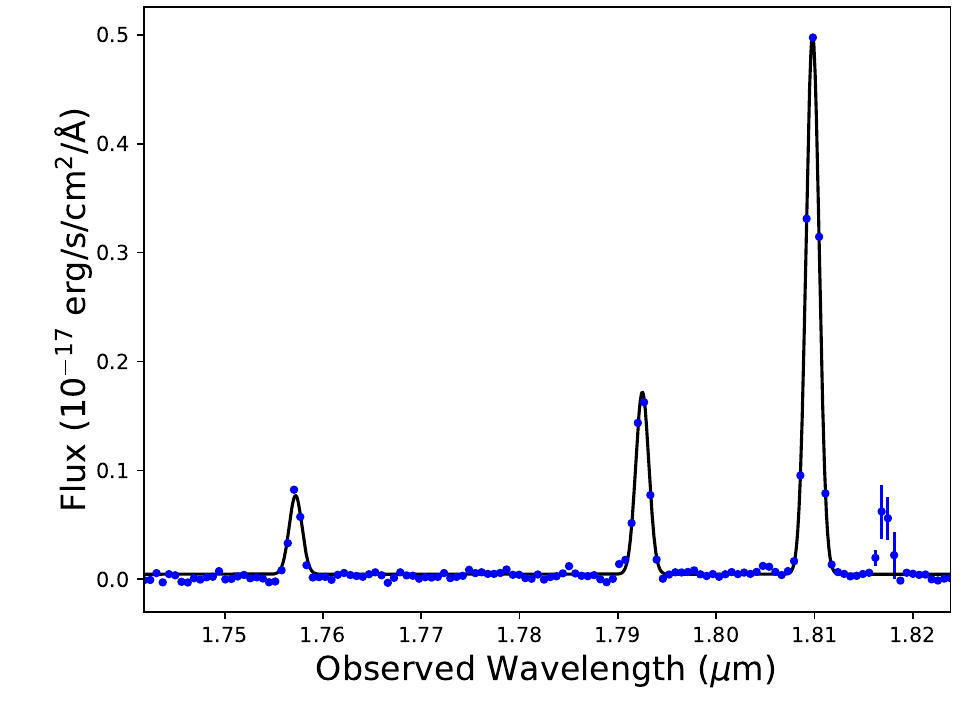}
    \includegraphics[width=0.33\linewidth]{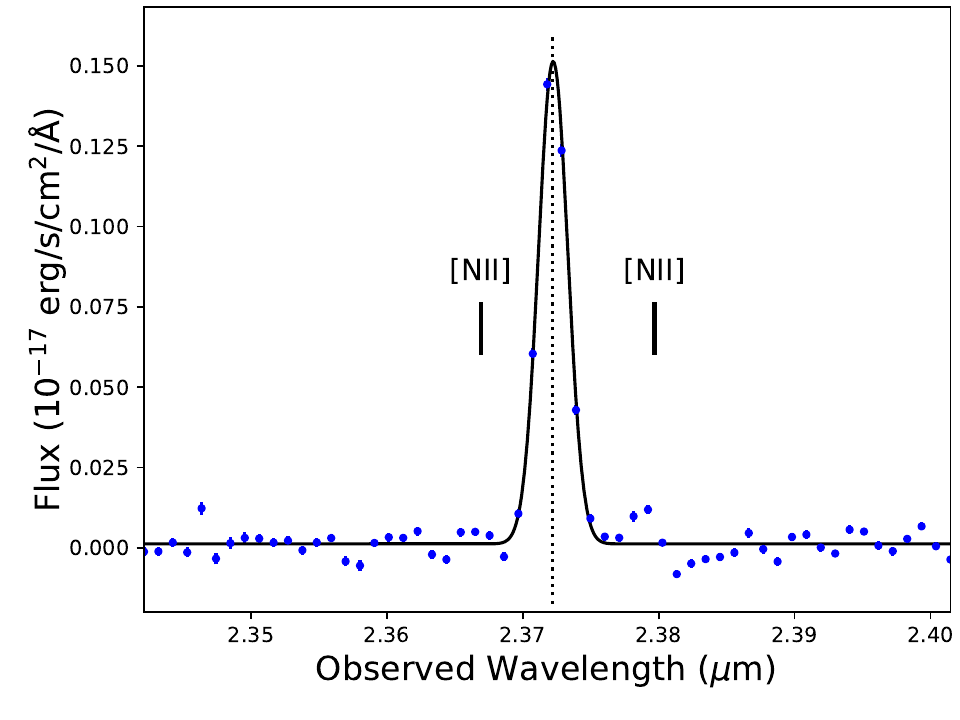}
    \includegraphics[width=0.33\linewidth]{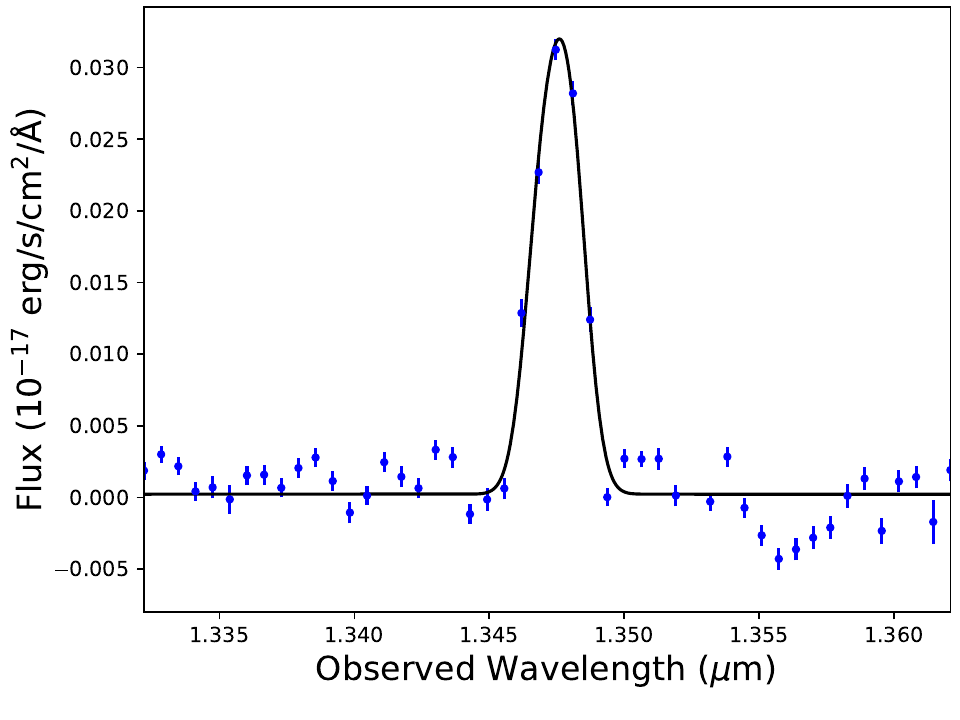}
    \includegraphics[width=0.33\linewidth]{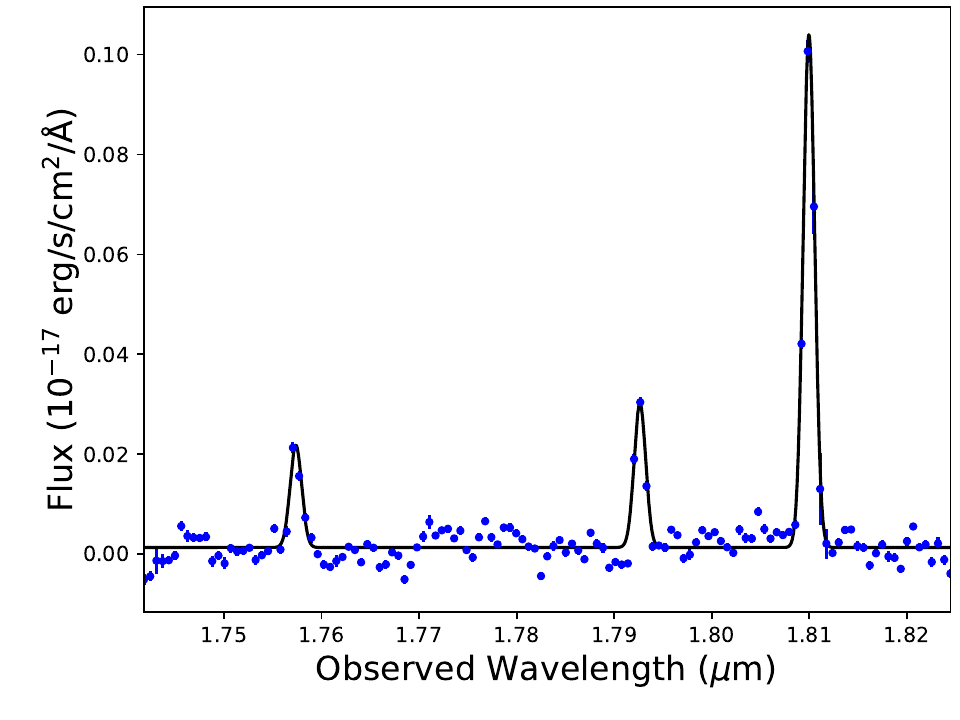}
    \includegraphics[width=0.33\linewidth]{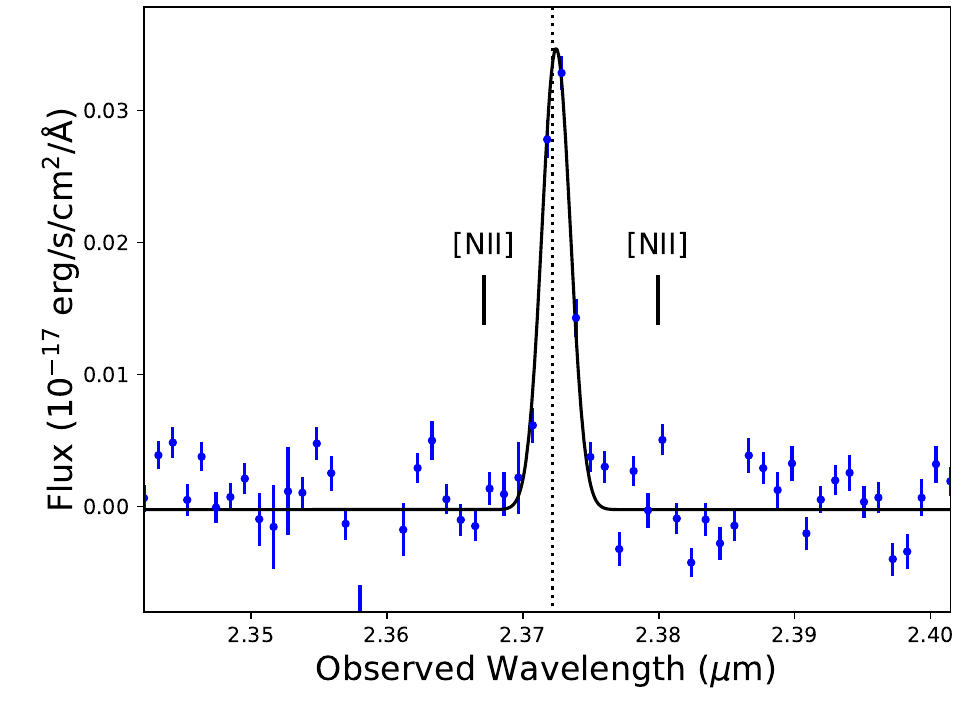}
    \caption{Emission line spectra of the host galaxy of GRB~050820A, taken from components A (top), B (middle) and C (bottom). The spectra are zoomed in on \oii\ (left), \hb\ and \oiii\ (centre), and \ha\ (right). In all cases the redshift and velocity dispersion were fixed to the best-fit values to \hb\ and the \oiii\ doublet. The best-fit velocity to the lines from component C was below the instrumental resolution, and it was therefore left as a free parameter for all lines, but the redshift was fixed to the best-fit value to \hb\ and \oiii. The location of the undetected \nii\ line doublet is indicated in the right-most panel in each row. The vertical dotted lines correspond to the observer-frame position of \ha\ for a systemic redshift $z=2.6133$ determined from the galaxy-integrated spectrum, which provides an indication of the relative velocity shift of each component. A summary of the best-fit parameters and line fluxes are given in Table~\ref{tab:linefluxes}.
    }
    \label{fig:grb050820A}
\end{minipage}
\end{figure*}

\begin{figure*}
\begin{minipage}[H]{1\textwidth}
    \includegraphics[width=0.33\linewidth]{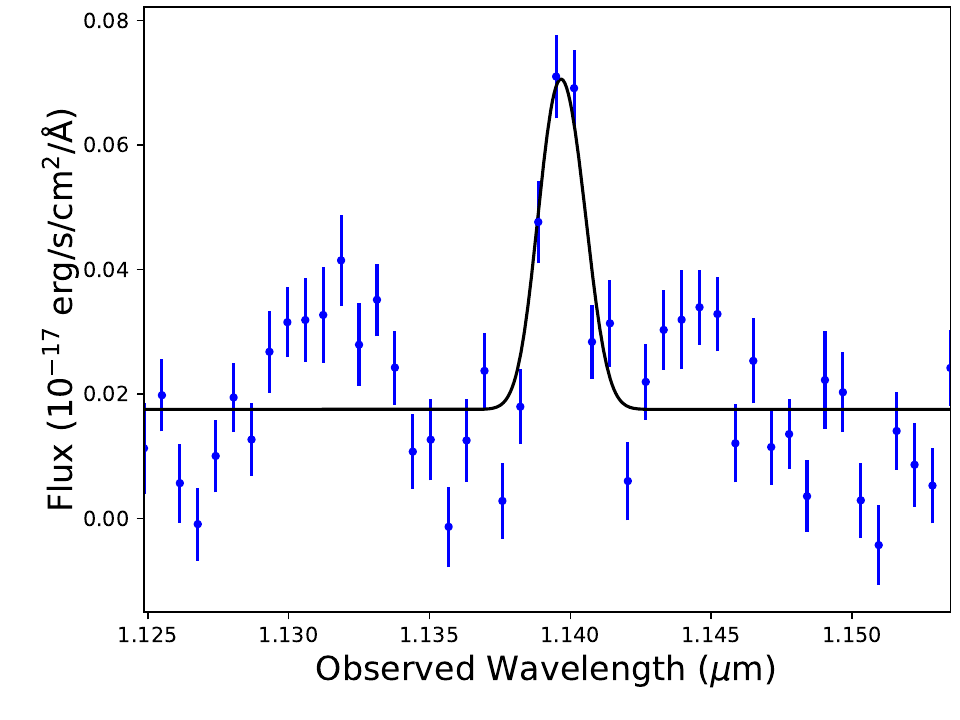}
    \includegraphics[width=0.33\linewidth]{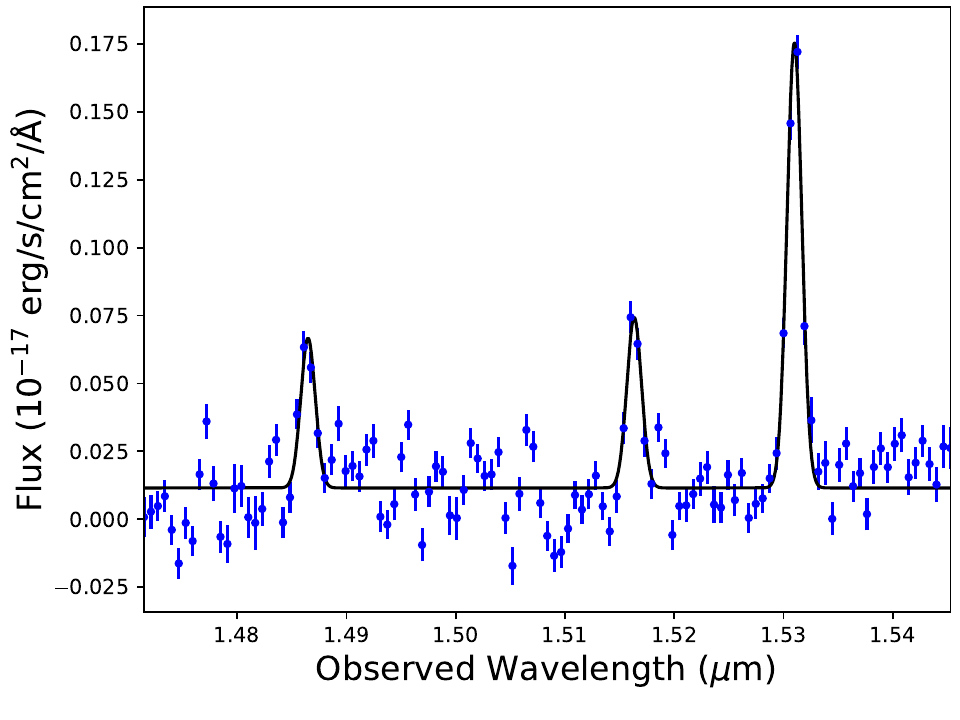}
    \includegraphics[width=0.33\linewidth]{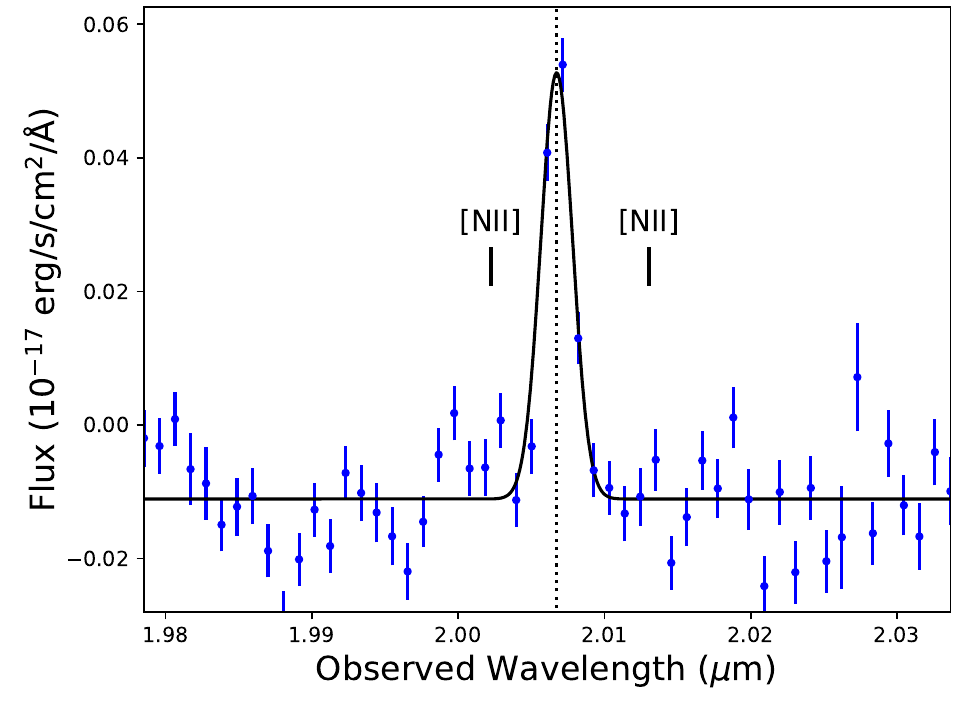}
    \includegraphics[width=0.33\linewidth]{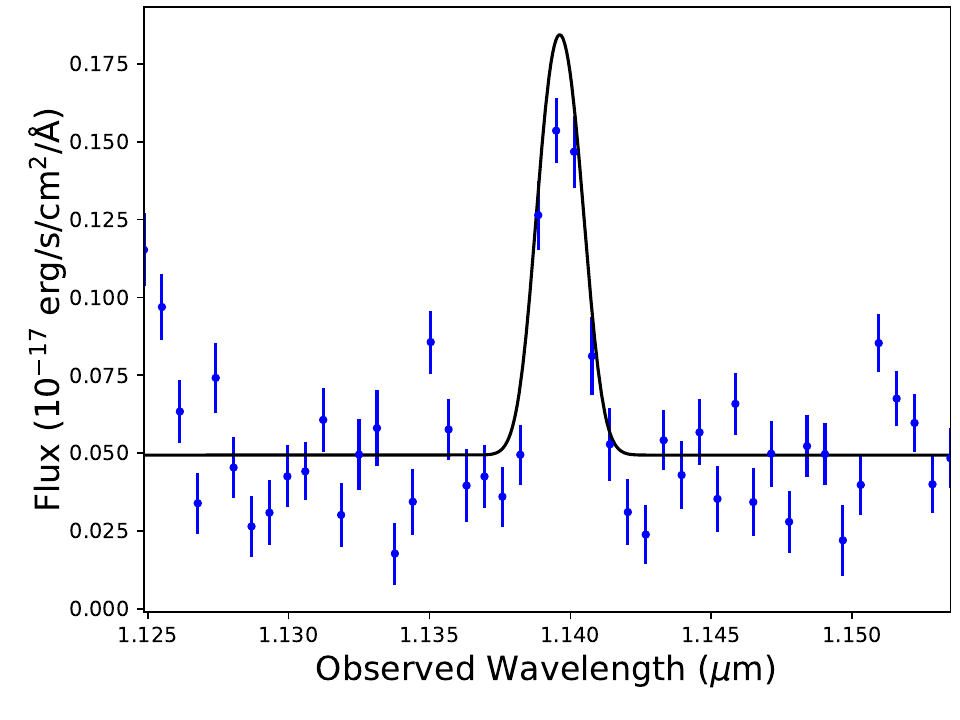}
    \includegraphics[width=0.33\linewidth]{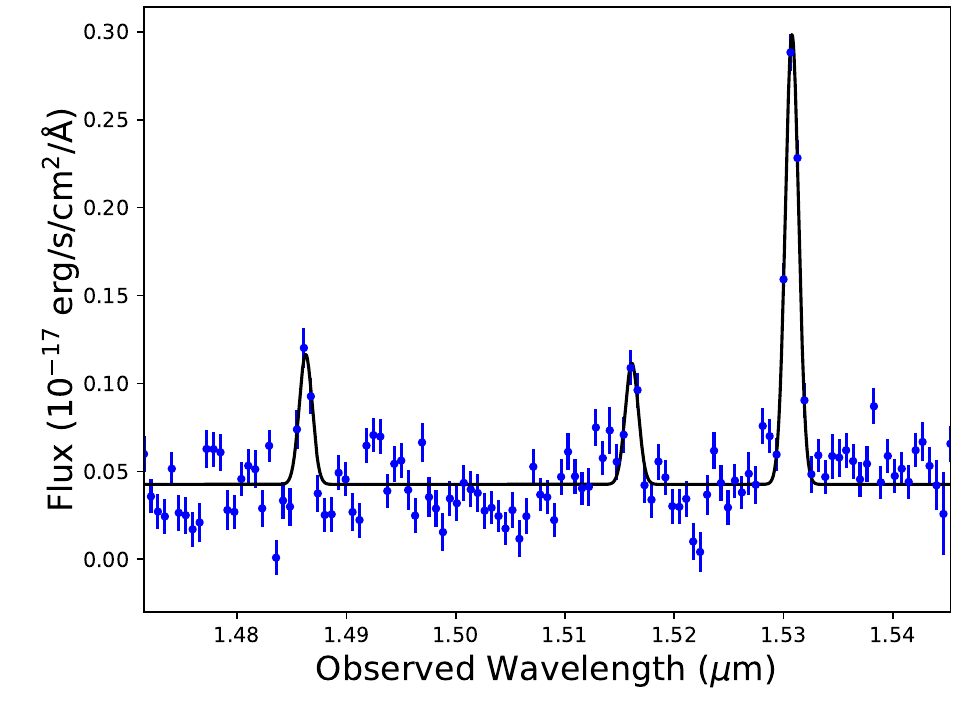}
    \includegraphics[width=0.33\linewidth]{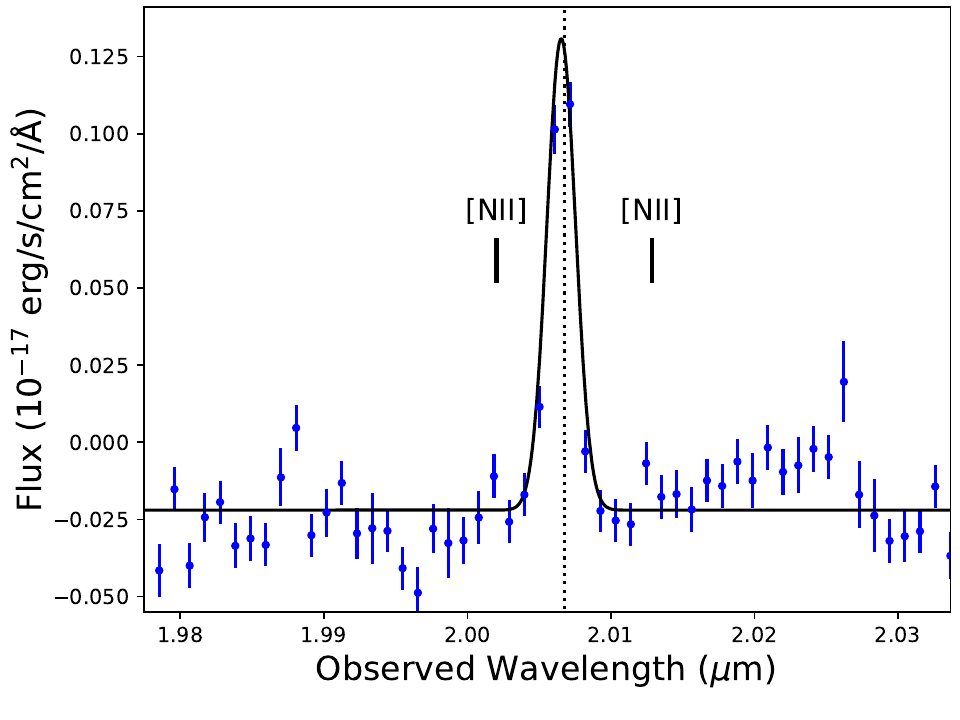}
    \includegraphics[width=0.33\linewidth]{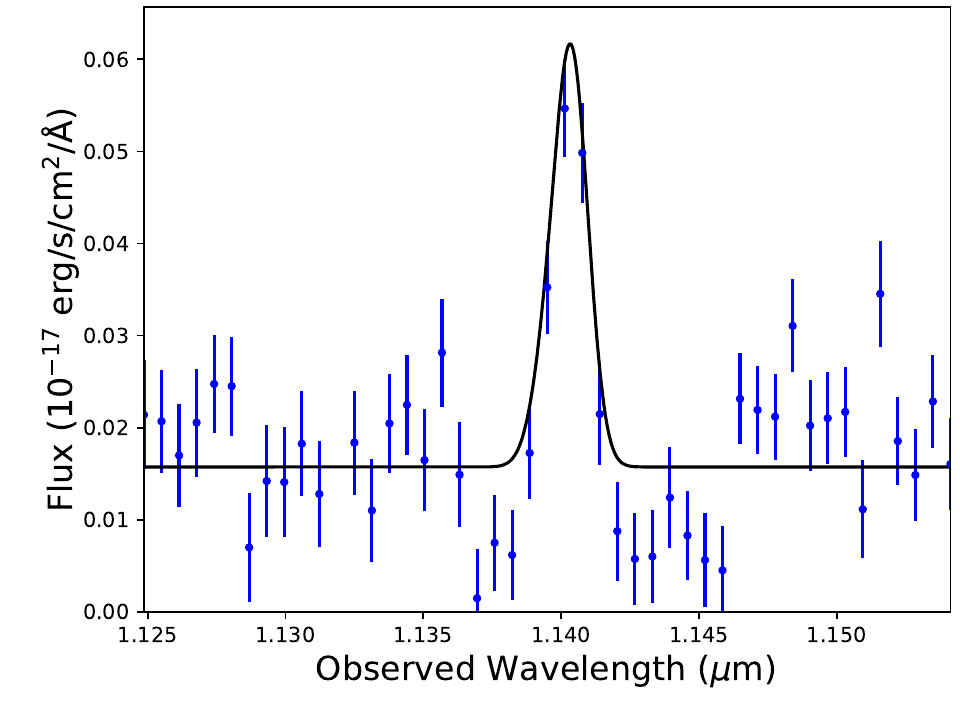}
    \includegraphics[width=0.33\linewidth]{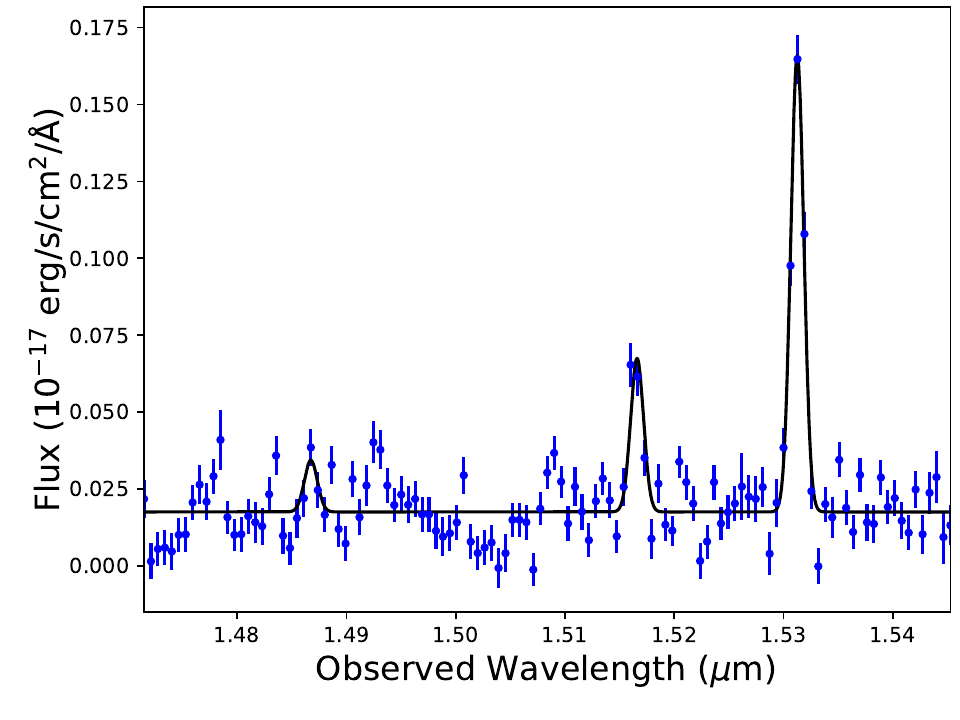}
    \includegraphics[width=0.33\linewidth]{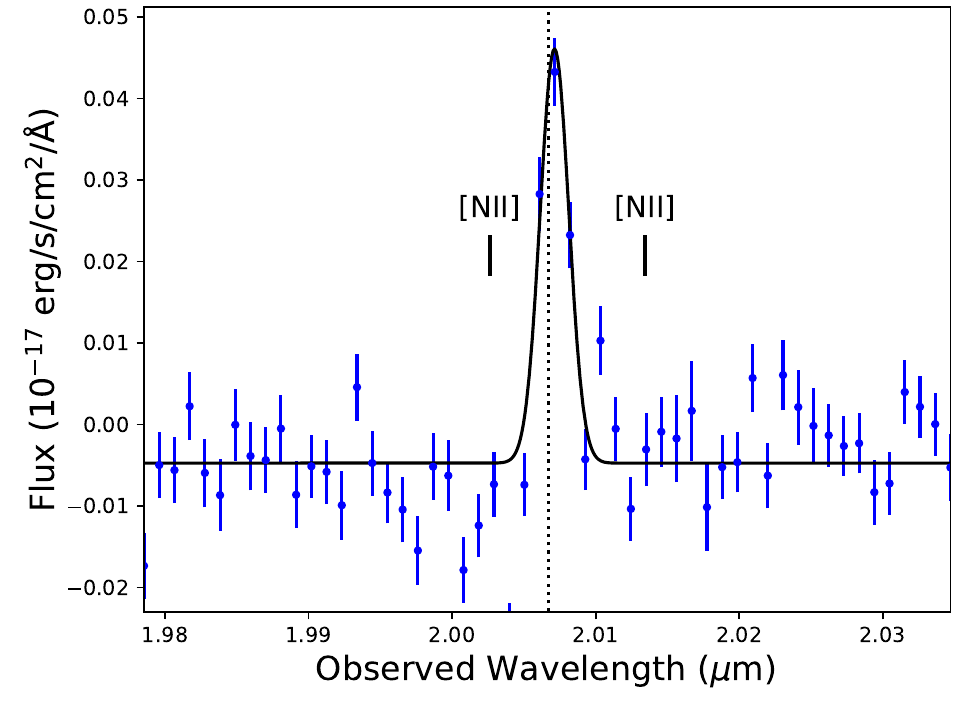}   
    \caption{Emission line spectra of the host galaxy of GRB~150403A, taken from components A (top), B (middle) and C (bottom). The spectra are zoomed in on \oii\ (left), \hb\ and \oiii\ (centre), and \ha\ (right). For component A, the redshift and velocity dispersion was fixed to the best-fit values to \hb\ and the \oiii\ doublet. The best-fit velocity to the lines from component B and C were below the instrumental resolution, and it was therefore left as a free parameter for all lines, but the redshift was fixed to the best-fit values fitted to \hb\ and \oiii. As in Fig.~\ref{fig:grb050820A}, the location of the undetected \nii\ line doublet is indicated in the right-most panel in each row, and the vertical dotted lines correspond to the observer-frame position of \ha\ for a systemic redshift $z=2.0570$ determined from the galaxy-integrated spectrum. A summary of the best-fit parameters and line fluxes are given in Table~\ref{tab:linefluxes}.
    }
    \label{fig:grb150403A}
\end{minipage}
\end{figure*}

\begin{figure*}
\begin{minipage}[H]{1\textwidth}
    \includegraphics[width=0.33\linewidth]{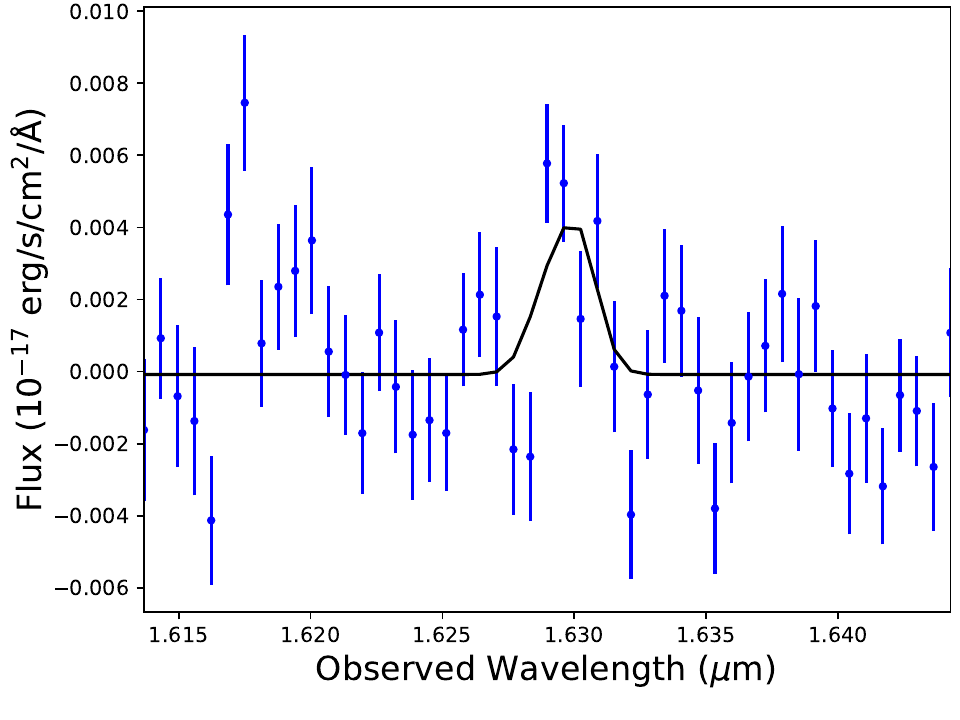}
    \includegraphics[width=0.33\linewidth]{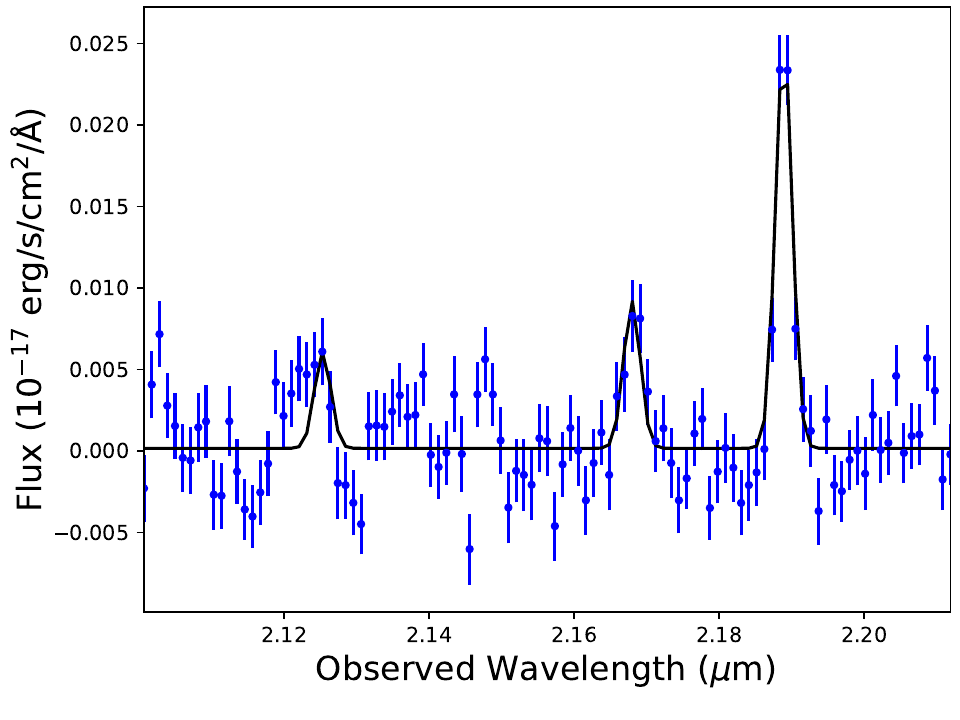}
    \includegraphics[width=0.33\linewidth]{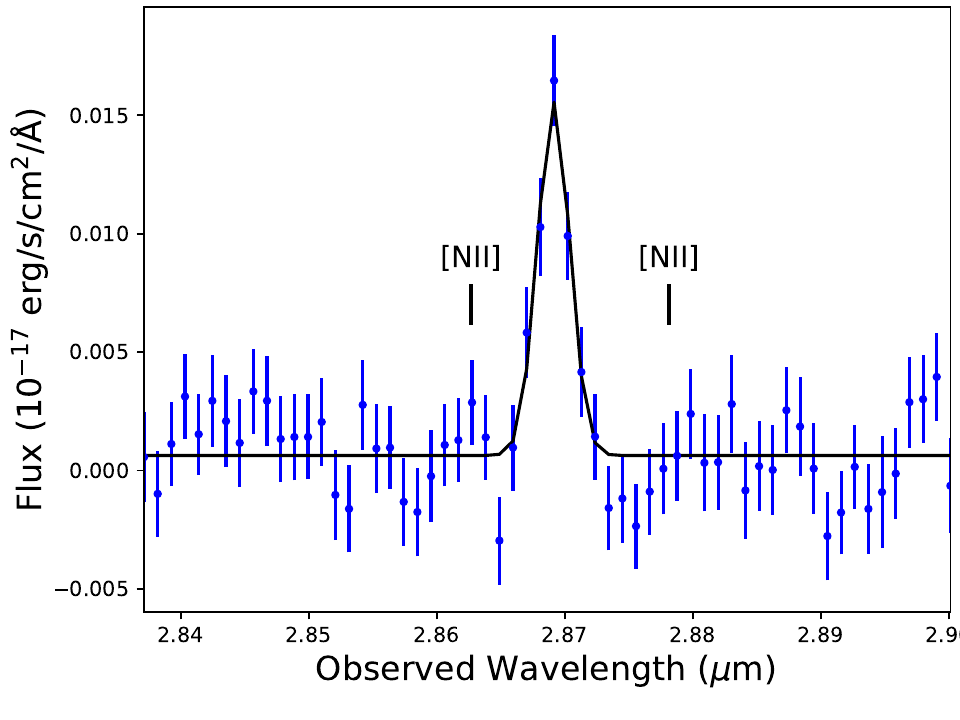}
    \caption{Spectrum of the host galaxy of GRB030323 (blue data points) zoomed in on the tentative \oii\ emission line doublet detection (left), \hb\ and \oiii\ emission lines (middle) and on \ha\ (right), with best-fit model overplotted (black line). The location of the undetected \nii\ line doublet is indicated in the right panel. The best-fit velocity dispersion and redshift fitted to \ha\ is $z=3.3710$ and $\sigma=80\pm 15$~km~s$^{-1}$, and these values were fixed in the fit to \hb, \oiii\ and \oii.}
    \label{fig:grb030323}
\end{minipage}
\end{figure*}

\begin{figure*}
\begin{minipage}[H]{1\textwidth}
    \includegraphics[width=0.33\linewidth]{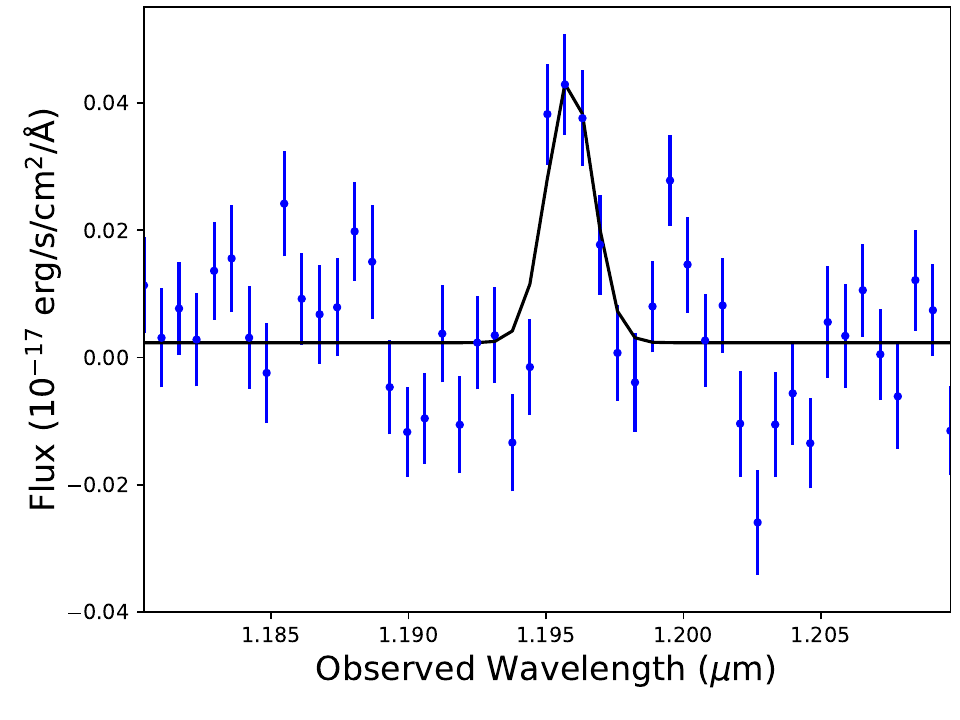}
    \includegraphics[width=0.33\linewidth]{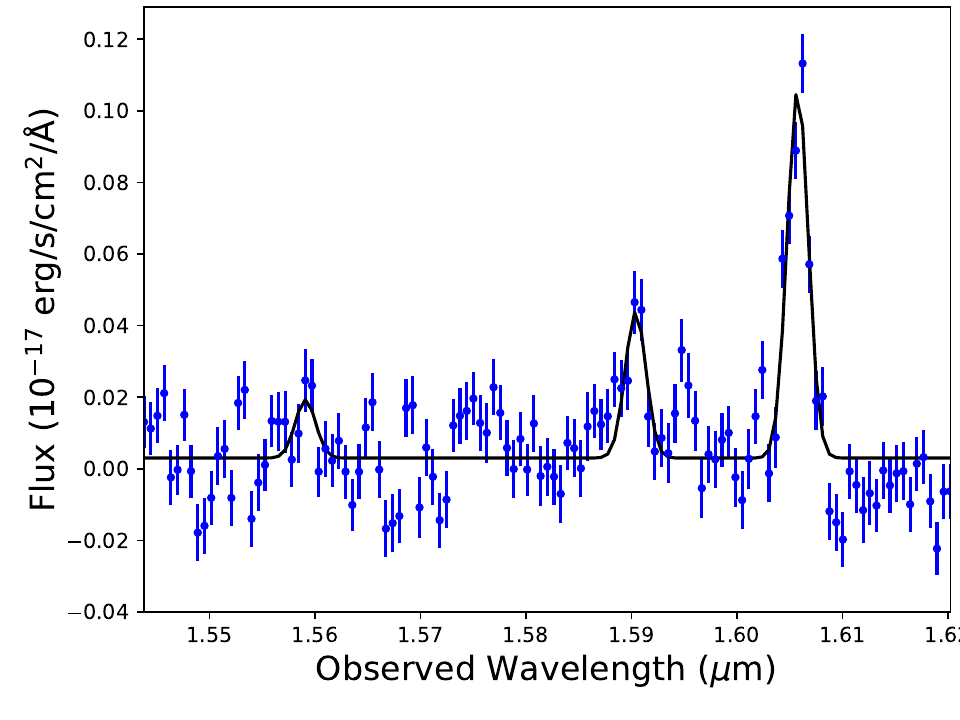}
    \includegraphics[width=0.33\linewidth]{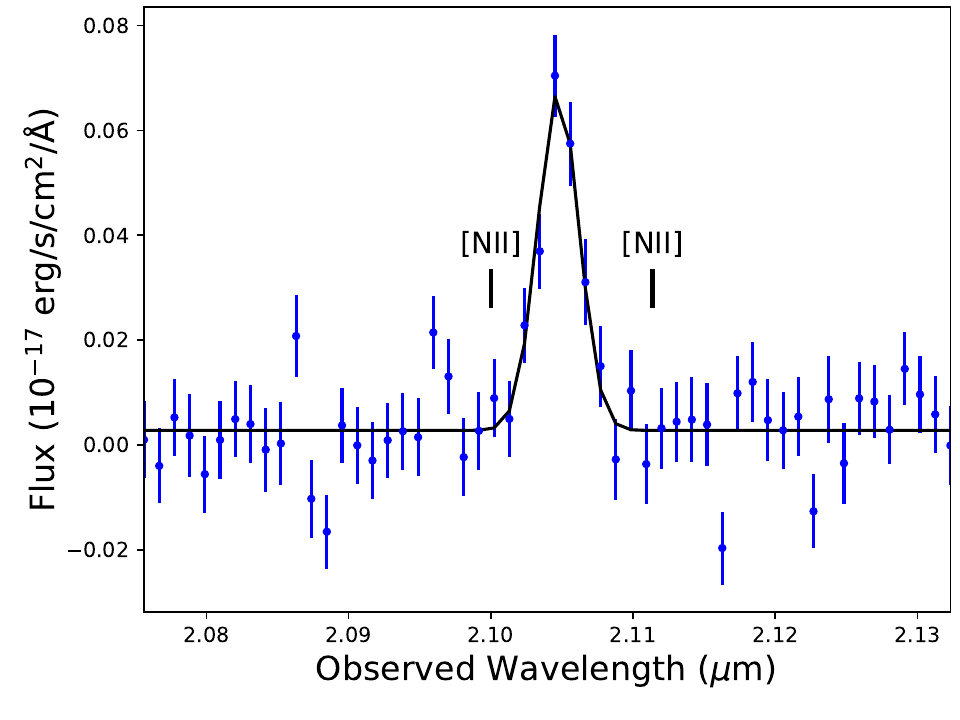}
    \caption{Detected emission lines (blue data points) from the host galaxy of GRB~080804 from \oii\ (left),  \hb\ and the \oiii\ doublet (middle), and \ha\ (right), with best-fit model (black line). The location of the undetected \nii\ line doublet is indicated in the right panel. The best-fit velocity dispersion and redshift from fits to \ha\ are $z=2.2065$ and $\sigma=148\pm23$~km~s$^{-1}$, and these best-fit parameters were frozen in the fits to \hb, \oiii\ and \oii.}
    \label{fig:grb080804}
\end{minipage}
\end{figure*}

\begin{figure*}
\begin{minipage}[H]{1\textwidth}
\centering
    \includegraphics[width=0.49\linewidth]{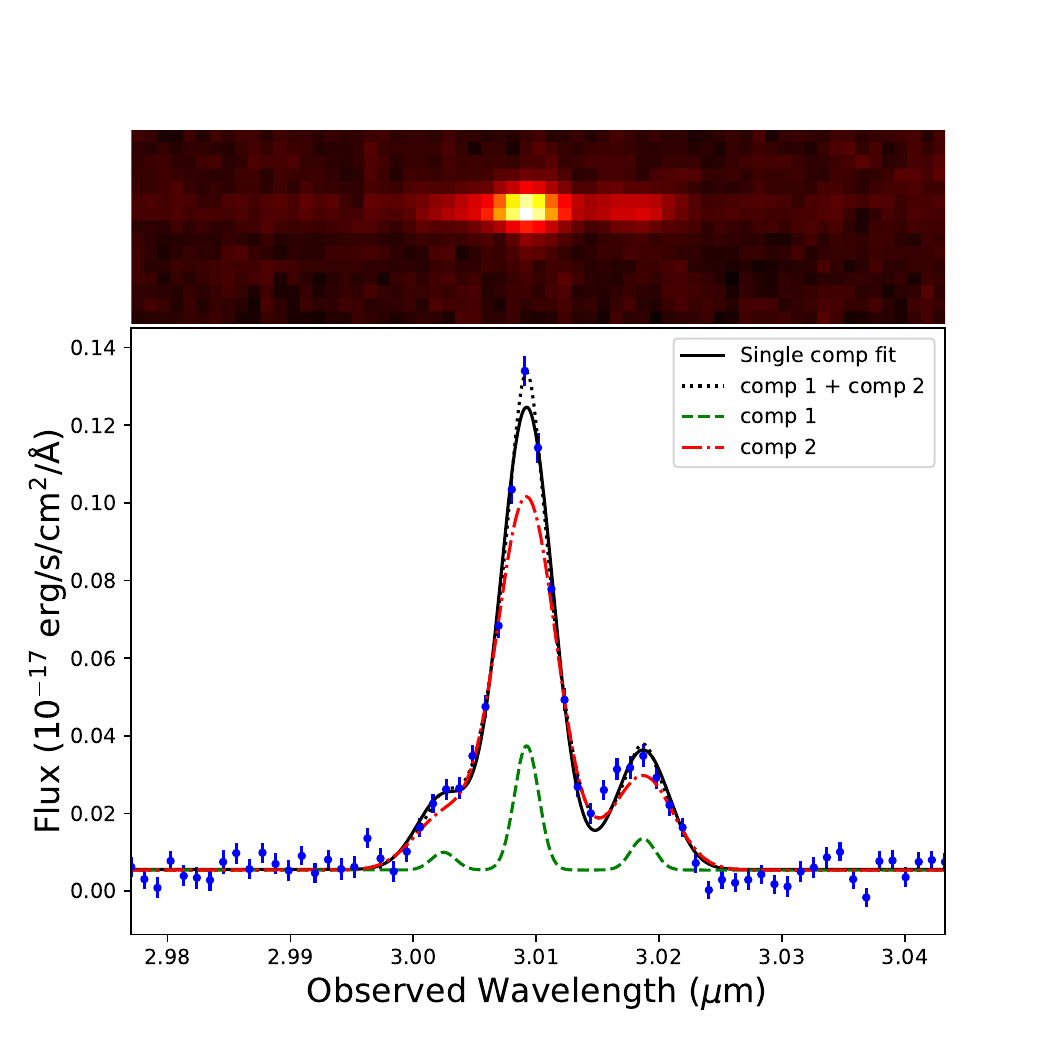}
    \caption{2D (top) and 1D (bottom) spectrum of the host galaxy of GRB~090323 zoomed in on the \ha\ and \nii\ lines and fitted with a single (black solid) and a two component model (black dotted), where in the latter case the two best-fit components are plotted with green dashed and red dot-dashed lines. In both fits the \nii\ and \ha\ lines were fitted simultaneously with the velocity dispersion of each component tied. The best-fit parameters for the single Gaussian fit (black solid) were $z=3.5844$ and $\sigma = 190\pm 5$~km~s$^{-1}$, and for the two component fit (black dotted) the best-fit redshift was unchanged but the velocity dispersion was $\sigma_1 = 221\pm 12$~km~s$^{-1}$ and $\sigma_2 <99$~km~s$^{-1}$ for the two respective components.
    }
    \label{fig:grb090323A-2Dspec}
\end{minipage}
\end{figure*}

\begin{figure*}
\begin{minipage}[H]{1\textwidth}
\centering
    \includegraphics[width=0.33\linewidth]{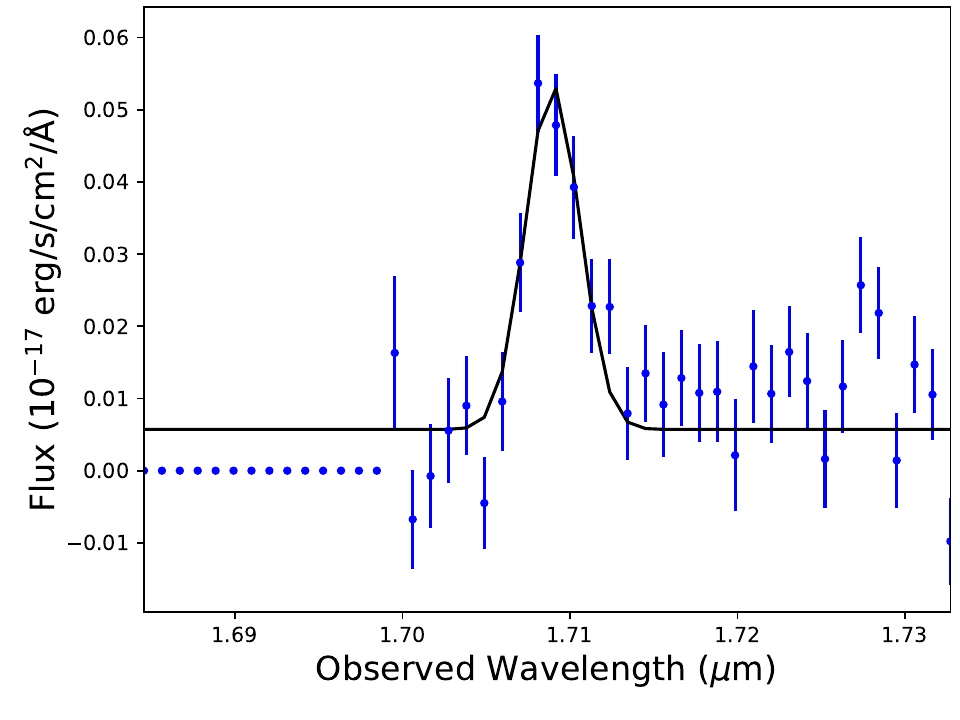}
    \includegraphics[width=0.33\linewidth]{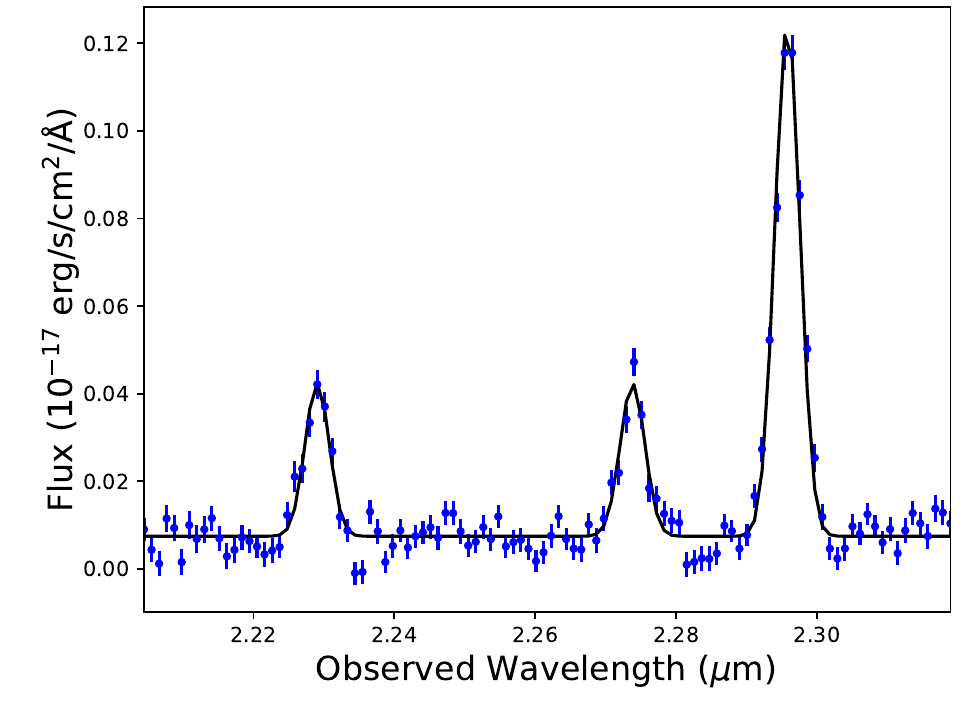}
    \includegraphics[width=0.33\linewidth]{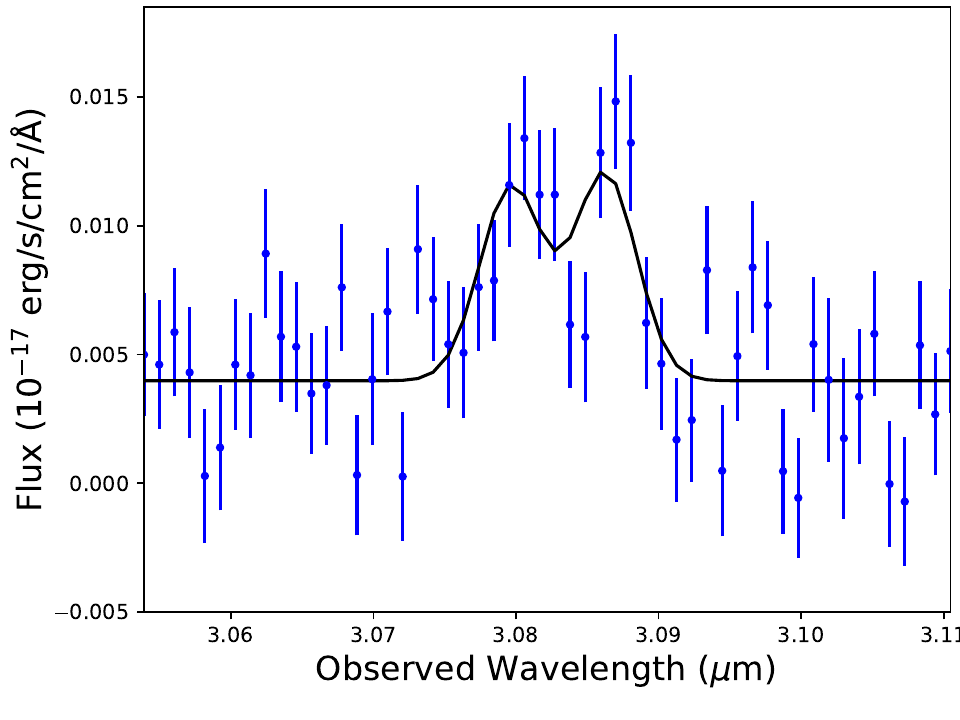}
    \caption{Emission line spectra of the host galaxy of GRB~090323, corresponding to \oii\ (left), \hb\ and \oiii\ (middle), and \sii\ (right). The line peak positions and velocity widths were kept fixed to the best-fit values fitted to the \ha\ and \nii\ lines (Fig.~\ref{fig:grb090323A-2Dspec}). Note, the data blueward of 1.71~$\mu$m shown in the fit to \oii\ all show zero flux because they lie below the effective lower-bound of the F170LP filter.}
    \label{fig:grb090323A}
\end{minipage}
\end{figure*}

\begin{figure*}
\begin{minipage}[H]{1\textwidth}
\centering
    \includegraphics[width=0.49\linewidth]{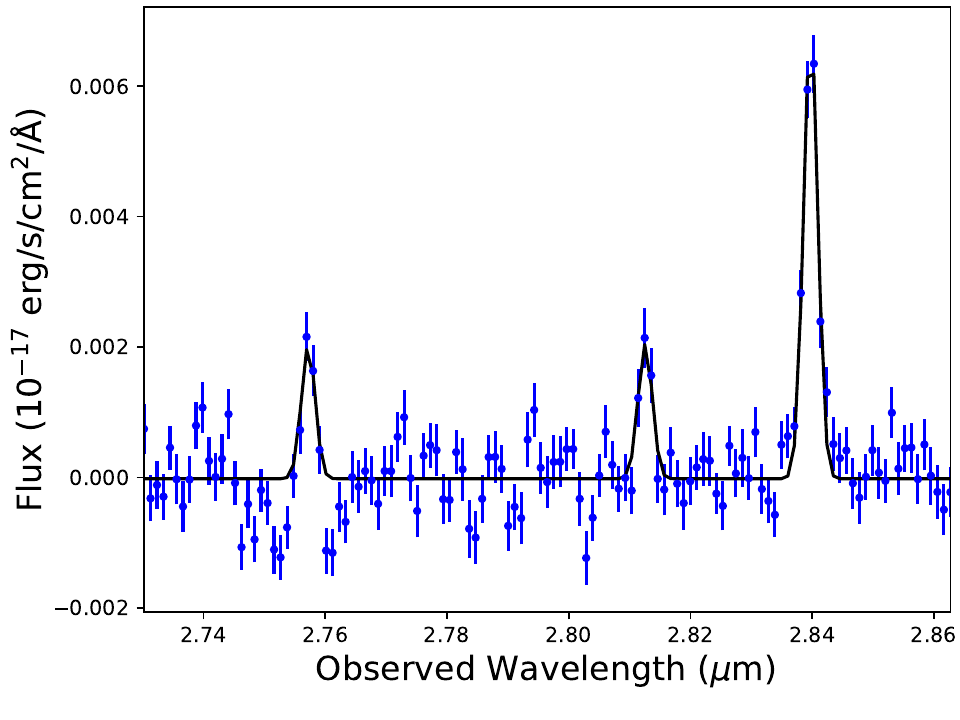}
    \includegraphics[width=0.49\linewidth]{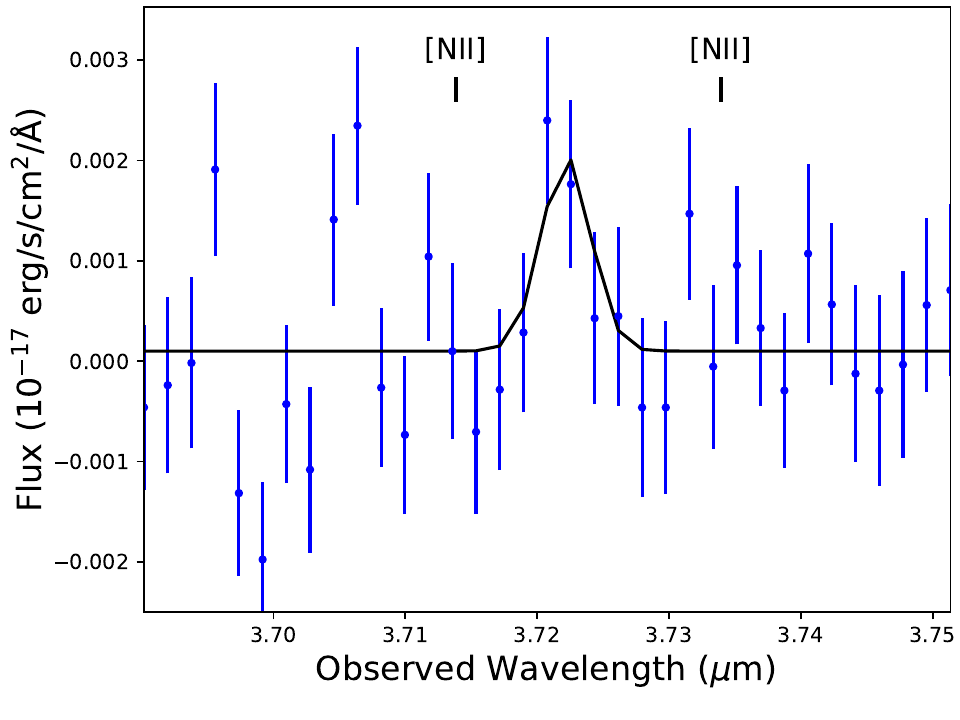}
    \caption{Emission line spectra of the host galaxy of GRB~100219A (blue data points) zoomed in on \hb\ and \oiii\ (left), and the tentative \ha\ emission line detection (right). The location of the undetected \nii\ line doublet is indicated in the right panel. The best-fit velocity dispersion and redshift from simultaneous fits to \hb\ and \oiii\ are $z=4.6698$ and $\sigma=66\pm 9$~km~s$^{-1}$. The line peak positions and velocity widths were kept fixed to the best-fit values fitted to the \hb\ and \oiii\ lines.}
    \label{fig:grb100219A}
\end{minipage}
\end{figure*}

\section{Stellar spectra of misidentified targets}
\begin{figure*}
\begin{minipage}[H]{1\textwidth}
    \includegraphics[width=0.5\linewidth]{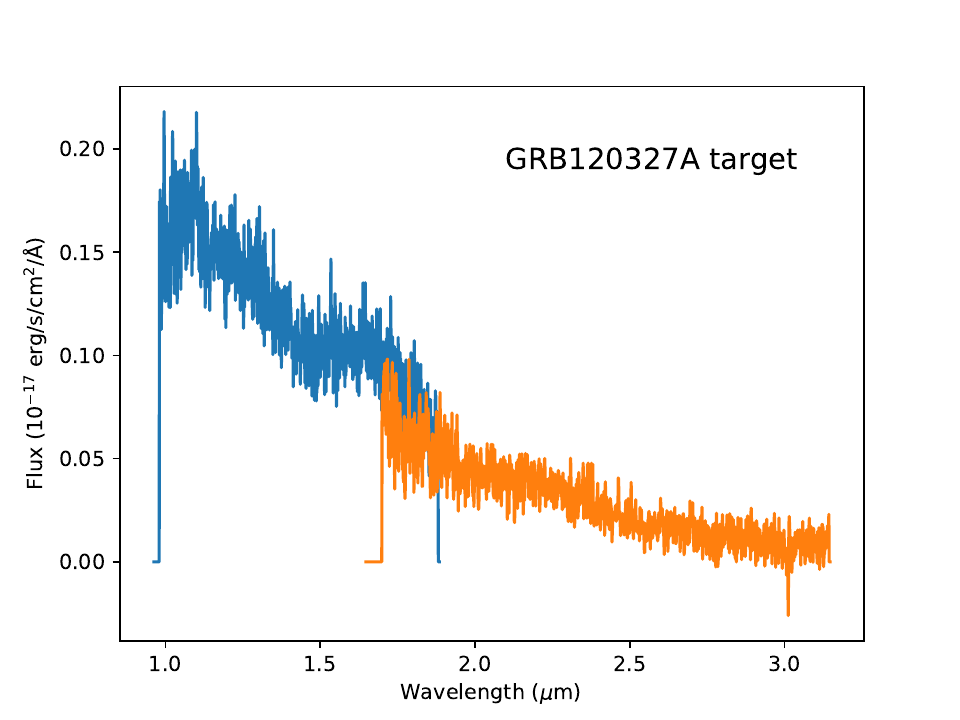}
    \includegraphics[width=0.5\linewidth]{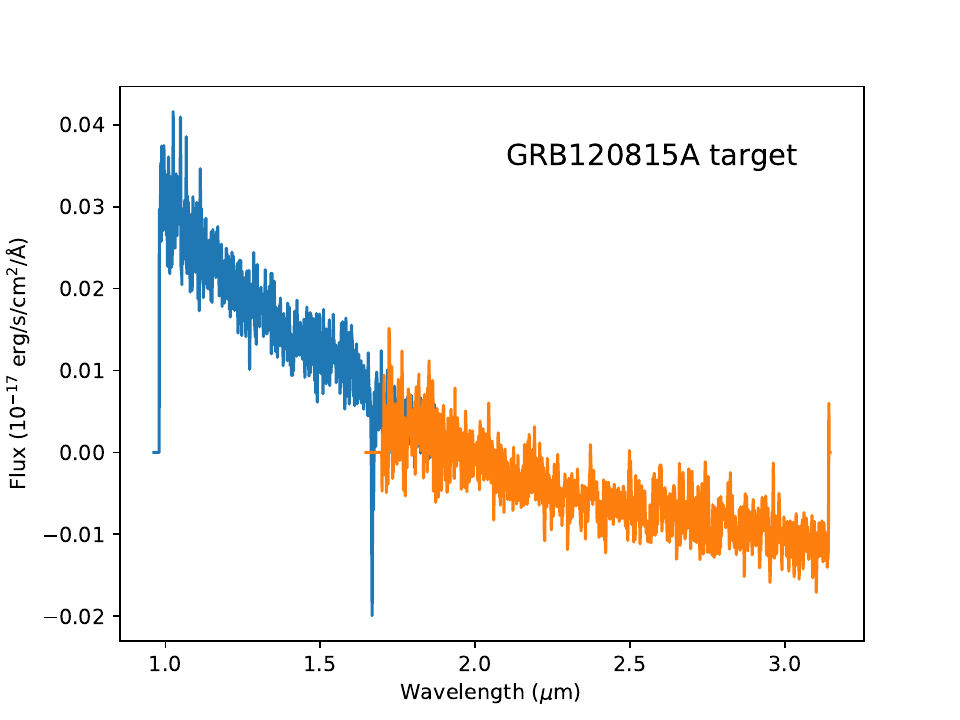}
    \caption{NIRspec spectra of the host galaxy candidates for GRB~120327A (left) and GRB~120815A (right). Both spectra have a continuum shape reminiscent of a blackbody, thus likey corresponding to unrelated foreground stars, with left having a peak temperature of $\sim 3500$~K, and the right-hand spectrum being consistent with a hotter star where the NIRSpec spectrum covers the Rayleigh-Jeans limits.}
    \label{fig:grb120327A-grb120815A_spec}
\end{minipage}
\end{figure*}

\section{QSO-DLA absorption and emission line metallicities}
\begin{table*}
\begin{center}
\begin{minipage}[H]{1\textwidth}
\caption{Absorption and emission line metallicities for QSO sample}\label{tab:QSOsample}
\end{minipage}
\begin{tabular}{|l|c|c|c|c|c|c|}
\hline
\hline
QSO & $z_{\rm abs}$ & \multicolumn{5}{c}{$12+\log({\rm O/H})$} \\
\cline{3-7}
 & & abs & NOX22 \Rtwothree & NOX \Rthree & LMC23 $\hat{R}$ & DKS16 \\
 \hline\hline
J0238+1636 & 0.5253 & $8.09\pm 0.40^a$ & $8.32\pm 0.01$ & $8.32\pm 0.02$ & $7.84\pm 0.05$ & \ldots \\	       
J0441-4313 & 0.1010 & $8.79\pm 0.15^a$ & $8.64\pm 0.02$ & $8.76\pm 0.01$ & $8.82\pm 0.01$ & $8.77\pm 0.01$ \\
J0830+2410 & 0.5263 & $8.20\pm 0.30^a$ & $8.22\pm 0.03$ & $8.02\pm 0.08$ & \ldots & \ldots \\		       
J0918+1636 & 2.583 &  $8.57\pm 0.05^b$ & $8.39\pm 0.07$ & $8.48\pm 0.07$ & $8.56\pm 0.07$ & \ldots \\	       
J0958+0549 & 0.6546 & $7.36\pm 0.23^c$ & $8.27\pm 0.04$ & $8.39\pm 0.02$ & $7.59\pm 0.05$ & \ldots \\	       
J1138+0139 & 0.6126 & $7.91\pm 0.16^c$ & $8.59\pm 0.15$ & $8.42\pm 0.06$ & $8.12\pm 0.12$ & \ldots \\	       
J1204+0953 & 0.6390 & $7.97\pm 0.16^c$ & \ldots & \ldots & $7.73\pm 0.29$ & \ldots \\	 		       
J1436-0051 & 0.7390 & $8.64\pm 0.12^d$ & $8.54\pm 0.03$ & $8.73\pm 0.02$ & $8.81\pm 0.02$ & \ldots \\	       
J1544+5912 & 0.0102 & $8.19\pm 0.33^e$ & $8.33\pm 0.04$ & $8.28\pm 0.08$ & $8.24\pm 0.13$ & $7.91\pm 0.17$ \\
J2222-0946 & 2.354 & $8.20\pm 0.05^{f,g}$ & $8.03\pm 0.05$ & $8.03\pm 0.09$ & $8.12\pm 0.07$ & \ldots \\
J2247-6015 & 2.33 & $7.97\pm 0.05^h$ & $8.38\pm 0.07$ & $8.20\pm 0.13$ & $8.12\pm 0.03$ & \ldots \\
\hline

\end{tabular}
\begin{minipage}[H]{0.90\textwidth}
$^\dag$ Absorption-based metallicity relative to solar and corrected for dust depletion. To convert to units of $\mbox{[M/H]}$, more commonly used in GRB absorption line studies, need to subtract the solar metallicity value 12+log(O/H)=8.69 \citep{ags+09}. \\
References: $^a$ \citet{ckr+05}; $^b$ \citet{fgc+13}; $^c$ \citet{rpt+16}; $^d$ \citet{sjy+16} ; $^e$ \citet{srd+04};\\$^f$ \citet{fll+10}; $^g$ \citet{kfl+13}; $^h$ \citet{bmk+13} 
\end{minipage}
\end{center}
\end{table*}

\section{Absorption line versus SST23 emission line metallicities}

\begin{table*}
\begin{center}
\begin{minipage}[H]{1\textwidth}
\caption{SST23 \Rtwothree, \Rthree, \Rtwo\ and \Othreetwo\ emission line metallicities for GRB host galaxy sample.}\label{tab:sst23metals}
\end{minipage}
\begin{tabular}{|l|c|c|c|c|}
\hline
\hline
GRB host & \multicolumn{4}{c}{$12+\log{\rm (O/H)}$}\\
\cline{2-5}
& \Rtwothree & \Rthree & \Rtwo & \Othreetwo \\
\hline\hline
030323 &	 $7.48\pm 0.45$ & $7.50\pm 0.42$ & $7.84\pm 0.23$ & $8.00\pm 0.16$ \\
050820A & 	 $7.93\pm 0.27$ & $7.87\pm 0.45$ & $8.32\pm 0.06$ & $8.32\pm 0.04$ \\
~~~component A & $7.90\pm 0.35$ & $7.86\pm 0.48$ & $8.27\pm 0.06$ & $8.27\pm 0.04$ \\
~~~component B & $7.80\pm 0.23$ & $7.68\pm 0.29$ & $8.29\pm 0.11$ & $8.29\pm 0.10$ \\
~~~component C & $7.80\pm 0.23$ & $7.82\pm 0.45$ & $8.30\pm 0.10$ & $8.29\pm 0.07$ \\
080804 &         $7.61\pm 0.42$ & $7.51\pm 0.41$ & $8.24\pm 0.17$ & $8.29\pm 0.05$ \\
090323 &         $7.22\pm 0.10$ & $7.09\pm 0.05$ & $8.16\pm 0.14$ & $8.35\pm 0.07$ \\
100219A & $7.10\pm 0.10$ & $7.12\pm 0.12$ & $7.60\pm 0.09$ & $7.85\pm 0.05$ \\
150403A &        $7.39\pm 0.21$ & $7.22\pm 0.11$ & $8.16\pm 0.23$ & $8.28\pm 0.13$ \\
~~~component A & $7.14\pm 0.14$ & $7.10\pm 0.10$ & $8.01\pm 0.12$ & $8.26\pm 0.10$ \\
~~~component B & $7.29\pm 0.17$ & $7.14\pm 0.12$ & $8.20\pm 0.09$ & $8.38\pm 0.03$ \\
~~~component C & $7.74\pm 0.30$ & $7.63\pm 0.29$ & $8.21\pm 0.20$ & $8.21\pm 0.15$ \\
\hline		                                                                     
120815A &	 $7.60\pm 0.44$ & $7.51\pm 0.41$ & $7.96\pm 0.46$ & $8.14\pm 0.27$ \\
121024A & 	 $7.33\pm 0.08$ & $7.27\pm 0.07$ & $7.92\pm 0.06$ & $8.04\pm 0.04$ \\
\hline

\end{tabular}
\begin{minipage}[H]{0.90\textwidth} 
\end{minipage}
\end{center}
\end{table*}

\begin{figure*}
\begin{minipage}[H]{1\textwidth}
    \includegraphics[width=0.5\linewidth]{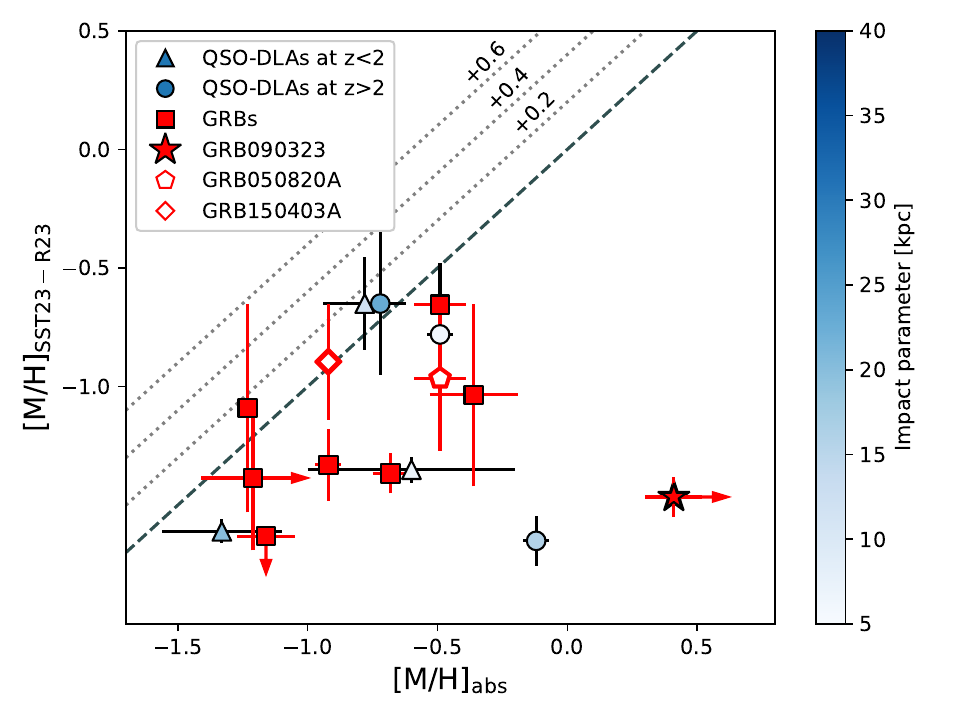}
    \includegraphics[width=0.5\linewidth]{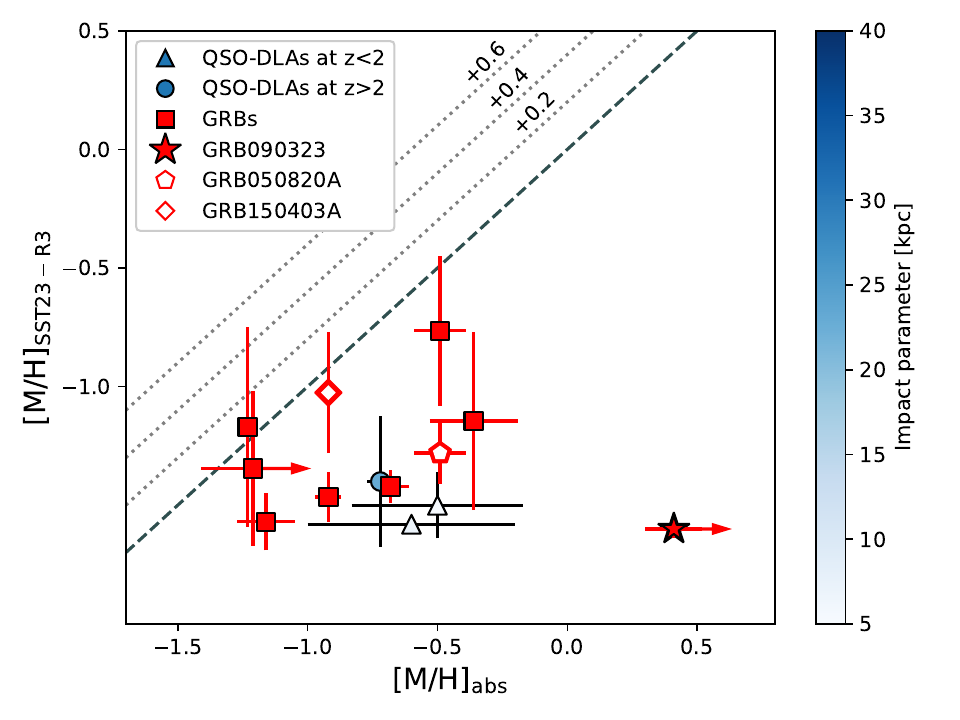}
    \includegraphics[width=0.5\linewidth]{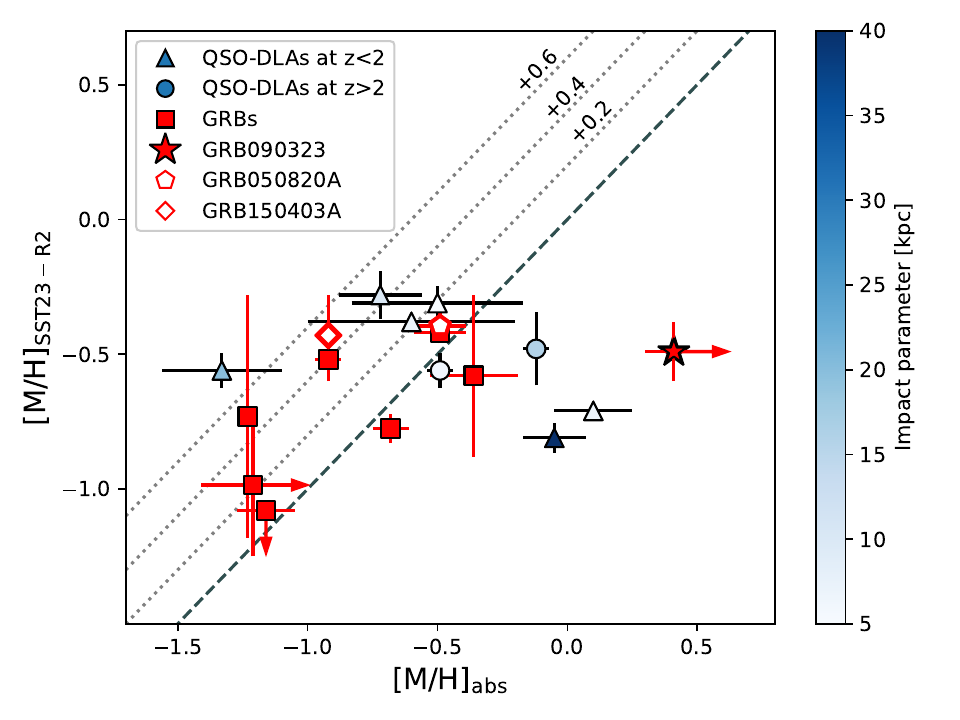}
    \includegraphics[width=0.5\linewidth]{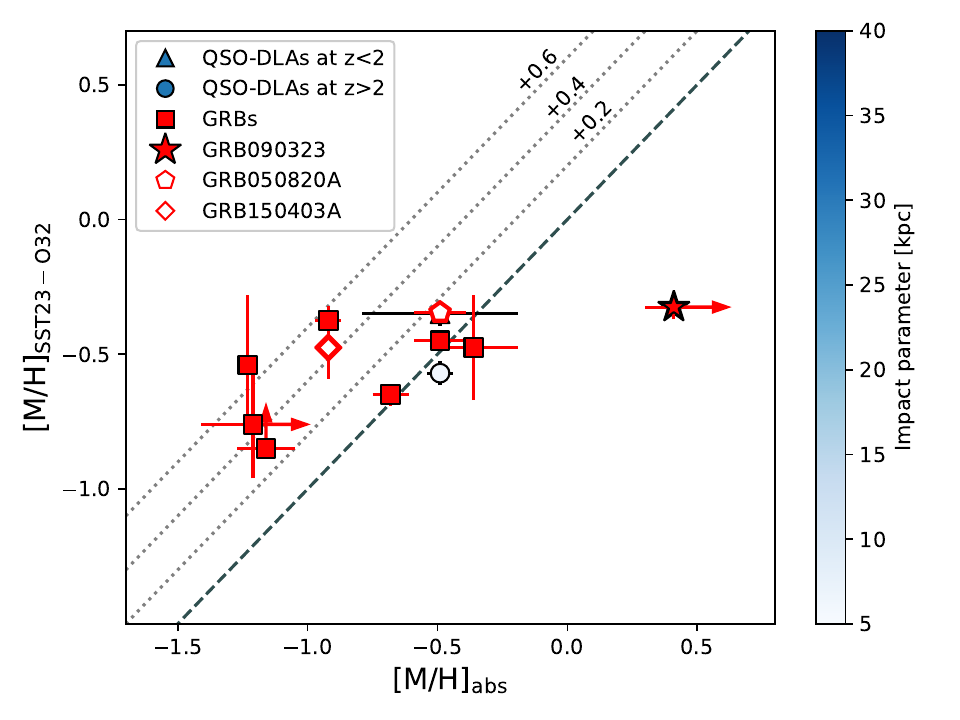}
    \caption{Similar to Figs.~\ref{fig:ZabsvsZem-NOX22} and \ref{fig:ZabsvsZem-LMC23} but now for the SST23 \Rtwothree\ (top left), \Rthree\ (top right), \Rtwo\ (bottom left) and \Othreetwo\ (bottom right) emission line metallicities.}
    \label{fig:ZabsvsZem-SST23}
\end{minipage}
\end{figure*}

\end{document}